\documentclass[%
 reprint,
superscriptaddress,   
 amsmath,amssymb,
 aps,     
prb,
]{revtex4-2}

\usepackage{graphicx}
\usepackage{dcolumn}
\usepackage{bm}
\usepackage{braket}  
\usepackage{hyperref}
\hypersetup{hypertex=true,
            colorlinks=true,
            linkcolor=blue,
            anchorcolor=blue,
            citecolor=blue}
\usepackage{lineno}

\begin{document}


\title{Experimental Implementation of Short-Path Non-adiabatic Geometric Gates in a Superconducting Circuit}

\author{Xin-Xin Yang}
\affiliation{CAS Key Laboratory of Quantum Information, University of Science and Technology of China, Hefei, Anhui 230026, China}
\affiliation{CAS Center for Excellence in Quantum Information and Quantum Physics, University of Science and Technology of China, Hefei, Anhui 230026, China}

\author{Liang-Liang Guo}
\affiliation{CAS Key Laboratory of Quantum Information, University of Science and Technology of China, Hefei, Anhui 230026, China}
\affiliation{CAS Center for Excellence in Quantum Information and Quantum Physics, University of Science and Technology of China, Hefei, Anhui 230026, China}

\author{Hai-Feng Zhang}
\affiliation{CAS Key Laboratory of Quantum Information, University of Science and Technology of China, Hefei, Anhui 230026, China}
\affiliation{CAS Center for Excellence in Quantum Information and Quantum Physics, University of Science and Technology of China, Hefei, Anhui 230026, China}

\author{Lei Du}
\affiliation{CAS Key Laboratory of Quantum Information, University of Science and Technology of China, Hefei, Anhui 230026, China}
\affiliation{CAS Center for Excellence in Quantum Information and Quantum Physics, University of Science and Technology of China, Hefei, Anhui 230026, China}

\author{Chi Zhang}
\affiliation{CAS Key Laboratory of Quantum Information, University of Science and Technology of China, Hefei, Anhui 230026, China}
\affiliation{CAS Center for Excellence in Quantum Information and Quantum Physics, University of Science and Technology of China, Hefei, Anhui 230026, China}

\author{Hao-Ran Tao}
\affiliation{CAS Key Laboratory of Quantum Information, University of Science and Technology of China, Hefei, Anhui 230026, China}
\affiliation{CAS Center for Excellence in Quantum Information and Quantum Physics, University of Science and Technology of China, Hefei, Anhui 230026, China}

\author{Yong Chen}
\affiliation{CAS Key Laboratory of Quantum Information, University of Science and Technology of China, Hefei, Anhui 230026, China}
\affiliation{CAS Center for Excellence in Quantum Information and Quantum Physics, University of Science and Technology of China, Hefei, Anhui 230026, China}

\author{Peng Duan}
\affiliation{CAS Key Laboratory of Quantum Information, University of Science and Technology of China, Hefei, Anhui 230026, China}
\affiliation{CAS Center for Excellence in Quantum Information and Quantum Physics, University of Science and Technology of China, Hefei, Anhui 230026, China}

\author{Zhi-Long Jia}
\affiliation{CAS Key Laboratory of Quantum Information, University of Science and Technology of China, Hefei, Anhui 230026, China}
\affiliation{CAS Center for Excellence in Quantum Information and Quantum Physics, University of Science and Technology of China, Hefei, Anhui 230026, China}

\author{Wei-Cheng Kong}
\affiliation{Origin Quantum Computing Company Limited, Hefei, Anhui 230088, China}

\author{Guo-Ping Guo}
\email{gpguo@ustc.edu.cn}
\affiliation{CAS Key Laboratory of Quantum Information, University of Science and Technology of China, Hefei, Anhui 230026, China}
\affiliation{CAS Center for Excellence in Quantum Information and Quantum Physics, University of Science and Technology of China, Hefei, Anhui 230026, China}
\affiliation{Origin Quantum Computing Company Limited, Hefei, Anhui 230088, China}

\date{\today}

\begin{abstract}
 The non-adiabatic geometric quantum computation (NGQC) has attracted a lot of attention for noise-resilient quantum control. However, previous implementations of NGQC require long evolution paths that make them more vulnerable to incoherent errors than their dynamical counterparts. In this work, we experimentally realize a universal short-path non-adiabatic geometric gate set (SP-NGQC) with a 2-times shorter evolution path on a superconducting quantum processor. Characterizing with both quantum process tomography and randomized benchmarking methods, we report an average single-qubit gate fidelity of $99.86\%$ and a two-qubit gate fidelity of $97.9\%$. Additionally, we demonstrate superior robustness of single-qubit SP-NGQC gate to Rabi frequency error in some certain parameter space by comparing their performance to those of the dynamical gates and the former NGQC gates.
\end{abstract}


\maketitle

Quantum computation now is entering the ‘noisy intermediate-scale quantum’ (NISQ) technology era {\cite{Preskill2018quantumcomputingin}}, with the fact that quantum processors are susceptible to environmental fluctuations and operational imperfections. To realize quantum logic surpassing the fault-tolerance threshold for large-scale quantum computation {\cite{548464,PhysRevLett.98.190504,PhysRevA.86.032324}}, a universal set of quantum gates, including arbitrary single-qubit gates and a non-trivial two-qubit gate {\cite{PhysRevLett.75.346,PhysRevLett.89.247902}}, is in great request with not only high gate fidelity but also robustness to ambient noise.

Recently, close attention is paid to the geometric phase due to its intrinsic noise-resilience features {\cite{doi:10.1098/rspa.1984.0023, PhysRevLett.52.2111,PhysRevLett.58.1593,ANANDAN1988171}}. Unlike the dynamical phase that comes from the time integral of energy, the geometric phase only depends on the evolution path and is immune to any deviation that does not change the enclosed area by the path, which suggests it is noise-resilient to certain types of errors {\cite{PhysRevLett.91.090404,PhysRevLett.92.020402,PhysRevA.72.020301,doi:10.1126/science.1149858,PhysRevLett.102.030404,PhysRevA.84.042335,PhysRevA.86.062322,PhysRevA.87.060303}}. With adiabatic cyclic evolutions, geometric {\cite{PhysRevA.87.032326,PhysRevLett.122.010503}} and holonomic {\cite{ZANARDI199994,jones2000geometric,doi:10.1126/science.1058835,PhysRevLett.95.130501,PhysRevA.87.052307,leroux2018non}} quantum gates using the pure geometric phases were the first ones demonstrated in physical systems. However, the long run time required by the adiabatic evolution makes quantum gates vulnerable to considerable environment-induced decoherence. Although some transitionless quantum driving algorithms have been put forward to speed up the adiabatic loops, an almost adiabatic process will also introduce unwanted control errors {\cite{bason2012high,Song_2016,zhang2015fast,zhou2017accelerated,kleissler2018universal,PhysRevLett.122.080501,PhysRevApplied.13.064012,doi:10.1063/5.0049967}}.

To overcome these drawbacks, non-adiabatic holonomic quantum computation (NHQC) based on the non-Abelian geometric phase {\cite{ANANDAN1988171,abdumalikov2013experimental,PhysRevLett.121.110501,Zhang_2019,PhysRevApplied.16.064003,PhysRevLett.110.190501,zu2014experimental,arroyo2014room,song2017continuous,sekiguchi2017optical,PhysRevLett.119.140503,nagata2018universal,Ishida.18,PhysRevApplied.11.014017,PhysRevApplied.14.054062}}, and further the non-adiabatic geometric quantum computation (NGQC) based on the Abelian geometric phase were demonstrated {\cite{PhysRevLett.58.1593,leibfried2003experimental,PhysRevLett.124.230503,zhao2021experimental}}. For the implementations of NHQC in superconducting transmon qubits {\cite{abdumalikov2013experimental,PhysRevLett.121.110501,Zhang_2019,PhysRevApplied.16.064003}}, they involve the lowest three energy levels. However, the relatively short coherence time and the small anharmonicity of the transmon qubits cause extra decoherence and leakage errors {\cite{PhysRevA.76.042319,barends2014superconducting}}. 

In contrast, for NGQC, the non-adiabatic Abelian geometric phases are generated by the cyclic evolution of quantum states with the removal of the dynamic phase. In other words, NGQC only involves two-level computational space with commonly used manipulations and retains the merits of robustness to noises. Lately, NGQC with a so-called orange-slice-shaped evolution path (labeled as NGQC in {$\rm{Fig.\ \ref{sample}(b)}$}) is demonstrated in superconducting circuits {\cite{PhysRevLett.124.230503,zhao2021experimental}}. However, such a long evolution path is time-wasting and exposes the qubit to more decoherent errors. To resolve this problem, some new NGQC schemes such as noncyclic evolutions {\cite{PhysRevResearch.2.043130,PhysRevApplied.14.064009, PhysRevLett.127.030502, https://doi.org/10.1002/qute.202100019}} or optimized evolution paths {\cite{PhysRevResearch.2.023295,Ding_2021,doi:10.1063/5.0071569}} have been put forward with shorter evolution paths. Nevertheless, these proposals require careful design and precise parameter control of the system’s Hamiltonian, which makes the experimental realization of short-path NGQC gates more challenging.

Here, we experimentally demonstrate a short-path scheme of NGQC (SP-NGQC) in a couple-fixed superconducting chip with a half-orange-slice-shaped evolution loop (labeled as SP-NGQC in {$\rm{Fig.\ \ref{sample}(b)}$}) that still satisfies the cyclic evolution and parallel transport conditions \cite{li2021high}. Simple and controllable all-microwave manipulations are used to realize the universal geometric quantum gates so that there’s no need to consider the versatile pulse distortions. In our experiment, we demonstrate some specific single-qubit non-adiabatic geometric gates with an average fidelity of $99.86(1)\%$ and the two-qubit non-adiabatic CZ gate with a fidelity of $97.9(3)\%$, which shows a convincing path to reliable and robust universal geometric quantum computation. Furthermore, we also investigate the noise-resilient feature of our short path geometric gates in contrast with dynamical gates and previous orange-slice-shaped NGQC gates. Especially, our scheme shows better performance to Rabi frequency error when the rotation angle is large or the rotating axis is close to the z-axis. 

Before specifying the experimental details, we first give an outline of constructing single-qubit SP-NGQC gates. Conventionally, we consider a general setup that a two-level qubit is driven by a classical microwave field. Hereafter, $\hbar$ is set equal to 1. In the interaction picture, the Hamiltonian under the rotating wave approximations gets
\begin{align} \label{singleH}
H(t)=\frac{1}{2}
\begin{pmatrix}
   \Delta(t)  && \Omega(t) e^{-i\varphi(t)}\\
   \Omega(t) e^{i\varphi(t)} && -\Delta(t) 
\end{pmatrix}
\end{align}
where $\Delta(t)=\omega_q-\omega(t)$ is the frequency difference between the drive $\omega(t)$ and the qubit $\omega_q$, $\Omega(t)$ is called the Rabi frequency which can be tuned by driving amplitude and $\varphi(t)$ is the phase of the drive. To realize a non-adiabatic single-qubit gate set through a short single evolution loop, we divide the evolution period $T$ into four intervals. In each interval, the microwave field has different amplitude and phases as follows to satisfy the cyclic evolution condition:
\begin{figure}[b]
\includegraphics{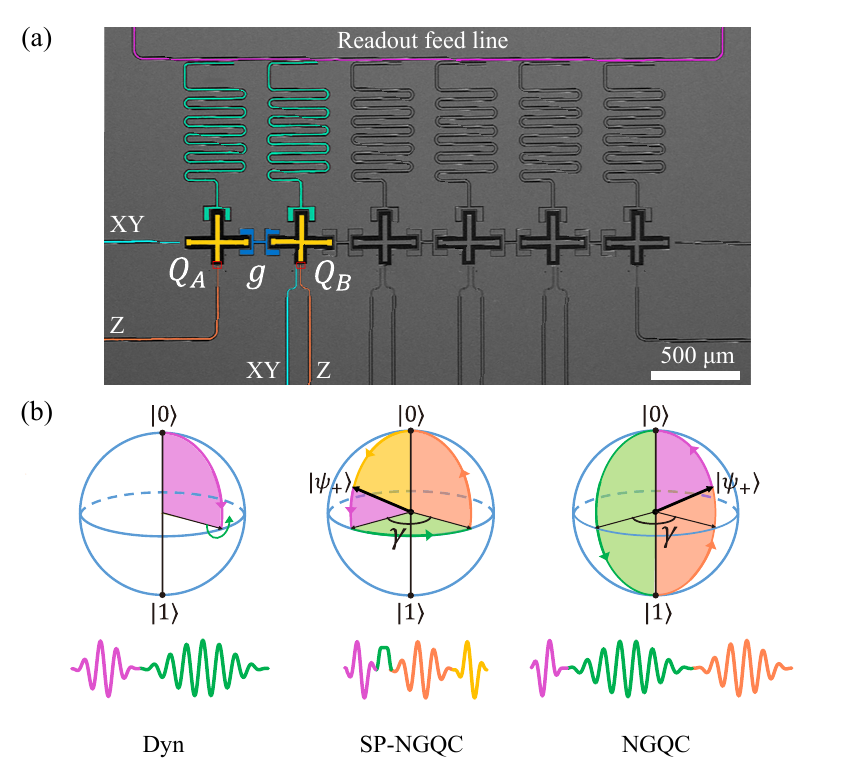}
\caption{\label{sample}(a) Electron microscope image of the six-qubits quantum processor. The first two capacitive-coupled qubits, $Q_A$ and $Q_B$ (false-colored), are used in this work. (b) Bloch sphere representations of the evolutionary trajectories to realize single-qubit dynamical and geometric gates and example pulse shapes for Hadamard gate. The dynamical Hadamard gate is implemented with a dynamical $Y/2$ rotation followed by a dynamical $X$ rotation.} 
\end{figure}
\begin{subequations}\label{3}
\begin{align}
&\int^{\tau_1}_{0}\Omega(t) dt=\frac{\pi}{2}-\theta,\quad \Delta(t)=0, \quad \varphi=\phi+\frac{\pi}{2},\quad t\in[0, \tau_1)  \\
\label{3b}
&\Omega(t)=0,\quad \int^{\tau_2}_{\tau_1}\Delta(t) dt=\gamma,\quad \varphi=\phi+\gamma-\frac{\pi}{2},\quad t\in[\tau_1, \tau_2)  \\   
&\int^{\tau_3}_{\tau_2}\Omega(t) dt=\frac{\pi}{2},\quad \Delta(t)=0, \quad \varphi=\phi+\gamma-\frac{\pi}{2},\quad t\in[\tau_2, \tau_3)  \\
\label{3d}
&\int^{T}_{\tau_3}\Omega(t) dt=\theta,\quad \Delta(t)=0, \quad \varphi=\phi+\frac{\pi}{2},\quad t\in[\tau_3, T].
\end{align}
\end{subequations}
The final evolution operator can be obtained as 
\begin{equation}\label{evolu}
\begin{aligned}
U(T) &= \cos\frac{\gamma}{2} -i\sin\frac{\gamma}{2}
\begin{pmatrix}
   \cos\theta  &&  \sin\theta e^{-i\phi}\\
   \sin\theta e^{i\phi}  && -\cos\theta
\end{pmatrix} \\
&= e^{-i\frac{\gamma}{2} \boldsymbol{n} \cdot \boldsymbol{\sigma}}.
\end{aligned}
\end{equation}  
The operator $U(T)$ represents rotation operations around the axis $\boldsymbol{n}=(\sin\theta \cos \phi, \sin\theta \sin\phi, \cos\theta)$ with an angle $\gamma$, where $\boldsymbol{\sigma}=(\sigma_x, \sigma_y, \sigma_z)$ are the Pauli operators. $\theta, \;\phi, \;\gamma$ are determined by the drive. Following the evolution, two orthogonal eigenstates $\ket{\psi_+}=\cos\frac{\theta}{2}\ket{0}+\sin\frac{\theta}{2}e^{i\phi}\ket{1}$ and $\ket{\psi_-}=\sin\frac{\theta}{2}e^{-i\phi}\ket{0}-\cos\frac{\theta}{2}\ket{1}$ of $U(T)$ undergo a cyclic half-orange-slice-shaped path with an enclosed solid angle equal to the rotation angle $\gamma$ (see the Bloch sphere labeled with SP-NGQC in {$\rm{Fig.\ \ref{sample}(b))}$}, resulting in a geometric phase $-\gamma/2$ ($\gamma/2$) on the quantum state $\ket{\psi_+}$ ($\ket{\psi_-}$) (Detailed calculation can be found in Appendix $\ref{geometry}$). A comparison of the gate time between the conventional dynamical gate (Dyn), the NGQC gate, and the SP-NGQC gate is shown in {$\rm{Fig.\ \ref{sample}(b)}$}. It can be seen that the SP-NGQC gate has a relatively short evolution path and the gate time can be further shortened to twice smaller than NGQC scheme without the green part by applying virtual Z gates {\cite{PhysRevA.96.022330}}. 

Our experiment is performed in a six-transmon-qubit-chain device {\cite{PhysRevApplied.16.024063}}. The two adjacent qubits used in this experiment and their coupling capacitance are shown in {$\rm{Fig.\ \ref{sample}(a)}$}. Each qubit equips a microwave line for driving, a flux bias line to tune the frequency, and a $\lambda/4$ resonator for individual and simultaneous readout. The transition frequency of $Q_A$ and $Q_B$ are $\omega_A/2\pi=5.511 \;{\rm GHz}$ and $\omega_B/2\pi=5.001 \;{\rm GHz}$, respectively. Moreover, the anharmonicities of the qubit are $\alpha_A/2\pi=-242.6 \;{\rm MHz}$ and $\alpha_B/2\pi=-250.0 \;{\rm MHz}$, respectively, ensuring a well-defined two-level system to encode the qubits. The fixed capacitive coupling strength $g_{\rm AB}/2\pi$ between two qubits is about 10 MHz. More details about the device parameters and the measuring circuits can be found in the Appendix {\ref{sec:level1}}.
\begin{figure}[htb]
\includegraphics{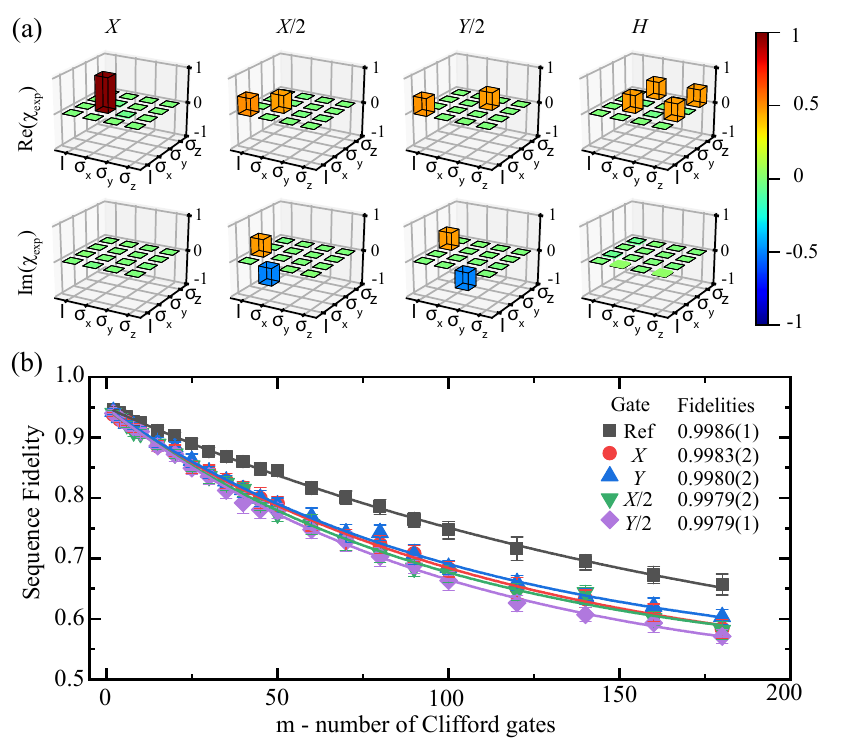}
\caption{\label{single}Characterization of the single-qubit SP-NGQC gates. (a) Bar charts of the real and imaginary part of $\chi_{\rm exp}$. Four specific gates: $X$, $X/2$, $Y/2$, and $H$ are shown with process fidelities of $99.5(1)\%$, $99.5(2)\%$, $98.8(4)\%$, and $99.0(5)\%$, respectively. The solid black outlines are $\chi_{\rm ideal}$ for the ideal gates. (b) RB of single-qubit SP-NGQC gates. Fitting the reference RB and interleaved RB decay curves gives the gate fidelity of four specific gates: $X$, $Y$, $X/2$, and $Y/2$.}
\end{figure}

The single-qubit SP-NGQC gates were performed on $Q_A$. To show the ability to construct a universal single-qubit gate set, we implement Hadamard gate ($H$), $\pi$, and $\pi/2$ rotations around both X and Y axes (denoted as $X$, $X/2$, $Y$, $Y/2$, respectively). Taking advantage of the virtual Z gate, the geometric gate is a composite of three rotations with a rotation angle no more than $\pi/2$. For simplicity, we fix the single interval's period $\tau$ to 20 ns with 5 ns buffer times both before and after it to prevent the microwave reflection. Then the fixed gate length of a geometric gate is 90 ns, and the magnitude of the microwave is tuned to control the rotation angle. The envelope of each pulse is cosine-shaped and the derivative removal by adiabatic gate (DRAG) correction is also used to suppress the leakage to the undesired energy levels, especially $\ket{2}$ state {\cite{PhysRevLett.103.110501}}.

We first use the quantum process tomography (QPT) method to characterize the performance of these single-qubit geometric gates {\cite{doi:10.1080/09500349708231894,PhysRevLett.93.080502,PhysRevLett.102.090502}}. The experimental process matrix $\chi_{\rm exp}$ of four specific geometric gates $X$, $X/2$, $Y/2$, and $H$ are shown in {$\rm{Fig.\ \ref{single}(a)}$} with an average gate fidelity of 99.3(1)$\%$. Considering that the process fidelities contain the state preparation and measurement (SPAM) errors, we then utilize another commonly used method, Clifford-based randomized benchmarking (RB) {\cite{PhysRevA.77.012307,PhysRevLett.102.090502,PhysRevLett.109.080505}}, to characterize the geometric gates. The experimentally measured ground state probability (the sequence fidelity) decays as a function of the number of single-qubit Cliffords m are shown in {$\rm{Fig.\ \ref{single}(b)}$}. The reference RB experiment gives an average gate fidelity $99.86(1)\%$ for the realized single-qubit gates in the Clifford group. The measured interleaved gate fidelities of the four specific gates $X$, $X/2$, $Y$, and $Y/2$ are $99.83(2)\%$, $99.79(2)\%$, $99.80(2)\%$, and $99.79(1)\%$, respectively. All the data are corrected for readout errors and the corresponding readout fidelity matrixes are listed in Appendix $\ref{readout}$. 

\begin{figure}[b] 
\includegraphics{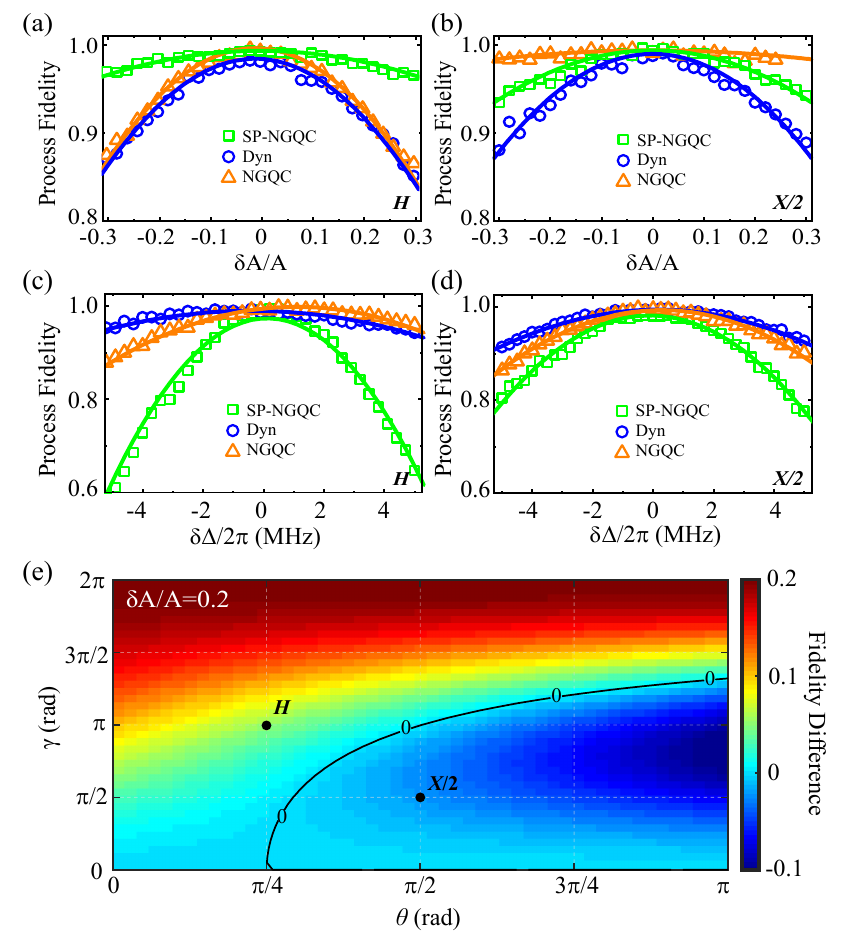}
\caption{\label{fig:epsart3}Noise-resilient feature of SP-NGQC single-qubit gates. $X/2$ and $H$ gates are chosen for rotating axes in different planes. (a) and (b) are respectively the process fidelities of quantum gates $H$ and $X/2$ as a function of Rabi frequency error $\delta A$. (c) and (d) are respectively the process fidelities of quantum gates $H$ and $X/2$ as a function of qubit-frequency shift error $\delta\Delta$. Performances of different evolution paths including the dynamic (Dyn), the NGQC, and the SP-NGQC are shown for comparison. The experimental results are also consistent with the numerical simulations (solid lines). (e) The landscape of process fidelity difference $F_{\rm SP-NGQC}-F_{\rm NGQC}$ versus $\theta$ and $\gamma$ where $\delta A/A=0.2$. Black dots show the working points for (a)-(b).}
\end{figure}
After demonstrating high-fidelity single-qubit SP-NGQC gates, we deliberate on the noise-resilient characteristic of these gates. All noise can be attributed to the effect on the axis of rotation and the angle of rotation. Here, we focus on two typical types of errors: Rabi frequency error through changing the microwave driving amplitude by an amount of $\delta A$, and frequency shift error by changing the microwave driving frequency by an amount of $\delta\Delta$. The Rabi frequency error is an intuitive manifestation of rotation angle error and the frequency shift error affects the rotation angle. As shown in {$\rm{Fig.\ \ref{fig:epsart3}} (a)-(d)$}, we compare the performance of both $X/2$ and $H$ gates with three evolution paths, evaluated by the process fidelities from QPT. These figures show that in the case of Rabi frequency error, the geometric gates are more robust than the dynamical gates, and for the $H$ gate, the SP-NGQC gate is the most robust one. However, for the frequency shift error, the SP-NGQC gates are not as good as the other two types of gates.

To explain the noise-resilient features of different types of gates, we theoretically calculate the fidelity changing with an extra Rabi frequency error term $\delta A$. For the frequency shift error $\delta\Delta$, the process fidelity of the gates is not as intuitive as Rabi frequency error to formulate, and theoretical results are obtained using master-equation numerical simulation (See the Appendix {\ref{secG}} for more details). The calculated fidelities are overlaid on {$\rm{Fig.\ \ref{fig:epsart3}}$}, which shows our theoretical formulas agree well with the experimental results. Since the Hamiltonian of SP-NGQC has $\sigma_z$ component, the frequency shift error changes the evolution path of SP-NGQC significantly, and our SP-NGQC scheme does not perform well under this error compared to the other two schemes. For the reason why our gates perform best for the H gate against the Rabi frequency error, we then calculate the fidelity difference $F_{\rm SP-NGQC}-F_{\rm NGQC}$ as a function of both $\theta$ and $\gamma$, which contains the universal single-qubit gates. As shown in {$\rm{Fig.\ \ref{fig:epsart3} (e)}$}, where $\delta A/A $ is fixed to 0.2, the SP-NGQC gates are better than the traditional NGQC gates when the rotating axis is close to the z-axis ($\theta$ is small) or the rotation angle $\gamma$ is large, which reflects the short-path advantage of our SP-NGQC gates.


In order to achieve a universal SP-NGQC, we also realize the non-trivial two-qubit geometric CZ gate similarly to the single-qubit case. To fully control the rotation angle and coupling strength between $\ket{11}$ and $\ket{20}$ states, we utilize the radio-frequency flux modulation method without extra z-pulse distortion correction {\cite{PhysRevA.97.022330, PhysRevApplied.10.034050}}. Here, $\ket{AB}$ denotes the state of $\ket{Q_A, Q_B}$. The frequency of $Q_A$ is modulated with a cosinoidal form: $\omega_A(t)={\bar{\omega}_A}+\varepsilon \cos{(2\nu t+2\Phi)}$, where ${\bar{\omega}_A}$ is the mean operating frequency, $\varepsilon$, $\nu$, and $\Phi$ are the modulation amplitude, frequency, and phase, respectively. The factor of two arises because $Q_A$ is at the sweet spot, and the frequency undergoes two cycles for each cycle of flux. Ignoring the higher-order oscillating terms, the obtained effective Hamiltonian in the interacting picture can be reduced to
\begin{equation}\label{H2}
\begin{aligned}
H=\frac{1}{2}
\begin{pmatrix}
   \Delta'  && g_{\rm eff} e^{i\beta}\\
   g_{\rm eff} e^{-i\beta} && -\Delta' 
\end{pmatrix}
\end{aligned}
\end{equation}
in the subspace $\{ \ket{11}$, $\ket{20}\}$, where $\Delta'=\lvert {\bar{\omega}_A}-\omega_B \rvert -\alpha_A-2\nu$ and $\beta=\Phi+\pi/2$. The effective coupling strength $g_{\rm eff}$ is equal to $2\sqrt{2}g_{\rm AB}J_1({\frac{\varepsilon}{2\nu}})$ and $J_1({\frac{\varepsilon}{2\nu}})$ is the first order Bessel function of the first kind.

Similar to the single-qubit Z rotation described by the Hamiltonian in {$\rm{Eq.\ ({\ref{singleH}})}$}, we can acquire a pure geometric phase $e^{-i\gamma'/2}$ on the state of $\ket{11}$. Thus, within the computational subspace {$\ket{00},\ket{01},\ket{10},\ket{11}$}, the resulting unitary operation corresponding to a controlled-phase gate with a geometric phase $\gamma'$ is:
\begin{equation}\label{U2}
\begin{aligned}
U_2(\gamma')=
\begin{pmatrix}
   1 && 0 && 0 && 0 \\
   0 && 1 && 0 && 0 \\
   0 && 0 && 1 && 0 \\
   0 && 0 && 0 && e^{-i\gamma'/2}
\end{pmatrix}.
\end{aligned}
\end{equation}
By setting $\gamma'=2\pi$, we can achieve a CZ gate. According to {$\rm{Eq.\ ({\ref{3}})}$}, a rotation around $z$ axis requires $\theta=0$ so that the time of the fourth segment corresponding to {$\rm{Eq.\ ({\ref{3d}})}$} is equal to zero. 

\begin{figure}[ht]
\includegraphics{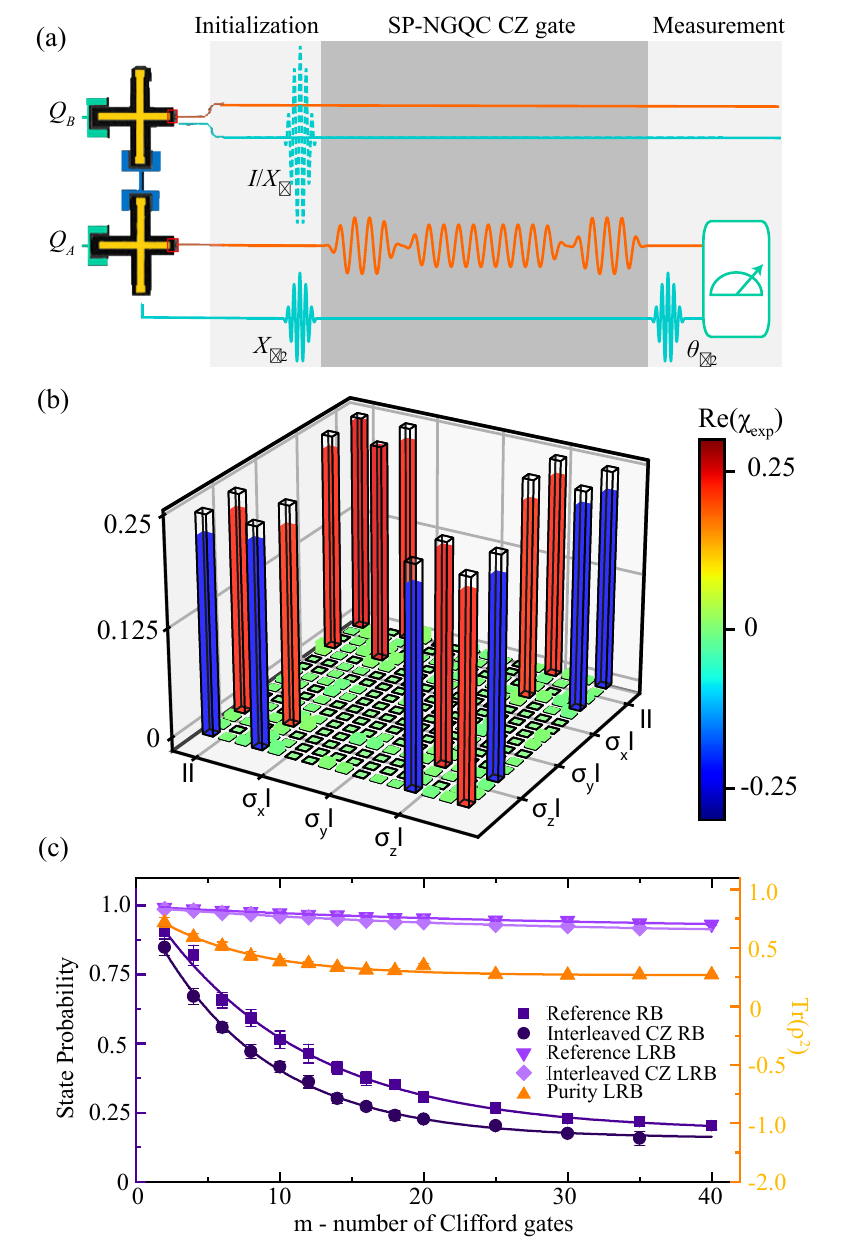}
\caption{\label{twoqubit}Two-qubit SP-NGQC CZ gate. (a) Experimental pulse sequence to acquire conditional phase of the geometric controlled-phase gate. The conditional phase is the phase difference in $Q_A$ due to $Q_B$ occupying either $\ket{0}$ or $\ket{1}$. (b) The real part of the experimental process matrix $\chi_{\rm exp}$ for the geometric CZ gate. The imaginary part is not shown, for theoretically its all elements are zero and experimentally they are smaller than 0.02. The solid black outlines are for the ideal CZ gate. The process fidelity is $94.2(2)\%$ after maximum likelihood estimation {\cite{PhysRevLett.93.080502}}. (c) Randomized benchmarking data for the geometric CZ gate. Fitting to both reference and interleaved RB curves gives the gate fidelity $97.9(3)\%$. Reference and interleaved LRB gives a leakage error of $0.13\%$. The Purity RB (right axis) shows an incoherent error of $3.7\%$.}
\end{figure}
Accordingly, the two-qubit geometric CZ gate is performed with three cosinoidal modulation drives applied in series. Each has a flat-topped Gaussian envelope with 20 ns rising and falling edges to suppress the undesired impact of a sudden phase change of the microwave modulation. The modulation frequency $\nu/{2\pi}= 80 \;{\rm MHz}$ and the modulation amplitude $\varepsilon/{2\pi}=107.8 \;{\rm MHz}$ lead to an efficient coupling strength $g_{\rm eff}/{2\pi} \approx 9 \;{\rm MHz}$. We can not realize zero coupling strength required by {$\rm{Eq.\ ({\ref{3b}})}$} using parametric driving. But CZ gate is quite special with $\gamma'=2\pi$, which corresponds to the solid angle of a hemisphere. Therefore, the trajectory on the Bloch sphere formed by any plane passing through the center of the sphere can meet that requirement, and $g_{\rm eff}/{2\pi}$ is not restricted to 0 in the second segment. The modulation amplitude is tuned to $100 \;{\rm MHz}$ at the second part causing $\Delta'/{2\pi}=7 \;{\rm MHz}$. The experimental sequence we use to acquire the conditional phase is shown in {$\rm{Fig.\ \ref{twoqubit}(a)}$}. Note that the first and third intervals are resonant operations with trajectories along the geodesic. According to the geodesic rule {\cite{PhysRevLett.60.2339}}, these operations do not accumulate dynamical phases and thus do not influence the final conditional phase. We first run the experiment with only the first and second segments and sweep the length of the second segment to find the working region where the conditional phase is close to $\pi$. Then within that region, we add the third segment and sweep the microwave’s phase and time to bring the state back to $\ket{11}$. Consequently, the total effective gate time is 189.375 ns. The detailed evolution path and the conditional phase against pulse times can be seen in Appendix {\ref{secI}}.

Likewise, we characterize the CZ gate with both QPT and RB methods. For QPT, the experimentally reconstructed process matrix of the CZ gate is shown in {$\rm{Fig.\ \ref{twoqubit}(b)}$} and indicates a process fidelity of $94.2(2)\%$. Besides, we extract the CZ gate fidelity $F_{\rm CZ}=97.9(3)\%$ from fitting both the reference and interleaved RB decay curves. To further investigate the error budgets, we first quantitatively extract the particular gate error from the QPT and get the error rates from decoherence $(4.3\%)$, dynamic ZZ coupling $(1.3\%)$, and SPAM $(0.3\%)$, respectively, following the error matrix method developed by Korotkov {\cite{korotkov2013error}}. To examine the leakage error, we perform leakage RB (LRB) (see {$\rm{Fig.\ \ref{twoqubit}(c)}$}) and extract a leakage rate $L^{\rm CZ}_1=0.13\%$ per CZ gate by fitting the state population in the computational subspace {$\ket{00},\ket{01},\ket{10},\ket{11}$} {\cite{PhysRevA.97.032306, PhysRevX.11.021058}}. We find that most residual leakage is introduced into the second excited state of $Q_A$, which indicates it may come from the residual pulse distortion in Z-control pulses of $Q_A$. We then perform purity RB (also see {$\rm{Fig.\ \ref{twoqubit}(c)}$}) to check the incoherent error {\cite{Wallman_2015}}. The Purity RB experiment is done by performing QST instead of measuring the state probability at the end of the sequence and gives the incoherent error $\epsilon=3.7\%$ per CZ gate, which is comparable with the results obtained from QPT. Taken together, the main error in the demonstrated two-qubit SP-NGQC CZ gate comes from decoherence in our qubits, which could be further improved by optimizing the device design and fabrication. Another way to improve it is to use a more complicated control optimization scheme (Appendix {\ref{secJ}}). Moreover, a quantum circuit with tunable couplers may also help to suppress these errors by reducing operating time \cite{PhysRevApplied.10.054062}.

In conclusion, we have experimentally realized single-qubit short-path non-adiabatic geometric gates with a 2-times shorter evolution path and fidelities above $99.86(1)\%$. We illustrate the significant advantages of our gates that are resilient to control amplitude noise, especially when the rotating axis is close to the z-axis or the rotation angle is large, by comparing the performance with the orange-slice-shaped geometric gates and the dynamical gates. Besides, we also demonstrate the two-qubit non-adiabatic geometric controlled-Z gate with all-microwave controls to prevent complicated flux pulse distortion calibrations. The CZ gate has a comparable fidelity of $97.9(3)\%$, mainly limited by the qubit decoherence time. Consequently, the shown universal geometric quantum gate set paves the way for high-fidelity robust geometric quantum computation for NISQ-era applications. Other experimental systems, such as trapped ions {\cite{doi:10.1063/1.5088164}} and semiconductor quantum dots {\cite{10.1093/nsr/nwy153}} can also benefit from the methods utilized here for the geometric realization of universal gates.

\begin{acknowledgments}
We thank Sai Li and Zheng-Yuan Xue for helpful theoretical discussion. We also thank Gang Cao and Hai-Ou Li for helpful discussions and improving the paper. This work is supported by the National Natural Science Foundation of China (Grants No. 12034018), and this work is partially carried out at the USTC Center for Micro and Nanoscale Research and Fabrication.
\end{acknowledgments}

\appendix
\section{\label{sec:level1}Experimental setup}
Our experiments are implemented on a six-transmon-qubit-chain device, which consists of six adjacent cross-shaped transmon qubits arranged in a linear array with nearly identical nearest-neighbor coupling strengths $g/{2\pi}\approx 10 \;{\rm MHz}$, as illustrated in {$\rm{Fig.\ \ref{sample}(a)}$}. The qubits in the device are frequency-tunable transmons, of which the frequencies can be adjusted indivisually by tuning the external magnetic field through the Z control line and the qubits can be driven through the XY control lines. Each qubit can be read out by the individual $\lambda/4$ resonators, where the resonators are coupled to the transmission line. In the experiments, we have performed short-path geometric gates with the first two adjacent qubits $Q_A$ and $Q_B$, whose main parameters are summarized and listed in Table \ref{tab:table1}. The other four qubits are biased far away from these two operation qubits and thus are nearly completely decoupled. Both qubits work at the sweet spots to maintain the best operating performance. 
\begin{table}
\centering
\renewcommand\arraystretch{1.5}
\caption{\label{tab:table1}%
Device parameters of the two operating qubits.
}
\begin{ruledtabular}
\begin{tabular}{ccc}
\textrm{Parameters}&
\textrm{$Q_A$}&
\textrm{$Q_B$}\\
\hline
Readout frequency (GHz) & 6.5455 & 6.4005\\
Qubit frequency (GHz) & 5.5114 & 5.0010 \\
Anharmonicity ($\alpha/{2\pi}$) (MHz) & -242.6 & -250.0 \\
$T_1 \;(\mu s)$ (sweet spot) & 11.5 & 22.3 \\
$T^*_2 \;(\mu s)$ (sweet spot) & 7.5 & 27.8 \\
$T_1 \;(\mu s)$ (CZ working point) & 10.6 & 22.3 \\
$T^*_2 \;(\mu s)$ (CZ working point) & 5.9 & 27.8 \\
Readout fidelity $\ket{0}$ ($F_0$)& $97\%$ & $96\%$ \\
Readout fidelity $\ket{1}$ ($F_1$) & $91\%$ & $90\%$ \\
Readout fidelity $\ket{2}$ ($F_2$) & $85\%$ & $\backslash$ \\
Qubit-qubit coupling strength $g_{\rm AB}/2\pi$ (MHz) &\multicolumn{2}{c}{9.5}\\
Qubit-readout dispersive shift $\chi_{\rm 01}$ (MHz) & 0.55 & 0.3
\end{tabular}
\end{ruledtabular}
\end{table}

\begin{figure*}
    \centering
    \includegraphics{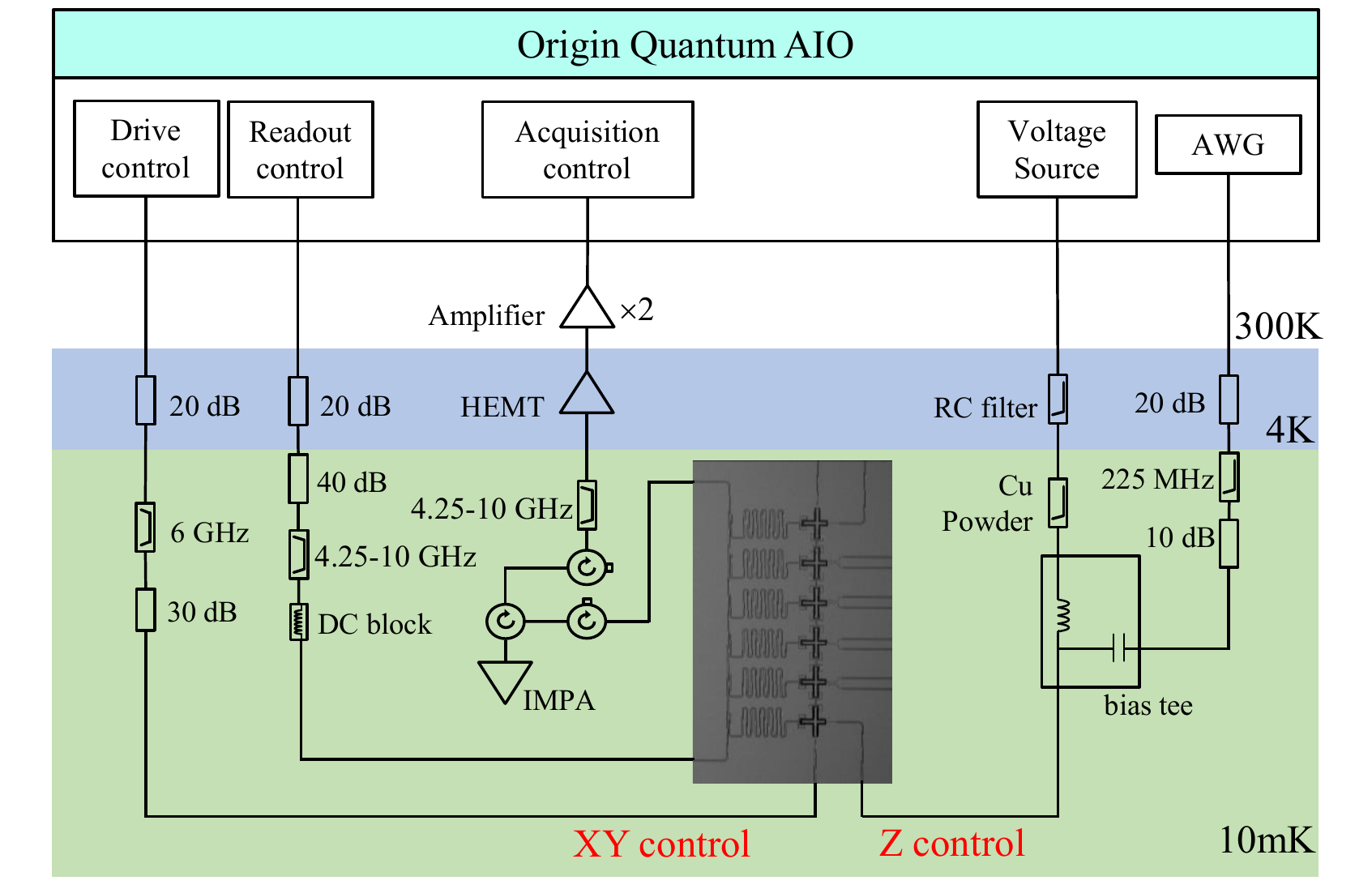}
    \caption{Details of wiring and circuit component. Testing equipments are supported by a highly-integrated quantum computer control system from OriginQ Inc. at room temperature. In the dilution refrigerator, attenuates and filters are installed at different stages to reduce noise. An IMPA and three amplifiers are used to acquire adequate signal-to-noise ratio.}
    \label{circuit}
\end{figure*}

The sample is cooled down to 10 mK within a dilution refrigerator of Oxford Triton XL. The wiring diagram and circuit components for control and readout of qubits are shown in {$\rm{Fig.\ {\ref{circuit}}}$}. In the dilution refrigerator, attenuators and filters are installed at different stages to reduce noise. The qubits are controlled by a highly integrated quantum control system, Quantum AIO from OriginQ Inc.$\;${\cite{OriginAIO}}. For qubit state readout, the readout signal after dual-quadrature up-conversion in Quantum AIO passes through the attenuators and filters and then reaches the quantum processor. The output signal firstly passes through two circulators, then is amplified by an impedance transformer parametric amplifier (IMPA) {\cite{Duan_2021}}, of which the noise as well as the noise from higher temperature stages has been blocked by the preceding of two circulators. The IMPA with an amplification gain of 15 dB and a bandwidth about 500 MHz allows high-fidelity single-shot measurements of the two qubits individually and simultaneously. Being magnified by a high electron mobility transistor (HEMT) amplifier at the 4K stage and two low noise amplifiers at room temperature respectively, the signal is finally captured and analyzed by the Quantum AIO. 
\section{Crosstalk}

Crosstalk between Z control lines of the two qubits is inevitable due to the ground plane return currents. For DC flux bias to control the working frequency of the qubits, we measure the crosstalk between the Z control lines and qubits to be approximately $10\%$. The crosstalk can be compensated by orthonormalising the Z bias lines through an inversion of the normalized qubit frequency response matrix
\begin{align}
M=
\begin{pmatrix}
   1.0000  && -0.1001\\
   0.1546  &&  1.0000
\end{pmatrix}.
\end{align}

\begin{figure}[htb] 
\includegraphics{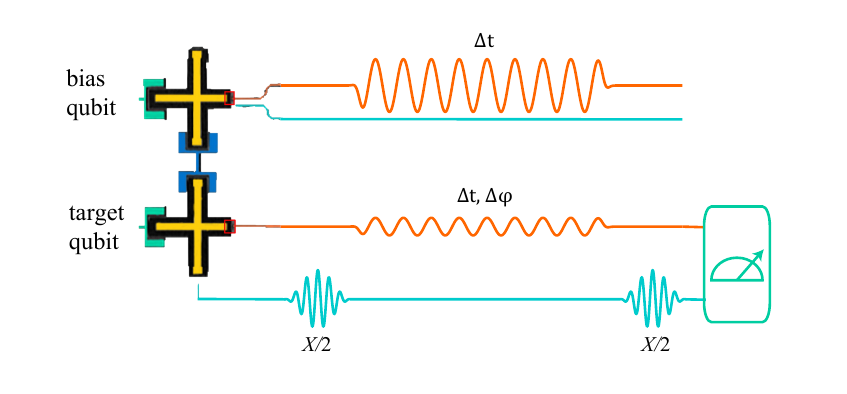}
\caption{\label{crosstalk}Experimental pulse sequence to measure the phase difference between the two-qubit Z control lines. For each $\Delta\varphi$, we sweep $\Delta{\rm t}$ to acquire the frequency of the target qubit under the parametric control. The frequency of the target qubit reaches a modulational maximum when $\Delta\varphi$ is tuned synchronous with the bias qubit.}
\end{figure}
For parametric driving of qubits, additional calibration of the phase difference between two control lines of qubits is needed. The experimental sequence to measure the phase difference is shown in {$\rm{Fig.\ \ref{crosstalk}}$} and the phase difference of the Z control lines of $Q_B$ with respect to $Q_A$ is 3.3681 rad at an 80 MHz parametric driving frequency. The driving frequency will influence the phase difference and slightly change the qubit frequency response matrix, as
\begin{align}
M=
\begin{pmatrix}
   1.0000  && -0.0974\\
   -0.1572  &&  1.0000
\end{pmatrix}
\end{align}
at 80 MHz.

\section{Readout calibration} \label{readout}

Due to the thermal fluctuations and qubit relaxation during measurements, there are non-negligible readout infidelities. The intrinsic probabilities for the single qubit can be inferred from the measured probabilities $P_M=(P_0,\; P_1)^T$ and the measurement fidelities are gotten by preparing the system in each computational basis state and simultaneously measure the assignment probability of the qubit for 10000 executions (Table. \ref{tab:table1}). The intrinsic occupation probabilities are computed as $P_i=F^{-1}\cdot P_M$ where 
\begin{align}
F=
\begin{pmatrix}
   F_0  && 1-F_1  \\
   1-F_0  &&  F_1 
\end{pmatrix}.
\end{align}

For two qubits, the readout fidelity matrix is then given by $F_{\rm Q_A} \otimes F_{\rm Q_B}$. However, we need to read and calibrate the second excited state of the qubit in leakage RB so the fidelity matrix should be expanded to $3\times3$ dimensions. On account of poor control of the $\ket{2}$ state of $Q_B$, we prepare and simultaneously measure the assignment probability $P_M=(P_{\rm 00},\; P_{\rm 01},\; P_{\rm 10},\; P_{\rm 11},\; P_{\rm 20},\; P_{\rm 21})^T$ of the two qubits, and take the sum probability of state $\ket{20}$ and $\ket{21}$ as leakage. The measured two-qubit readout fidelity matrix is shown as
\begin{align}
    \bordermatrix{%
        & \ket{00} & \ket{01} & \ket{10} & \ket{11} & \ket{20} &\ket{21} \cr
        \bra{00} & 0.912 & 0.085 & 0.088& 0.008 & 0.035 & 0.005 \cr
        \bra{01} & 0.059 & 0.888 & 0.005 & 0.085 & 0.002 & 0.049 \cr
        \bra{10} & 0.026 & 0.002 & 0.834 & 0.077 & 0.107 & 0.008 \cr
        \bra{11} & 0.001 & 0.023 & 0.057 & 0.812 & 0.004 & 0.023 \cr
        \bra{20} & 0.001 & 0.000 & 0.013 & 0.003 & 0.819 & 0.145 \cr
        \bra{21} & 0.000 & 0.002 & 0.001 & 0.014 & 0.031 & 0.769
}.
\end{align}

\section{The geometry of SP-NGQC} \label{geometry}

Non-adiabatic Abelian geometric phases are generated by the cyclic evolution and parallel transport of a state subspace in the Hilbert space. Following the evolution path described by {$\rm{Eq.\ (\ref{3})}$}, the two orthogonal bases $\ket{\psi_{+}}$ and $\ket{\psi_{-}}$ can undergo cyclic evolution,
\begin{align}
\ket{\psi_{+}}  \rightarrow U(T)\ket{\psi_{+}} = e^{-i\gamma/2}\ket{\psi_{+}} \\
\ket{\psi_{-}}  \rightarrow U(T)\ket{\psi_{-}} = e^{i\gamma/2}\ket{\psi_{+}},
\end{align}
and they respectively acquire phases $-\gamma/2$ and $\gamma/2$. Driven by the designed Hamiltonian, the two orthogonal bases can evolve along the geodesic on the Bloch sphere. Similar to the parallel transport of a vector, no dynamical phase is accumulated for the state moving along the geodesic lines. Quantitatively, the dynamical phase $\gamma_d$ accumulated during the evolution path is calculated by
\begin{align}
\gamma_d = \int_{\tau} \bra{\psi_{\pm}(t)}H(t)\ket{\psi_{\pm}(t)} dt = 0
\end{align}
for each segment of the evolution path, where $\ket{\psi_{\pm}(t)}=U(t)\ket{\psi_{\pm}}$ with $U(t)$ being the evolution operator and $H(t)$ is {$\rm{Eq.\ (1)}$} in the main text. Therefore, only geometric phases $-\gamma/2$ and $\gamma/2$ are accumulated in the process. Using the Bloch sphere representation for the evolution of the basis $\ket{\psi_{+}}$, $\gamma$ is proportional to the solid angle enclosed by the half-orange-slice-shaped loop, shown in {$\rm{Fig.\ \ref{sample}(b)}$}. This corresponds to an essential feature of non-adiabatic Abelian geometric phases, i.e., the geometric phase is equal to half of the solid angle subtended by a curve traced on a sphere \cite{PhysRevLett.58.1593}.

\section{NGQC with orange-slice Loops}

We present the details in implementing previous NGQC with orange-slice-shaped loops \cite{PhysRevLett.124.230503,zhao2021experimental}. The evolution path is divided into three intervals with resonant drive which has different amplitudes and phases satisfying
\begin{subequations}
\begin{align}
&\int^{\tau_1}_{0}\Omega(t) dt=\theta, \quad \varphi=\phi-\frac{\pi}{2},\quad t\in[0, \tau_1)  \\
&\int^{\tau_2}_{\tau_1}\Omega(t) dt=\pi, \quad \varphi=\phi+\gamma+\frac{\pi}{2},\quad t\in[\tau_1, \tau_2)  \\
&\int^{T}_{\tau_2}\Omega(t) dt=\pi-\theta, \quad \varphi=\phi-\frac{\pi}{2},\quad t\in[\tau_2, T].
\end{align}
\end{subequations}
The final evolution operator can be obtained as $U(T) = e^{i\gamma} \boldsymbol{n} \cdot \boldsymbol{\sigma}$, which corresponds to a rotation operation around the axis $\boldsymbol{n}$ by an angle $-2\gamma$.

\begin{figure}[htb] 
\includegraphics{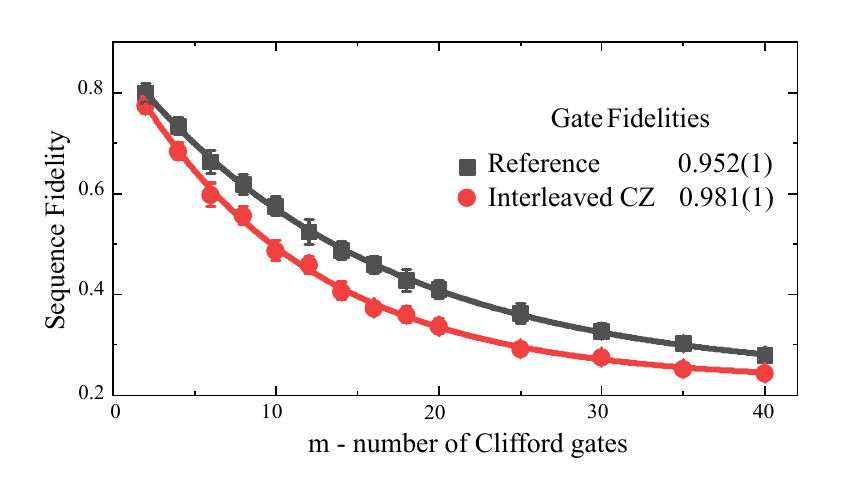}
\caption{\label{fig:NGQC}Reference and interleaved RB results for two-qubit NGQC CZ gate. Fitting to both reference and interleaved RB curves gives the gate fidelity $98.1(1)\%$}.
\end{figure}

The performance of single-qubit gates and its gate robustness against amplitude error is shown in {$\rm{Fig.\ \ref{fig:epsart3}(a)-(b)}$}. And the same-loop two-qubit CZ gate has the fidelity of $98.1(1)\%$ using the RB method (shown in {$\rm{Fig.\ \ref{fig:NGQC}}$}). The fidelity is slightly higher than the short-path CZ gate because the effective gate length of the orange-slice-shaped CZ gate is 85 ns which suffers less incoherent errors.

\section{Quantum process tomography} \label{QPT}

Quantum process tomography (QPT) are used to characterize both the single-qubit and two-qubit gates \cite{doi:10.1080/09500349708231894,PhysRevLett.93.080502,PhysRevLett.102.090502}. For N-qubit gate, we first prepare a set of initial states {$\{\ket{0},\;\ket{1},\; (\ket{0}-i\ket{1})/\sqrt{2},\; (\ket{0}+\ket{1})/\sqrt{2}\}^{\otimes N}$}, and measure with standard quantum state tomography (QST) with prerotations $\{I,\; X/2, \;Y/2\}^{\otimes N}$ to get the density matrix {$\{\rho_i\}$} of input states. Then a specific non-adiabatic geometric gate is applied following the initial states' preparation. Finally, we measure the output states with QST and reconstruct the output states {$\{\rho_f\}$}. By mapping between the input states and output states, we can determine the process matrix $\chi_{\rm exp}$ of the geometric gate through {$\rho_f=\sum_{\rm m,n}\chi_{\rm mn} E_m \rho_i E_n^{\dagger}$}, where the basis operators $E_m$ and $E_n$ are chosen from the set $\{I, \;\sigma_x, \;\sigma_y, \;\sigma_z\}^{\otimes N}$ with $\sigma_x$, $\sigma_y$, and $\sigma_z$ being Pauli operators.

Based on the fact that the experiments are disturbed by various noises such as coherent errors due to imperfect control, decoherence error, state preparation and measurement (SPAM) error, and so on, the experimental process matrix $\chi_{\rm exp}$ is different with the corresponding ideal process matrix $\chi_{\rm ideal}$ and the difference is evaluated by the process fidelity $F_p={\rm Tr}(\chi_{\rm exp}\chi_{\rm ideal})$. Each QPT experiment for the specific gate is repeated four times for the sake of eliminating the measurement uncertainty. The error analysis in QPT is obtained from bootstrap resampling.

For the two-qubit gates, we follow a method developed by Korotkov {\cite{korotkov2013error}} to further distinguish the error sources. First, we extract the SPAM error by comparing the experimentally reconstructed input states {$\{\rho_i\}$} with ideal input states {$\{\rho_i^{\rm ideal}\}$}. Similarly, we can get a process matrix of SPAM $\chi_{\rm exp}^{\rm SPAM}$ whose corresponding theoretical matrix $\chi^I$ is equal to the perfect identity operation with only one non-zero element $\chi_{\rm II,II}^I=1$. Therefore, the SPAM error is calculated as $1-{\rm Tr}(\chi_{\rm exp}^{\rm SPAM}\chi^I)=0.3\%$. To facilitate error analysis, we then take another representation, the error matrix $\chi^{err}$ by factoring out the desired unitary operation, $U=U_{\rm CZ}$ in this paper, from the standard process matrix $\chi_{\rm exp}$. We extract $\chi^{err}$ from the standard $\chi_{\rm exp}$ matrix with the relations:
\begin{align}
\chi^{err} = T \chi_{\rm exp} T^\dagger, \; T_{\rm mn}={\rm Tr}(E_m^{\dagger}E_nU^\dagger)/d
\end{align}
where $U=U_{\rm CZ}, \;d=2^2$ for two-qubit CZ gate. In the ideal case, the error matrix is equal to $\chi^I$; otherwise, the only one large element $\chi_{\rm II,II}^{err}$ reflects the process fidelity $F_p=Tr(\chi_{\rm exp}\chi_{\rm ideal})={\rm Tr}(\chi^{err}\chi^{I})$ and other non-zero elements indicate the imperfections of the gate. The imaginary parts of the elements along the left column and top row correspond to unitary imperfections, while the real parts of the elements come from decoherence error. We extract the infidelity induced by the dynamic ZZ coupling with $\epsilon_{\rm ZZ}=({\rm Im}(\chi_{\rm II,ZZ}^{err}))^2/F_p=1.3\%$. And the decoherence error is assessed by
\begin{align}
\epsilon_{\rm dec}=t_G(\frac{1}{2T_1^A}+\frac{1}{2T_1^B}+\frac{1}{2T_{\varphi}^A}+\frac{1}{2T_{\varphi}^B})=4.3\%
\end{align}
where $t_G$ is the gate duration, $T_1$ is the energy relaxation time, $T_{\varphi}$ is the pure dephasing time, and the qubits are labeled as A and B.

\section{Robustness of SP-NGQC gates}\label{secG}

\begin{figure}
    \centering
    \includegraphics{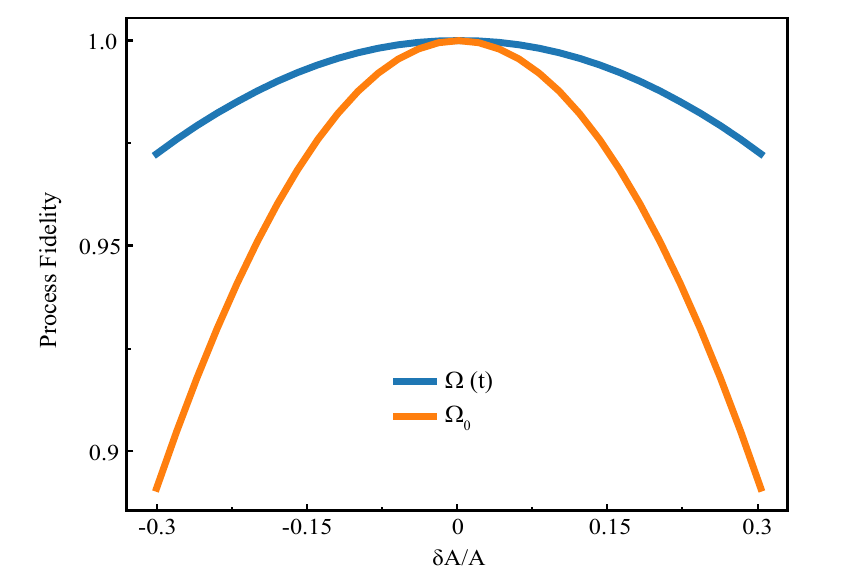}
    \caption{Process fidelity of SP-NGQC H gate against Rabi frequency error $\delta A/A$. $\Omega_0$ is equal to the $\Omega(t)_{max}$. The difference between process fidelities is two times.}
    \label{compare}
\end{figure}

To explain the noise resilient features of different types of gates, we add a Rabi frequency error term $\epsilon = \delta A/A$ satisfying $\lvert \epsilon \rvert \ll 1$ into the Hamiltonian in {$\rm{Eq.\ (1)}$} in the main text as following:
\begin{align} \label{hamiltonian}
H=\frac{1}{2}
\begin{pmatrix}
   \Delta(t)  && (1+\epsilon)\Omega_0 e^{-i\varphi}\\
   (1+\epsilon)\Omega_0 e^{i\varphi} && -\Delta(t) 
\end{pmatrix}.
\end{align}
Here, we fix the Rabi frequency to $\Omega_0$ because when we calculate the fidelity, we care about the integration of the Hamiltonian instead of the Hamiltonian itself. An example of fidelity difference between the time-dependent Rabi frequency and time-independent Rabi frequency is shown in {$\rm{Fig.\ \ref{compare}}$}, where $\Omega(t)_{max}=\Omega_0$. The process fidelities differ by a factor of two because the total power of cosine envelope we use is half that of the time independent one by integration. So we can utilize the constant Rabi frequency by setting it to half of the maximum $\Omega(t)$.

The fidelity used to characterize the performance is defined as 
\begin{align} \label{fidelity}
F=\frac{\lvert Tr(U_{\rm{ideal}}^{\dagger}U) \rvert}{Tr(U_{\rm{ideal}}^{\dagger}U_{\rm{ideal}})}
\end{align}
where $U_{\rm{ideal}}$ is the ideal gate operator and $U$ is calculated through the Hamiltonian in  {$\rm{Eq.\ \ref{hamiltonian}}$}. We note that in the case of unitary evolution, the fidelity in {$\rm{Eq.\ \ref{fidelity}}$} is the square root of the process fidelity defined with $\chi$-matrix ($F_p={\rm Tr}(\chi_{\rm exp}\chi_{\rm ideal})$). Then we get the process fidelity of the SP-NGQC as $F_{\rm SP-NGQC}=[1+\cos\gamma-(\cos\gamma-1)\cos{(\epsilon \theta)}]/2$ where only up to the second order of error $\epsilon$ is considered. Similarly, the process fidelities of NGQC and Dyn can be expressed as $F_{\rm NGQC}=1-[\frac{\pi^2}{4}(1-\cos{\frac{\gamma}{2}})-\frac{\theta (\pi-\theta)}{2}{\sin^{2}{\frac{\gamma}{2}}}]{\epsilon}^2$, $F_{\rm Dyn}=\cos({\gamma \epsilon/2})$ for $X$ and $Y$ gates and $F_{\rm Dyn}=\lvert (\sin{\pi \epsilon})/[4 \sin{(\pi\epsilon/4)}] \rvert$ for $H$ gate, respectively. The calculated fidelities are overlaid on {$\rm{Fig.\ \ref{fig:epsart3}(a)-(b)}$} in the main text, which shows our theoretical formulas agree well with the experimental results. 

\begin{figure}[htb] 
\includegraphics{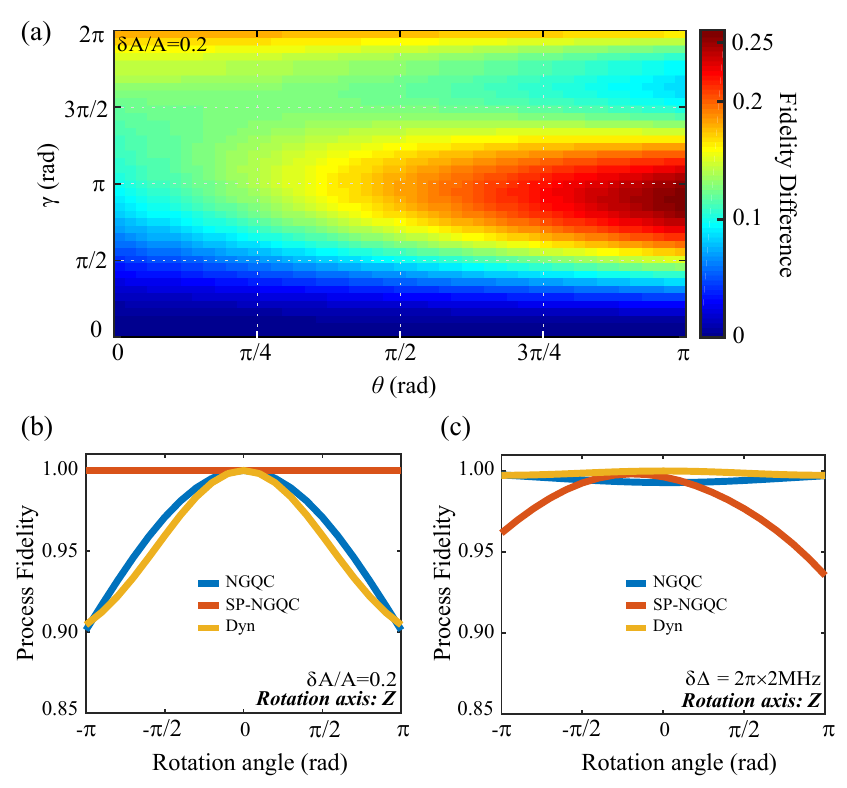}
\caption{\label{fig:error}(a) The landscape of process fidelity difference $F_{\rm{SP-NGQC}}-F_{\rm{Dyn}}$ against Rabi frequency error, where $\delta A/A=0.2$. The azimuth angle $\varphi$ is 0 that rotation axis is in the $X-Z$ plane. The dynamical gates are realized by composite $Y-X-Y$ sequences. $\theta=0$ corresponds to the $Z$ rotation and $\theta=\pi/2$ is $X$ rotation. The SP-NGQC scheme performs much better than the dynamical scheme. (b) The performance of three schemes' $Z$ rotation against Rabi frequency error, where $\delta A/A=0.2$. (c) The performance of three schemes' $Z$ rotation against frequency shift error, where $\delta\Delta=2\pi\times2\;{\rm MHz}$.}
\end{figure}

In the main text {$\rm{Fig.\ \ref{fig:epsart3}(e)}$}, we compare the performance of SP-NGQC and NGQC scheme over the full single-qubit gate space against Rabi frequency error. Here, we also compare the performance of SP-NGQC and dynamical scheme for some special gates, shown in {$\rm{Fig.\ \ref{fig:error}(a)}$}. The dynamical rotation of any rotation axis can be decomposed into $Y-X-Y$ composite pulse sequence and its process fidelity is affected by three parameters, $\theta$, $\varphi$ for rotation axis and $\gamma$ for rotation angle. Here we choose $\varphi=0$ for an example to show the superior robustness of SP-NGQC scheme against Rabi frequency error than dynamical scheme. For the sake of completeness, the error robustness of Z rotations against the Rabi frequency error and frequency shift error are shown in {$\rm{Fig.\ \ref{fig:error}(b)-(c)}$}.

In the main text, we also investigate the influence of frequency detuning error {$\delta\Delta$} to the fidelities of single-qubit SP-NGQC gates, NGQC gates, and dynamical gates. We numerically simulate the system by adding a qubit frequency shift error term to the Hamiltonian:
\begin{align} 
H=\frac{1}{2}
\begin{pmatrix}
   \Delta(t)+\delta\Delta  && \Omega_0 e^{-i\varphi}\\
   \Omega_0 e^{i\varphi} && -\Delta(t) - \delta\Delta
\end{pmatrix}
\end{align}
and the results match well in {$\rm{Fig.\ \ref{fig:epsart3}(c)-(d)}$}. Considering the frequency shift error, dynamical gates work better because the frequency shifts strongly affect the global change of the evolution path, especially the intervals in {$\rm{Eq.\ (2b)}$} of SP-NGQC scheme, thus giving a limitation to the SP-NGQC scheme. In experimental conditions, the frequency can be accurately manipulated in our superconducting system with fluctuations at kHz level. In practical experiments, one can choose an appropriate scheme against the dominant error in the system.

\section{Randomized benchmarking} \label{RandomB}

We also use another conventional method, Clifford-based randomized benchmarking (RB) to characterize the non-adiabatic geometric quantum gates \cite{PhysRevA.77.012307,PhysRevLett.102.090502,PhysRevLett.109.080505}. In the single-qubit reference RB experiment, the state is prepared in $\ket{0}$ and follows a series of randomly chosen Clifford gates from the single-qubit Clifford group. Finally, a reverse gate is applied to bring the qubit back to $\ket{0}$. We measure the survival probability of state $\ket{0}$ (the sequence fidelity) as the number of single-qubit Cliffords m increases. The whole experiment is repeated for $k=30$ different sequences to get the average sequence fidelity. The maximum number of Cliffords is restricted by the measuring instrument with at most 30 $\mu s$ driving time. In the interleaved RB experiment, we insert a specific gate $G$ after each randomly chosen Cliffords and a similar reverse gate is applied to invert the whole sequence. The transformation of gate errors to a depolarizing channel leads to an exponential decay of the ground state population towards the maximally mixed state, where the decay rate is a measure of the average gate fidelity. Therefore, we fit both reference and interleaved sequence fidelity curves to $F=Ap^m+B$ with different decay rates $p_{\rm ref}$ and $p_{\rm int}$. SPAM errors are absorbed in parameters $A$ and $B$ and thus do not affect the extracted fidelities. The average gate fidelity is given by $F_{\rm ref}=1-(1-p_{\rm ref})(d-1)/d$ where $d=2^N$ for N qubits. The specific fidelity for gate $G$ can be calculated by $F_{\rm gate}=1-(1-p_{\rm int}/p_{\rm ref})(d-1)/d$.

For two-qubit Clifford-based RB, the experiment is similar, but with the randomly chosen Clifford gates from the two-qubit Clifford group instead. We cannot simply estimate the two-qubit gate error per Cliffords through the reference RB because two qubit Cliffords contain both single and two qubit gates. Also, we get the specific fidelity by interleaved RB and $F_{\rm gate}=1-(1-p_{\rm int}/p_{\rm ref})(d-1)/d$.

\subsection{leakage RB}

We estimate the average leakage error of our short-path non-adiabatic geometric CZ gate from the reference and interleaved RB experiment by fitting the population in the computational subspace $P_{\chi_1}\equiv P_{\rm 00}+P_{\rm 01}+ P_{\rm 10}+P_{\rm 11}$ {\cite{PhysRevA.97.032306,PhysRevX.11.021058}}, instead of the $P_{\rm 00}$, to an exponential model:
\begin{align}
P_{\rm {\chi_1},ref} = A_{\rm ref}+B_{\rm ref}{({\lambda}_{\rm 1,ref})}^{m} \\
P_{\rm {\chi_1},int} = A_{\rm int}+B_{\rm int}{({\lambda}_{\rm 1,int})}^{m}.
\end{align}
Then, we calculate the average leakage rates $L_{\rm ref}$ and $L_{\rm int}$ per Cliffords as follows:
\begin{align}
L_{\rm 1,ref} = (1-A_{\rm ref})(1-{\lambda}_{\rm 1,ref}) \\
L_{\rm 1,int} = (1-A_{\rm int})(1-{\lambda}_{\rm 1,int}).
\end{align}
The average leakage rate per CZ gate is subsequently obtained by:
\begin{align}
L_1^{\rm CZ} = 1-\frac{1-L_{\rm 1,int}}{1-L_{\rm 1,ref}}.
\end{align}
The leakage rate $L_{\rm CZ}$ per CZ gate is extracted as $0.13\%$. Analyzing the residual state populations, we find that most leakage probabilities are lying on the $\ket{20}$,  which indicates it may come from the residual pulse distortion in Z-control pulses of $Q_A$.

\subsection{purity RB}
We utilize purity RB to distinguish between coherent errors due to mistakes in our calibration, versus incoherent error due to noise in the qubit's environment \cite{Wallman_2015}. The experiment is done by performing QST to determine the state of the qubit after the random sequence, instead of just measuring the state probabilities. The purity of the state is defined as $P=tr({\rho}^2)$. Starting in the ground state, the purity satisfies
\begin{align}
P = (1-\frac{1}{d}){\gamma}^{2m} +\frac{1}{d}
\end{align}
after m Cliffords with a pure depolarizing noise $\gamma$. Accordingly, we fit the data to $A{\gamma}^{2m}+B$ and the incoherent error per CZ gate is estimated as $\epsilon=\frac{3}{4}(1-{\gamma}^{\frac{2}{3}})=3.7\%$. This value is comparable with the results obtained from QPT,  demonstrating that our gate is dominated by incoherent errors.

\section{Two-qubit SP-NGQC CZ gate} \label{secI}
\begin{figure}
    \centering
    \includegraphics{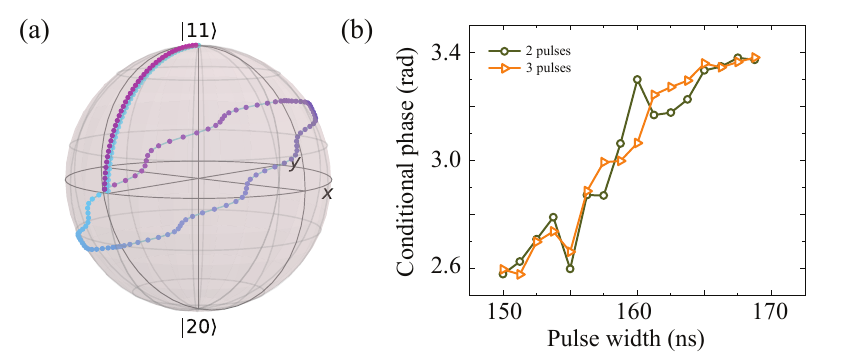}
    \caption{(a) The evolution trajectory of our CZ gate in the $\ket{11}-\ket{20}$ space. The evolutionary time sequence is marked by the dots' color from light blue to dark violet. (b) Conditional phase of the Cphase gate versus the second segment's pulse width with (without) the final half-$\pi$ pulse. The final half-$\pi$ pulse does not influence the conditional phase through geodesic principle.}
    \label{trajectory}
\end{figure}

We used the parametric driving method to change the coupling strength between the two qubits. For the first interval, we choose $\nu,\;\varepsilon$ to satisfy the resonance condition $\lvert {\bar{\omega}_A}-\omega_B \rvert -\alpha_A=2\nu$, the Hamiltonian in the subspace can be written as:
\begin{equation}\label{H21}
\begin{aligned}
H=\frac{1}{2}
\begin{pmatrix}
   0  && g_{\rm eff} e^{i\beta}\\
   g_{\rm eff} e^{-i\beta} && 0 
\end{pmatrix}
\end{aligned}
\end{equation}
under the rotation frame, where $\beta=\Phi+\pi/2$. And the rotation frame is selected. Once we change the parametric modulation to $\omega'_A(t)={\bar{\omega}'_A}+\varepsilon' \cos{(2\nu' t+2\beta')}$ in the second segment, the Hamiltonian in the same rotating frame turns into
\begin{equation}\label{H22}
\begin{aligned}
H=\frac{1}{2}
\begin{pmatrix}
    H_{11}  && g_{\rm eff} e^{i\beta}\\
   g_{\rm eff} e^{-i\beta} && H_{20} 
\end{pmatrix}
\end{aligned}
\end{equation}

\begin{equation}\label{H11}
\begin{aligned}
H_{11}& =\omega_A(t)-\omega'_A(t)\\
& =\varepsilon'-\varepsilon+\varepsilon'\cos{(2\nu' t+2\beta')}-\varepsilon\cos{(2\nu t+2\beta)}
\end{aligned}
\end{equation}
\begin{equation}\label{H20}
\begin{aligned}
H_{20}=-H_{11}.
\end{aligned}
\end{equation}
In the experiment, we choose $\nu'$ and $\beta'$ to remain constant and change the modulation amplitude $\varepsilon$. And the third interval's Hamiltonian is the same as the first one. The final evolution trajectory is shown in {$\rm{Fig.\ \ref{trajectory}(a)}$}. The trajectory approximately forms a hemisphere with subtle oscillation which comes from $H_{11}$ and $H_{20}$. The greater the modulation amplitude $\varepsilon'$ differs from $\varepsilon$, the corresponding $\Delta'/{2\pi}$ is also larger, which will lead to more severe oscillation and destroy the geometric phase.

As shown in {$\rm{Fig.\ \ref{trajectory}(b)}$}, the conditional phase acquired by the pulse sequence shown in {$\rm{Fig.\ \ref{twoqubit}(a)}$} with(without) the final half-$\pi$ pulse remains nearly the same because the final segment is moving along the geodesic to bring the state back to $\ket{11}$, which does not change the final geometric phase.

\section{An alternative solution for two-qubit SP-NGQC CZ gate} \label{secJ}

We theoretically present another solution for the two-qubit short-path geometric CZ gate that regards the three intervals of the CZ gate as a whole one. Considering a typical parametric modulation function $F(t)=A(t)\sin{[(\omega_B-\omega_A+\delta(t)+\alpha_A)t+\beta(t)]}$ where $A(t)$, $\delta(t)$ and $\beta(t)$ indicate the strength, frequency detuning and phase of the modulated field, respectively, the frequency of $Q_A$ is modulated as $\omega_A(t) = \omega_A + \dot{F}(t)$. Unitary transformation $U=U_1 \times U_2$ is carried out with
\begin{widetext}
\begin{flalign}
    & U_1 = \rm{exp}\left[{i
    {\begin{pmatrix}
	 0 &   & \\
	  & \omega_A & \\
	  &   & 2\omega_A-\alpha_A
	 \end{pmatrix}}\otimes{\begin{pmatrix}
	 1 &   & \\
	  & 1 & \\
	  &   & 1
	 \end{pmatrix}}t+i{\begin{pmatrix}
	 1 &   & \\
	  & 1 & \\
	  &   & 1
	 \end{pmatrix}}\otimes{\begin{pmatrix}
	 0 &   & \\
	  & \omega_B & \\
	  &   & 2\omega_B-\alpha_B
	 \end{pmatrix}}t}\right] \\
	  & U_2 = \rm{exp}\left[{i
    {\begin{pmatrix}
	 0 &   & \\
	  & F(t) & \\
	  &   & 2F(t)
	 \end{pmatrix}}\otimes{\begin{pmatrix}
	 1 &   & \\
	  & 1 & \\
	  &   & 1
	 \end{pmatrix}}t}\right].
\end{flalign}
\end{widetext}
After the unitary transformation and utilizing Jacobi-Anger expansion, we can get an effective Hamiltonian 
\begin{align}
    H = \sqrt{2}g_{\rm AB}J_1[A(t)]
    \begin{pmatrix}
       0                          && e^{-i[\delta(t)t+\beta(t)]}  \\
       e^{i[\delta(t)t+\beta(t)]} && 0
    \end{pmatrix}
\end{align}
in the subspace $\{\ket{11}$, $\ket{20}\}$ by applying the rotating-wave approximation which ignores the high-frequency oscillation terms. Here $J_1[A(t)]$ is the first order Bessel function of the first kind. Now we apply a second unitary transformation by $U_3={\rm{exp}}[-i\delta(t)t\sigma_z/2]$. In the new frame, the Hamiltonian can be rewritten as 
\begin{align}
    H = \frac{1}{2}
    \begin{pmatrix}
       \dot{\delta}(t)t+\delta(t) && ge^{-i\beta(t)}  \\
       ge^{i\beta(t)} && -\dot{\delta}(t)t-\delta(t)
    \end{pmatrix}
\end{align}
where $g=2\sqrt{2}g_{\rm AB}J_1[A(t)]$ is the effective coupling strength between the two qubits. This Hamiltonian is the same as {$\rm{Eq.\ (\ref{singleH})}$}. By adjusting $\delta(t),\;A(t)$, and $\beta(t)$ to follow the cyclic evolution conditions shown in {$\rm{Eq.\ (\ref{3})}$}, we can realize the geometric phase gate $\begin{pmatrix} \rm{exp}(-\frac{i\gamma'}{2}) & 0 \\ 0 & \rm{exp}(\frac{i\gamma'}{2})\end{pmatrix}$ in the subspace $\{\ket{11}$, $\ket{20}\}$. Setting $\gamma'=2\pi$, it is a CZ gate in the two-qubit computational space. The actual flux pulse applied to the Z control line is calculated by
\begin{align}
    V(t) = f^{-1}(\dot{F}(t))
\end{align}
where $f$ is the nonlinear frequency response of the transmon qubit. 

\begin{figure}
    \centering
    \includegraphics{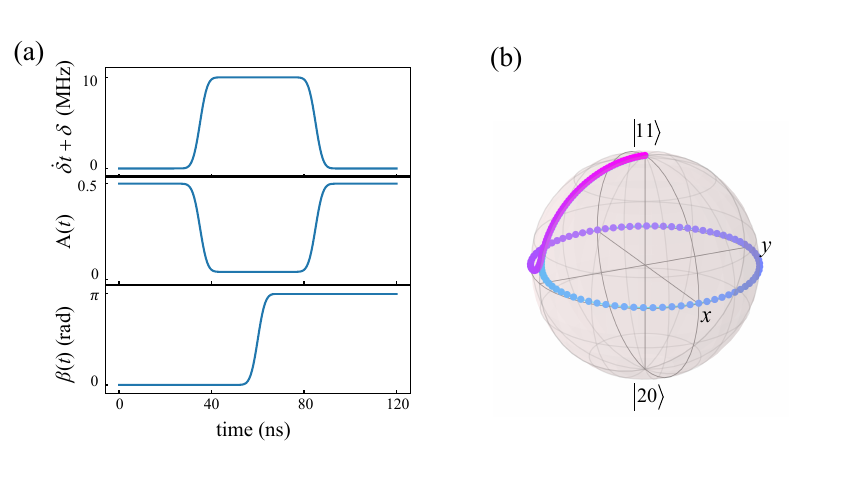}
    \caption{(a) An example pulse parameters based on {$\rm{Eq.\ (\ref{pulse})}$}. $\Delta_0$ is chosen to be $2\pi \times 10\; \rm{MHz}$. $A_0$ and offset make the on-off ratio of the effective coupling strength to be 10, large enough for a pure Z rotation. $\beta$ is changed between 0 and $\pi$ to make the state back to $\ket{11}$. (b) Evolution of the $\ket{11}$ according to the pulse parameters shown in (a).The evolutionary time sequence is marked by the dots' color from light blue to dark violet.}
    \label{waveform}
\end{figure}
As an example, we utilize the flat-top Gaussian shape to smoothly change the $\delta(t),\;A(t)$, and $\beta(t)$ to guarantee the existence of the derivative of $F(t)$. To satisfy the conditions for non-adiabatic geometric CZ gate, we constrain the parameters to
\begin{subequations} \label{pulse}
\begin{align} 
&\dot{\delta}(t)t+\delta(t)=\frac{\Delta_0}{2}[erf(\frac{t-t_b}{\sqrt{2}\sigma})-erf(\frac{t-t_b-t_c}{\sqrt{2}\sigma})] \\
&A(t) = A_0-\frac{A_0}{2}[erf(\frac{t-t_b-t_c}{\sqrt{2}\sigma})-erf(\frac{t-t_b}{\sqrt{2}\sigma})] + offset  \\
&\beta(t)=\left\{
             \begin{array}{lr}
             0   \\
             \frac{\pi}{2}[erf(\frac{t-t_{b1}}{\sqrt{2}\sigma})-erf(\frac{t-t_{b1}-t_{c1}}{\sqrt{2}\sigma})] \\
             0  
             \end{array}
             \right.
\end{align}
\end{subequations}
where $t_b$ is the rotation time around $x-y$ axes in the $\{\ket{11}$, $\ket{20}\}$ subspace which is controlled by $A_0$, $t_c$ is the rotation time around $z$ axis which is controlled by $\Delta$, $t_{b1}$ and $t_{c1}$ are chosen to make the phase transition happen in the time interval of $z$ rotation, and $\sigma$ determines the ramping rate of the parameters. A little offset is added to $A(t)$ to ensure that the pulse amplitude is non-zero and $g$ is small compared to $\Delta$. The pulse and the evolution trajectory are shown in {$\rm{Fig.\ \ref{waveform}}$} \cite{JOHANSSON20121760,JOHANSSON20131234}.

In principle, this solution can reach a fidelity higher than $99\%$ with shorter gate time than the CZ gate described in the main text. However, this solution works at a flux bias far from the sweet spot where the qubit suffers more from the $1/f$ noise. Also, the solution changes the parametric modulation frequency during the evolution, which requires a real-time crosstalk correction. To acquire an ideal value of $\gamma'$, it requires adjusting $\delta(t),\;A(t)$, and $\beta(t)$ with multiple parameters simultaneously, indicating a quite complicated control optimization scheme which is more susceptible to control errors, thus we didn't show it experimentally in the main text. 

\nocite{*}
\bibliography{apssamp}

\providecommand{\noopsort}[1]{}\providecommand{\singleletter}[1]{#1}%
\begin{thebibliography}{89}%
\makeatletter
\providecommand \@ifxundefined [1]{%
 \@ifx{#1\undefined}
}%
\providecommand \@ifnum [1]{%
 \ifnum #1\expandafter \@firstoftwo
 \else \expandafter \@secondoftwo
 \fi
}%
\providecommand \@ifx [1]{%
 \ifx #1\expandafter \@firstoftwo
 \else \expandafter \@secondoftwo
 \fi
}%
\providecommand \natexlab [1]{#1}%
\providecommand \enquote  [1]{``#1''}%
\providecommand \bibnamefont  [1]{#1}%
\providecommand \bibfnamefont [1]{#1}%
\providecommand \citenamefont [1]{#1}%
\providecommand \href@noop [0]{\@secondoftwo}%
\providecommand \href [0]{\begingroup \@sanitize@url \@href}%
\providecommand \@href[1]{\@@startlink{#1}\@@href}%
\providecommand \@@href[1]{\endgroup#1\@@endlink}%
\providecommand \@sanitize@url [0]{\catcode `\\12\catcode `\$12\catcode
  `\&12\catcode `\#12\catcode `\^12\catcode `\_12\catcode `\%12\relax}%
\providecommand \@@startlink[1]{}%
\providecommand \@@endlink[0]{}%
\providecommand \url  [0]{\begingroup\@sanitize@url \@url }%
\providecommand \@url [1]{\endgroup\@href {#1}{\urlprefix }}%
\providecommand \urlprefix  [0]{URL }%
\providecommand \Eprint [0]{\href }%
\providecommand \doibase [0]{https://doi.org/}%
\providecommand \selectlanguage [0]{\@gobble}%
\providecommand \bibinfo  [0]{\@secondoftwo}%
\providecommand \bibfield  [0]{\@secondoftwo}%
\providecommand \translation [1]{[#1]}%
\providecommand \BibitemOpen [0]{}%
\providecommand \bibitemStop [0]{}%
\providecommand \bibitemNoStop [0]{.\EOS\space}%
\providecommand \EOS [0]{\spacefactor3000\relax}%
\providecommand \BibitemShut  [1]{\csname bibitem#1\endcsname}%
\let\auto@bib@innerbib\@empty
\bibitem [{\citenamefont {Preskill}(2018)}]{Preskill2018quantumcomputingin}%
  \BibitemOpen
  \bibfield  {author} {\bibinfo {author} {\bibfnamefont {J.}~\bibnamefont
  {Preskill}},\ }\bibfield  {title} {\bibinfo {title} {Quantum {C}omputing in
  the {NISQ} era and beyond},\ }\href
  {https://doi.org/10.22331/q-2018-08-06-79} {\bibfield  {journal} {\bibinfo
  {journal} {{Quantum}}\ }\textbf {\bibinfo {volume} {2}},\ \bibinfo {pages}
  {79} (\bibinfo {year} {2018})}\BibitemShut {NoStop}%
\bibitem [{\citenamefont {Shor}(1996)}]{548464}%
  \BibitemOpen
  \bibfield  {author} {\bibinfo {author} {\bibfnamefont {P.}~\bibnamefont
  {Shor}},\ }\bibfield  {title} {\bibinfo {title} {Fault-tolerant quantum
  computation},\ }in\ \href {https://doi.org/10.1109/SFCS.1996.548464} {\emph
  {\bibinfo {booktitle} {Proceedings of 37th Conference on Foundations of
  Computer Science}}}\ (\bibinfo {year} {1996})\ pp.\ \bibinfo {pages}
  {56--65}\BibitemShut {NoStop}%
\bibitem [{\citenamefont {Raussendorf}\ and\ \citenamefont
  {Harrington}(2007)}]{PhysRevLett.98.190504}%
  \BibitemOpen
  \bibfield  {author} {\bibinfo {author} {\bibfnamefont {R.}~\bibnamefont
  {Raussendorf}}\ and\ \bibinfo {author} {\bibfnamefont {J.}~\bibnamefont
  {Harrington}},\ }\bibfield  {title} {\bibinfo {title} {Fault-tolerant quantum
  computation with high threshold in two dimensions},\ }\href
  {https://doi.org/10.1103/PhysRevLett.98.190504} {\bibfield  {journal}
  {\bibinfo  {journal} {Phys. Rev. Lett.}\ }\textbf {\bibinfo {volume} {98}},\
  \bibinfo {pages} {190504} (\bibinfo {year} {2007})}\BibitemShut {NoStop}%
\bibitem [{\citenamefont {Fowler}\ \emph {et~al.}(2012)\citenamefont {Fowler},
  \citenamefont {Mariantoni}, \citenamefont {Martinis},\ and\ \citenamefont
  {Cleland}}]{PhysRevA.86.032324}%
  \BibitemOpen
  \bibfield  {author} {\bibinfo {author} {\bibfnamefont {A.~G.}\ \bibnamefont
  {Fowler}}, \bibinfo {author} {\bibfnamefont {M.}~\bibnamefont {Mariantoni}},
  \bibinfo {author} {\bibfnamefont {J.~M.}\ \bibnamefont {Martinis}},\ and\
  \bibinfo {author} {\bibfnamefont {A.~N.}\ \bibnamefont {Cleland}},\
  }\bibfield  {title} {\bibinfo {title} {Surface codes: Towards practical
  large-scale quantum computation},\ }\href
  {https://doi.org/10.1103/PhysRevA.86.032324} {\bibfield  {journal} {\bibinfo
  {journal} {Phys. Rev. A}\ }\textbf {\bibinfo {volume} {86}},\ \bibinfo
  {pages} {032324} (\bibinfo {year} {2012})}\BibitemShut {NoStop}%
\bibitem [{\citenamefont {Lloyd}(1995)}]{PhysRevLett.75.346}%
  \BibitemOpen
  \bibfield  {author} {\bibinfo {author} {\bibfnamefont {S.}~\bibnamefont
  {Lloyd}},\ }\bibfield  {title} {\bibinfo {title} {Almost any quantum logic
  gate is universal},\ }\href {https://doi.org/10.1103/PhysRevLett.75.346}
  {\bibfield  {journal} {\bibinfo  {journal} {Phys. Rev. Lett.}\ }\textbf
  {\bibinfo {volume} {75}},\ \bibinfo {pages} {346} (\bibinfo {year}
  {1995})}\BibitemShut {NoStop}%
\bibitem [{\citenamefont {Bremner}\ \emph {et~al.}(2002)\citenamefont
  {Bremner}, \citenamefont {Dawson}, \citenamefont {Dodd}, \citenamefont
  {Gilchrist}, \citenamefont {Harrow}, \citenamefont {Mortimer}, \citenamefont
  {Nielsen},\ and\ \citenamefont {Osborne}}]{PhysRevLett.89.247902}%
  \BibitemOpen
  \bibfield  {author} {\bibinfo {author} {\bibfnamefont {M.~J.}\ \bibnamefont
  {Bremner}}, \bibinfo {author} {\bibfnamefont {C.~M.}\ \bibnamefont {Dawson}},
  \bibinfo {author} {\bibfnamefont {J.~L.}\ \bibnamefont {Dodd}}, \bibinfo
  {author} {\bibfnamefont {A.}~\bibnamefont {Gilchrist}}, \bibinfo {author}
  {\bibfnamefont {A.~W.}\ \bibnamefont {Harrow}}, \bibinfo {author}
  {\bibfnamefont {D.}~\bibnamefont {Mortimer}}, \bibinfo {author}
  {\bibfnamefont {M.~A.}\ \bibnamefont {Nielsen}},\ and\ \bibinfo {author}
  {\bibfnamefont {T.~J.}\ \bibnamefont {Osborne}},\ }\bibfield  {title}
  {\bibinfo {title} {Practical scheme for quantum computation with any
  two-qubit entangling gate},\ }\href
  {https://doi.org/10.1103/PhysRevLett.89.247902} {\bibfield  {journal}
  {\bibinfo  {journal} {Phys. Rev. Lett.}\ }\textbf {\bibinfo {volume} {89}},\
  \bibinfo {pages} {247902} (\bibinfo {year} {2002})}\BibitemShut {NoStop}%
\bibitem [{\citenamefont {Berry}(1984)}]{doi:10.1098/rspa.1984.0023}%
  \BibitemOpen
  \bibfield  {author} {\bibinfo {author} {\bibfnamefont {M.~V.}\ \bibnamefont
  {Berry}},\ }\bibfield  {title} {\bibinfo {title} {Quantal phase factors
  accompanying adiabatic changes},\ }\href
  {https://doi.org/10.1098/rspa.1984.0023} {\bibfield  {journal} {\bibinfo
  {journal} {Proceedings of the Royal Society of London. A. Mathematical and
  Physical Sciences}\ }\textbf {\bibinfo {volume} {392}},\ \bibinfo {pages}
  {45} (\bibinfo {year} {1984})}\BibitemShut {NoStop}%
\bibitem [{\citenamefont {Wilczek}\ and\ \citenamefont
  {Zee}(1984)}]{PhysRevLett.52.2111}%
  \BibitemOpen
  \bibfield  {author} {\bibinfo {author} {\bibfnamefont {F.}~\bibnamefont
  {Wilczek}}\ and\ \bibinfo {author} {\bibfnamefont {A.}~\bibnamefont {Zee}},\
  }\bibfield  {title} {\bibinfo {title} {Appearance of gauge structure in
  simple dynamical systems},\ }\href
  {https://doi.org/10.1103/PhysRevLett.52.2111} {\bibfield  {journal} {\bibinfo
   {journal} {Phys. Rev. Lett.}\ }\textbf {\bibinfo {volume} {52}},\ \bibinfo
  {pages} {2111} (\bibinfo {year} {1984})}\BibitemShut {NoStop}%
\bibitem [{\citenamefont {Aharonov}\ and\ \citenamefont
  {Anandan}(1987)}]{PhysRevLett.58.1593}%
  \BibitemOpen
  \bibfield  {author} {\bibinfo {author} {\bibfnamefont {Y.}~\bibnamefont
  {Aharonov}}\ and\ \bibinfo {author} {\bibfnamefont {J.}~\bibnamefont
  {Anandan}},\ }\bibfield  {title} {\bibinfo {title} {Phase change during a
  cyclic quantum evolution},\ }\href
  {https://doi.org/10.1103/PhysRevLett.58.1593} {\bibfield  {journal} {\bibinfo
   {journal} {Phys. Rev. Lett.}\ }\textbf {\bibinfo {volume} {58}},\ \bibinfo
  {pages} {1593} (\bibinfo {year} {1987})}\BibitemShut {NoStop}%
\bibitem [{\citenamefont {Anandan}(1988)}]{ANANDAN1988171}%
  \BibitemOpen
  \bibfield  {author} {\bibinfo {author} {\bibfnamefont {J.}~\bibnamefont
  {Anandan}},\ }\bibfield  {title} {\bibinfo {title} {Non-adiabatic non-abelian
  geometric phase},\ }\href
  {https://doi.org/https://doi.org/10.1016/0375-9601(88)91010-9} {\bibfield
  {journal} {\bibinfo  {journal} {Physics Letters A}\ }\textbf {\bibinfo
  {volume} {133}},\ \bibinfo {pages} {171} (\bibinfo {year}
  {1988})}\BibitemShut {NoStop}%
\bibitem [{\citenamefont {De~Chiara}\ and\ \citenamefont
  {Palma}(2003)}]{PhysRevLett.91.090404}%
  \BibitemOpen
  \bibfield  {author} {\bibinfo {author} {\bibfnamefont {G.}~\bibnamefont
  {De~Chiara}}\ and\ \bibinfo {author} {\bibfnamefont {G.~M.}\ \bibnamefont
  {Palma}},\ }\bibfield  {title} {\bibinfo {title} {Berry phase for a spin
  $1/2$ particle in a classical fluctuating field},\ }\href
  {https://doi.org/10.1103/PhysRevLett.91.090404} {\bibfield  {journal}
  {\bibinfo  {journal} {Phys. Rev. Lett.}\ }\textbf {\bibinfo {volume} {91}},\
  \bibinfo {pages} {090404} (\bibinfo {year} {2003})}\BibitemShut {NoStop}%
\bibitem [{\citenamefont {Carollo}\ \emph {et~al.}(2004)\citenamefont
  {Carollo}, \citenamefont {Fuentes-Guridi}, \citenamefont {Santos},\ and\
  \citenamefont {Vedral}}]{PhysRevLett.92.020402}%
  \BibitemOpen
  \bibfield  {author} {\bibinfo {author} {\bibfnamefont {A.}~\bibnamefont
  {Carollo}}, \bibinfo {author} {\bibfnamefont {I.}~\bibnamefont
  {Fuentes-Guridi}}, \bibinfo {author} {\bibfnamefont {M.~F. m.~c.}\
  \bibnamefont {Santos}},\ and\ \bibinfo {author} {\bibfnamefont
  {V.}~\bibnamefont {Vedral}},\ }\bibfield  {title} {\bibinfo {title}
  {Spin-$1/2$ geometric phase driven by decohering quantum fields},\ }\href
  {https://doi.org/10.1103/PhysRevLett.92.020402} {\bibfield  {journal}
  {\bibinfo  {journal} {Phys. Rev. Lett.}\ }\textbf {\bibinfo {volume} {92}},\
  \bibinfo {pages} {020402} (\bibinfo {year} {2004})}\BibitemShut {NoStop}%
\bibitem [{\citenamefont {Zhu}\ and\ \citenamefont
  {Zanardi}(2005)}]{PhysRevA.72.020301}%
  \BibitemOpen
  \bibfield  {author} {\bibinfo {author} {\bibfnamefont {S.-L.}\ \bibnamefont
  {Zhu}}\ and\ \bibinfo {author} {\bibfnamefont {P.}~\bibnamefont {Zanardi}},\
  }\bibfield  {title} {\bibinfo {title} {Geometric quantum gates that are
  robust against stochastic control errors},\ }\href
  {https://doi.org/10.1103/PhysRevA.72.020301} {\bibfield  {journal} {\bibinfo
  {journal} {Phys. Rev. A}\ }\textbf {\bibinfo {volume} {72}},\ \bibinfo
  {pages} {020301} (\bibinfo {year} {2005})}\BibitemShut {NoStop}%
\bibitem [{\citenamefont {Leek}\ \emph {et~al.}(2007)\citenamefont {Leek},
  \citenamefont {Fink}, \citenamefont {Blais}, \citenamefont {Bianchetti},
  \citenamefont {Göppl}, \citenamefont {Gambetta}, \citenamefont {Schuster},
  \citenamefont {Frunzio}, \citenamefont {Schoelkopf},\ and\ \citenamefont
  {Wallraff}}]{doi:10.1126/science.1149858}%
  \BibitemOpen
  \bibfield  {author} {\bibinfo {author} {\bibfnamefont {P.~J.}\ \bibnamefont
  {Leek}}, \bibinfo {author} {\bibfnamefont {J.~M.}\ \bibnamefont {Fink}},
  \bibinfo {author} {\bibfnamefont {A.}~\bibnamefont {Blais}}, \bibinfo
  {author} {\bibfnamefont {R.}~\bibnamefont {Bianchetti}}, \bibinfo {author}
  {\bibfnamefont {M.}~\bibnamefont {Göppl}}, \bibinfo {author} {\bibfnamefont
  {J.~M.}\ \bibnamefont {Gambetta}}, \bibinfo {author} {\bibfnamefont {D.~I.}\
  \bibnamefont {Schuster}}, \bibinfo {author} {\bibfnamefont {L.}~\bibnamefont
  {Frunzio}}, \bibinfo {author} {\bibfnamefont {R.~J.}\ \bibnamefont
  {Schoelkopf}},\ and\ \bibinfo {author} {\bibfnamefont {A.}~\bibnamefont
  {Wallraff}},\ }\bibfield  {title} {\bibinfo {title} {Observation of berry's
  phase in a solid-state qubit},\ }\href
  {https://doi.org/10.1126/science.1149858} {\bibfield  {journal} {\bibinfo
  {journal} {Science}\ }\textbf {\bibinfo {volume} {318}},\ \bibinfo {pages}
  {1889} (\bibinfo {year} {2007})}\BibitemShut {NoStop}%
\bibitem [{\citenamefont {Filipp}\ \emph {et~al.}(2009)\citenamefont {Filipp},
  \citenamefont {Klepp}, \citenamefont {Hasegawa}, \citenamefont
  {Plonka-Spehr}, \citenamefont {Schmidt}, \citenamefont {Geltenbort},\ and\
  \citenamefont {Rauch}}]{PhysRevLett.102.030404}%
  \BibitemOpen
  \bibfield  {author} {\bibinfo {author} {\bibfnamefont {S.}~\bibnamefont
  {Filipp}}, \bibinfo {author} {\bibfnamefont {J.}~\bibnamefont {Klepp}},
  \bibinfo {author} {\bibfnamefont {Y.}~\bibnamefont {Hasegawa}}, \bibinfo
  {author} {\bibfnamefont {C.}~\bibnamefont {Plonka-Spehr}}, \bibinfo {author}
  {\bibfnamefont {U.}~\bibnamefont {Schmidt}}, \bibinfo {author} {\bibfnamefont
  {P.}~\bibnamefont {Geltenbort}},\ and\ \bibinfo {author} {\bibfnamefont
  {H.}~\bibnamefont {Rauch}},\ }\bibfield  {title} {\bibinfo {title}
  {Experimental demonstration of the stability of berry's phase for a
  spin-$1/2$ particle},\ }\href
  {https://doi.org/10.1103/PhysRevLett.102.030404} {\bibfield  {journal}
  {\bibinfo  {journal} {Phys. Rev. Lett.}\ }\textbf {\bibinfo {volume} {102}},\
  \bibinfo {pages} {030404} (\bibinfo {year} {2009})}\BibitemShut {NoStop}%
\bibitem [{\citenamefont {Thomas}\ \emph {et~al.}(2011)\citenamefont {Thomas},
  \citenamefont {Lababidi},\ and\ \citenamefont {Tian}}]{PhysRevA.84.042335}%
  \BibitemOpen
  \bibfield  {author} {\bibinfo {author} {\bibfnamefont {J.~T.}\ \bibnamefont
  {Thomas}}, \bibinfo {author} {\bibfnamefont {M.}~\bibnamefont {Lababidi}},\
  and\ \bibinfo {author} {\bibfnamefont {M.}~\bibnamefont {Tian}},\ }\bibfield
  {title} {\bibinfo {title} {Robustness of single-qubit geometric gate against
  systematic error},\ }\href {https://doi.org/10.1103/PhysRevA.84.042335}
  {\bibfield  {journal} {\bibinfo  {journal} {Phys. Rev. A}\ }\textbf {\bibinfo
  {volume} {84}},\ \bibinfo {pages} {042335} (\bibinfo {year}
  {2011})}\BibitemShut {NoStop}%
\bibitem [{\citenamefont {Johansson}\ \emph
  {et~al.}(2012{\natexlab{a}})\citenamefont {Johansson}, \citenamefont
  {Sj\"oqvist}, \citenamefont {Andersson}, \citenamefont {Ericsson},
  \citenamefont {Hessmo}, \citenamefont {Singh},\ and\ \citenamefont
  {Tong}}]{PhysRevA.86.062322}%
  \BibitemOpen
  \bibfield  {author} {\bibinfo {author} {\bibfnamefont {M.}~\bibnamefont
  {Johansson}}, \bibinfo {author} {\bibfnamefont {E.}~\bibnamefont
  {Sj\"oqvist}}, \bibinfo {author} {\bibfnamefont {L.~M.}\ \bibnamefont
  {Andersson}}, \bibinfo {author} {\bibfnamefont {M.}~\bibnamefont {Ericsson}},
  \bibinfo {author} {\bibfnamefont {B.}~\bibnamefont {Hessmo}}, \bibinfo
  {author} {\bibfnamefont {K.}~\bibnamefont {Singh}},\ and\ \bibinfo {author}
  {\bibfnamefont {D.~M.}\ \bibnamefont {Tong}},\ }\bibfield  {title} {\bibinfo
  {title} {Robustness of nonadiabatic holonomic gates},\ }\href
  {https://doi.org/10.1103/PhysRevA.86.062322} {\bibfield  {journal} {\bibinfo
  {journal} {Phys. Rev. A}\ }\textbf {\bibinfo {volume} {86}},\ \bibinfo
  {pages} {062322} (\bibinfo {year} {2012}{\natexlab{a}})}\BibitemShut
  {NoStop}%
\bibitem [{\citenamefont {Berger}\ \emph {et~al.}(2013)\citenamefont {Berger},
  \citenamefont {Pechal}, \citenamefont {Abdumalikov}, \citenamefont {Eichler},
  \citenamefont {Steffen}, \citenamefont {Fedorov}, \citenamefont {Wallraff},\
  and\ \citenamefont {Filipp}}]{PhysRevA.87.060303}%
  \BibitemOpen
  \bibfield  {author} {\bibinfo {author} {\bibfnamefont {S.}~\bibnamefont
  {Berger}}, \bibinfo {author} {\bibfnamefont {M.}~\bibnamefont {Pechal}},
  \bibinfo {author} {\bibfnamefont {A.~A.}\ \bibnamefont {Abdumalikov}},
  \bibinfo {author} {\bibfnamefont {C.}~\bibnamefont {Eichler}}, \bibinfo
  {author} {\bibfnamefont {L.}~\bibnamefont {Steffen}}, \bibinfo {author}
  {\bibfnamefont {A.}~\bibnamefont {Fedorov}}, \bibinfo {author} {\bibfnamefont
  {A.}~\bibnamefont {Wallraff}},\ and\ \bibinfo {author} {\bibfnamefont
  {S.}~\bibnamefont {Filipp}},\ }\bibfield  {title} {\bibinfo {title}
  {Exploring the effect of noise on the berry phase},\ }\href
  {https://doi.org/10.1103/PhysRevA.87.060303} {\bibfield  {journal} {\bibinfo
  {journal} {Phys. Rev. A}\ }\textbf {\bibinfo {volume} {87}},\ \bibinfo
  {pages} {060303} (\bibinfo {year} {2013})}\BibitemShut {NoStop}%
\bibitem [{\citenamefont {Wu}\ \emph {et~al.}(2013)\citenamefont {Wu},
  \citenamefont {Gauger}, \citenamefont {George}, \citenamefont {M\"ott\"onen},
  \citenamefont {Riemann}, \citenamefont {Abrosimov}, \citenamefont {Becker},
  \citenamefont {Pohl}, \citenamefont {Itoh}, \citenamefont {Thewalt},\ and\
  \citenamefont {Morton}}]{PhysRevA.87.032326}%
  \BibitemOpen
  \bibfield  {author} {\bibinfo {author} {\bibfnamefont {H.}~\bibnamefont
  {Wu}}, \bibinfo {author} {\bibfnamefont {E.~M.}\ \bibnamefont {Gauger}},
  \bibinfo {author} {\bibfnamefont {R.~E.}\ \bibnamefont {George}}, \bibinfo
  {author} {\bibfnamefont {M.}~\bibnamefont {M\"ott\"onen}}, \bibinfo {author}
  {\bibfnamefont {H.}~\bibnamefont {Riemann}}, \bibinfo {author} {\bibfnamefont
  {N.~V.}\ \bibnamefont {Abrosimov}}, \bibinfo {author} {\bibfnamefont
  {P.}~\bibnamefont {Becker}}, \bibinfo {author} {\bibfnamefont {H.-J.}\
  \bibnamefont {Pohl}}, \bibinfo {author} {\bibfnamefont {K.~M.}\ \bibnamefont
  {Itoh}}, \bibinfo {author} {\bibfnamefont {M.~L.~W.}\ \bibnamefont
  {Thewalt}},\ and\ \bibinfo {author} {\bibfnamefont {J.~J.~L.}\ \bibnamefont
  {Morton}},\ }\bibfield  {title} {\bibinfo {title} {Geometric phase gates with
  adiabatic control in electron spin resonance},\ }\href
  {https://doi.org/10.1103/PhysRevA.87.032326} {\bibfield  {journal} {\bibinfo
  {journal} {Phys. Rev. A}\ }\textbf {\bibinfo {volume} {87}},\ \bibinfo
  {pages} {032326} (\bibinfo {year} {2013})}\BibitemShut {NoStop}%
\bibitem [{\citenamefont {Huang}\ \emph {et~al.}(2019)\citenamefont {Huang},
  \citenamefont {Wu}, \citenamefont {Wang}, \citenamefont {Hou}, \citenamefont
  {Wang}, \citenamefont {Zhang}, \citenamefont {Lian}, \citenamefont {Liu},
  \citenamefont {Wang}, \citenamefont {Zhang}, \citenamefont {He},
  \citenamefont {Chang}, \citenamefont {Xu},\ and\ \citenamefont
  {Duan}}]{PhysRevLett.122.010503}%
  \BibitemOpen
  \bibfield  {author} {\bibinfo {author} {\bibfnamefont {Y.-Y.}\ \bibnamefont
  {Huang}}, \bibinfo {author} {\bibfnamefont {Y.-K.}\ \bibnamefont {Wu}},
  \bibinfo {author} {\bibfnamefont {F.}~\bibnamefont {Wang}}, \bibinfo {author}
  {\bibfnamefont {P.-Y.}\ \bibnamefont {Hou}}, \bibinfo {author} {\bibfnamefont
  {W.-B.}\ \bibnamefont {Wang}}, \bibinfo {author} {\bibfnamefont {W.-G.}\
  \bibnamefont {Zhang}}, \bibinfo {author} {\bibfnamefont {W.-Q.}\ \bibnamefont
  {Lian}}, \bibinfo {author} {\bibfnamefont {Y.-Q.}\ \bibnamefont {Liu}},
  \bibinfo {author} {\bibfnamefont {H.-Y.}\ \bibnamefont {Wang}}, \bibinfo
  {author} {\bibfnamefont {H.-Y.}\ \bibnamefont {Zhang}}, \bibinfo {author}
  {\bibfnamefont {L.}~\bibnamefont {He}}, \bibinfo {author} {\bibfnamefont
  {X.-Y.}\ \bibnamefont {Chang}}, \bibinfo {author} {\bibfnamefont
  {Y.}~\bibnamefont {Xu}},\ and\ \bibinfo {author} {\bibfnamefont {L.-M.}\
  \bibnamefont {Duan}},\ }\bibfield  {title} {\bibinfo {title} {Experimental
  realization of robust geometric quantum gates with solid-state spins},\
  }\href {https://doi.org/10.1103/PhysRevLett.122.010503} {\bibfield  {journal}
  {\bibinfo  {journal} {Phys. Rev. Lett.}\ }\textbf {\bibinfo {volume} {122}},\
  \bibinfo {pages} {010503} (\bibinfo {year} {2019})}\BibitemShut {NoStop}%
\bibitem [{\citenamefont {Zanardi}\ and\ \citenamefont
  {Rasetti}(1999)}]{ZANARDI199994}%
  \BibitemOpen
  \bibfield  {author} {\bibinfo {author} {\bibfnamefont {P.}~\bibnamefont
  {Zanardi}}\ and\ \bibinfo {author} {\bibfnamefont {M.}~\bibnamefont
  {Rasetti}},\ }\bibfield  {title} {\bibinfo {title} {Holonomic quantum
  computation},\ }\href
  {https://doi.org/https://doi.org/10.1016/S0375-9601(99)00803-8} {\bibfield
  {journal} {\bibinfo  {journal} {Physics Letters A}\ }\textbf {\bibinfo
  {volume} {264}},\ \bibinfo {pages} {94} (\bibinfo {year} {1999})}\BibitemShut
  {NoStop}%
\bibitem [{\citenamefont {Jones}\ \emph {et~al.}(2000)\citenamefont {Jones},
  \citenamefont {Vedral}, \citenamefont {Ekert},\ and\ \citenamefont
  {Castagnoli}}]{jones2000geometric}%
  \BibitemOpen
  \bibfield  {author} {\bibinfo {author} {\bibfnamefont {J.~A.}\ \bibnamefont
  {Jones}}, \bibinfo {author} {\bibfnamefont {V.}~\bibnamefont {Vedral}},
  \bibinfo {author} {\bibfnamefont {A.}~\bibnamefont {Ekert}},\ and\ \bibinfo
  {author} {\bibfnamefont {G.}~\bibnamefont {Castagnoli}},\ }\bibfield  {title}
  {\bibinfo {title} {Geometric quantum computation using nuclear magnetic
  resonance},\ }\href {https://doi.org/10.1038/35002528} {\bibfield  {journal}
  {\bibinfo  {journal} {Nature}\ }\textbf {\bibinfo {volume} {403}},\ \bibinfo
  {pages} {869} (\bibinfo {year} {2000})}\BibitemShut {NoStop}%
\bibitem [{\citenamefont {Duan}\ \emph {et~al.}(2001)\citenamefont {Duan},
  \citenamefont {Cirac},\ and\ \citenamefont
  {Zoller}}]{doi:10.1126/science.1058835}%
  \BibitemOpen
  \bibfield  {author} {\bibinfo {author} {\bibfnamefont {L.-M.}\ \bibnamefont
  {Duan}}, \bibinfo {author} {\bibfnamefont {J.~I.}\ \bibnamefont {Cirac}},\
  and\ \bibinfo {author} {\bibfnamefont {P.}~\bibnamefont {Zoller}},\
  }\bibfield  {title} {\bibinfo {title} {Geometric manipulation of trapped ions
  for quantum computation},\ }\href {https://doi.org/10.1126/science.1058835}
  {\bibfield  {journal} {\bibinfo  {journal} {Science}\ }\textbf {\bibinfo
  {volume} {292}},\ \bibinfo {pages} {1695} (\bibinfo {year}
  {2001})}\BibitemShut {NoStop}%
\bibitem [{\citenamefont {Wu}\ \emph {et~al.}(2005)\citenamefont {Wu},
  \citenamefont {Zanardi},\ and\ \citenamefont
  {Lidar}}]{PhysRevLett.95.130501}%
  \BibitemOpen
  \bibfield  {author} {\bibinfo {author} {\bibfnamefont {L.-A.}\ \bibnamefont
  {Wu}}, \bibinfo {author} {\bibfnamefont {P.}~\bibnamefont {Zanardi}},\ and\
  \bibinfo {author} {\bibfnamefont {D.~A.}\ \bibnamefont {Lidar}},\ }\bibfield
  {title} {\bibinfo {title} {Holonomic quantum computation in decoherence-free
  subspaces},\ }\href {https://doi.org/10.1103/PhysRevLett.95.130501}
  {\bibfield  {journal} {\bibinfo  {journal} {Phys. Rev. Lett.}\ }\textbf
  {\bibinfo {volume} {95}},\ \bibinfo {pages} {130501} (\bibinfo {year}
  {2005})}\BibitemShut {NoStop}%
\bibitem [{\citenamefont {Toyoda}\ \emph {et~al.}(2013)\citenamefont {Toyoda},
  \citenamefont {Uchida}, \citenamefont {Noguchi}, \citenamefont {Haze},\ and\
  \citenamefont {Urabe}}]{PhysRevA.87.052307}%
  \BibitemOpen
  \bibfield  {author} {\bibinfo {author} {\bibfnamefont {K.}~\bibnamefont
  {Toyoda}}, \bibinfo {author} {\bibfnamefont {K.}~\bibnamefont {Uchida}},
  \bibinfo {author} {\bibfnamefont {A.}~\bibnamefont {Noguchi}}, \bibinfo
  {author} {\bibfnamefont {S.}~\bibnamefont {Haze}},\ and\ \bibinfo {author}
  {\bibfnamefont {S.}~\bibnamefont {Urabe}},\ }\bibfield  {title} {\bibinfo
  {title} {Realization of holonomic single-qubit operations},\ }\href
  {https://doi.org/10.1103/PhysRevA.87.052307} {\bibfield  {journal} {\bibinfo
  {journal} {Phys. Rev. A}\ }\textbf {\bibinfo {volume} {87}},\ \bibinfo
  {pages} {052307} (\bibinfo {year} {2013})}\BibitemShut {NoStop}%
\bibitem [{\citenamefont {Leroux}\ \emph {et~al.}(2018)\citenamefont {Leroux},
  \citenamefont {Pandey}, \citenamefont {Rehbi}, \citenamefont {Chevy},
  \citenamefont {Miniatura}, \citenamefont {Gr{\'e}maud},\ and\ \citenamefont
  {Wilkowski}}]{leroux2018non}%
  \BibitemOpen
  \bibfield  {author} {\bibinfo {author} {\bibfnamefont {F.}~\bibnamefont
  {Leroux}}, \bibinfo {author} {\bibfnamefont {K.}~\bibnamefont {Pandey}},
  \bibinfo {author} {\bibfnamefont {R.}~\bibnamefont {Rehbi}}, \bibinfo
  {author} {\bibfnamefont {F.}~\bibnamefont {Chevy}}, \bibinfo {author}
  {\bibfnamefont {C.}~\bibnamefont {Miniatura}}, \bibinfo {author}
  {\bibfnamefont {B.}~\bibnamefont {Gr{\'e}maud}},\ and\ \bibinfo {author}
  {\bibfnamefont {D.}~\bibnamefont {Wilkowski}},\ }\bibfield  {title} {\bibinfo
  {title} {Non-abelian adiabatic geometric transformations in a cold strontium
  gas},\ }\href {https://doi.org/10.1038/s41467-018-05865-3} {\bibfield
  {journal} {\bibinfo  {journal} {Nat. Commun.}\ }\textbf {\bibinfo {volume}
  {9}},\ \bibinfo {pages} {3580} (\bibinfo {year} {2018})}\BibitemShut
  {NoStop}%
\bibitem [{\citenamefont {Bason}\ \emph {et~al.}(2012)\citenamefont {Bason},
  \citenamefont {Viteau}, \citenamefont {Malossi}, \citenamefont {Huillery},
  \citenamefont {Arimondo}, \citenamefont {Ciampini}, \citenamefont {Fazio},
  \citenamefont {Giovannetti}, \citenamefont {Mannella},\ and\ \citenamefont
  {Morsch}}]{bason2012high}%
  \BibitemOpen
  \bibfield  {author} {\bibinfo {author} {\bibfnamefont {M.~G.}\ \bibnamefont
  {Bason}}, \bibinfo {author} {\bibfnamefont {M.}~\bibnamefont {Viteau}},
  \bibinfo {author} {\bibfnamefont {N.}~\bibnamefont {Malossi}}, \bibinfo
  {author} {\bibfnamefont {P.}~\bibnamefont {Huillery}}, \bibinfo {author}
  {\bibfnamefont {E.}~\bibnamefont {Arimondo}}, \bibinfo {author}
  {\bibfnamefont {D.}~\bibnamefont {Ciampini}}, \bibinfo {author}
  {\bibfnamefont {R.}~\bibnamefont {Fazio}}, \bibinfo {author} {\bibfnamefont
  {V.}~\bibnamefont {Giovannetti}}, \bibinfo {author} {\bibfnamefont
  {R.}~\bibnamefont {Mannella}},\ and\ \bibinfo {author} {\bibfnamefont
  {O.}~\bibnamefont {Morsch}},\ }\bibfield  {title} {\bibinfo {title}
  {High-fidelity quantum driving},\ }\href {https://doi.org/10.1038/nphys2170}
  {\bibfield  {journal} {\bibinfo  {journal} {Nat. Physics}\ }\textbf {\bibinfo
  {volume} {8}},\ \bibinfo {pages} {147} (\bibinfo {year} {2012})}\BibitemShut
  {NoStop}%
\bibitem [{\citenamefont {Song}\ \emph {et~al.}(2016)\citenamefont {Song},
  \citenamefont {Zhang}, \citenamefont {Ai}, \citenamefont {Qiu},\ and\
  \citenamefont {Deng}}]{Song_2016}%
  \BibitemOpen
  \bibfield  {author} {\bibinfo {author} {\bibfnamefont {X.-K.}\ \bibnamefont
  {Song}}, \bibinfo {author} {\bibfnamefont {H.}~\bibnamefont {Zhang}},
  \bibinfo {author} {\bibfnamefont {Q.}~\bibnamefont {Ai}}, \bibinfo {author}
  {\bibfnamefont {J.}~\bibnamefont {Qiu}},\ and\ \bibinfo {author}
  {\bibfnamefont {F.-G.}\ \bibnamefont {Deng}},\ }\bibfield  {title} {\bibinfo
  {title} {Shortcuts to adiabatic holonomic quantum computation in
  decoherence-free subspace with transitionless quantum driving algorithm},\
  }\href {https://doi.org/10.1088/1367-2630/18/2/023001} {\bibfield  {journal}
  {\bibinfo  {journal} {New Journal of Physics}\ }\textbf {\bibinfo {volume}
  {18}},\ \bibinfo {pages} {023001} (\bibinfo {year} {2016})}\BibitemShut
  {NoStop}%
\bibitem [{\citenamefont {Zhang}\ \emph {et~al.}(2016)\citenamefont {Zhang},
  \citenamefont {Kyaw}, \citenamefont {Tong}, \citenamefont {Sj{\"o}qvist},\
  and\ \citenamefont {Kwek}}]{zhang2015fast}%
  \BibitemOpen
  \bibfield  {author} {\bibinfo {author} {\bibfnamefont {J.}~\bibnamefont
  {Zhang}}, \bibinfo {author} {\bibfnamefont {T.~H.}\ \bibnamefont {Kyaw}},
  \bibinfo {author} {\bibfnamefont {D.}~\bibnamefont {Tong}}, \bibinfo {author}
  {\bibfnamefont {E.}~\bibnamefont {Sj{\"o}qvist}},\ and\ \bibinfo {author}
  {\bibfnamefont {L.-C.}\ \bibnamefont {Kwek}},\ }\bibfield  {title} {\bibinfo
  {title} {Fast non-abelian geometric gates via transitionless quantum
  driving},\ }\href {https://doi.org/10.1038/srep18414} {\bibfield  {journal}
  {\bibinfo  {journal} {Scientific reports}\ }\textbf {\bibinfo {volume} {5}},\
  \bibinfo {pages} {1} (\bibinfo {year} {2016})}\BibitemShut {NoStop}%
\bibitem [{\citenamefont {Zhou}\ \emph
  {et~al.}(2017{\natexlab{a}})\citenamefont {Zhou}, \citenamefont {Baksic},
  \citenamefont {Ribeiro}, \citenamefont {Yale}, \citenamefont {Heremans},
  \citenamefont {Jerger}, \citenamefont {Auer}, \citenamefont {Burkard},
  \citenamefont {Clerk},\ and\ \citenamefont
  {Awschalom}}]{zhou2017accelerated}%
  \BibitemOpen
  \bibfield  {author} {\bibinfo {author} {\bibfnamefont {B.~B.}\ \bibnamefont
  {Zhou}}, \bibinfo {author} {\bibfnamefont {A.}~\bibnamefont {Baksic}},
  \bibinfo {author} {\bibfnamefont {H.}~\bibnamefont {Ribeiro}}, \bibinfo
  {author} {\bibfnamefont {C.~G.}\ \bibnamefont {Yale}}, \bibinfo {author}
  {\bibfnamefont {F.~J.}\ \bibnamefont {Heremans}}, \bibinfo {author}
  {\bibfnamefont {P.~C.}\ \bibnamefont {Jerger}}, \bibinfo {author}
  {\bibfnamefont {A.}~\bibnamefont {Auer}}, \bibinfo {author} {\bibfnamefont
  {G.}~\bibnamefont {Burkard}}, \bibinfo {author} {\bibfnamefont {A.~A.}\
  \bibnamefont {Clerk}},\ and\ \bibinfo {author} {\bibfnamefont {D.~D.}\
  \bibnamefont {Awschalom}},\ }\bibfield  {title} {\bibinfo {title}
  {Accelerated quantum control using superadiabatic dynamics in a solid-state
  lambda system},\ }\href {https://doi.org/10.1038/nphys3967} {\bibfield
  {journal} {\bibinfo  {journal} {Nat. Physics}\ }\textbf {\bibinfo {volume}
  {13}},\ \bibinfo {pages} {330} (\bibinfo {year}
  {2017}{\natexlab{a}})}\BibitemShut {NoStop}%
\bibitem [{\citenamefont {Klei{\ss}ler}\ \emph {et~al.}(2018)\citenamefont
  {Klei{\ss}ler}, \citenamefont {Lazariev},\ and\ \citenamefont
  {Arroyo-Camejo}}]{kleissler2018universal}%
  \BibitemOpen
  \bibfield  {author} {\bibinfo {author} {\bibfnamefont {F.}~\bibnamefont
  {Klei{\ss}ler}}, \bibinfo {author} {\bibfnamefont {A.}~\bibnamefont
  {Lazariev}},\ and\ \bibinfo {author} {\bibfnamefont {S.}~\bibnamefont
  {Arroyo-Camejo}},\ }\bibfield  {title} {\bibinfo {title} {Universal,
  high-fidelity quantum gates based on superadiabatic, geometric phases on a
  solid-state spin-qubit at room temperature},\ }\href
  {https://doi.org/10.1038/s41534-018-0098-7} {\bibfield  {journal} {\bibinfo
  {journal} {npj Quantum Information}\ }\textbf {\bibinfo {volume} {4}},\
  \bibinfo {pages} {1} (\bibinfo {year} {2018})}\BibitemShut {NoStop}%
\bibitem [{\citenamefont {Yan}\ \emph {et~al.}(2019)\citenamefont {Yan},
  \citenamefont {Liu}, \citenamefont {Xu}, \citenamefont {Song}, \citenamefont
  {Liu}, \citenamefont {Zhang}, \citenamefont {Deng}, \citenamefont {Yan},
  \citenamefont {Rong}, \citenamefont {Huang}, \citenamefont {Yung},
  \citenamefont {Chen},\ and\ \citenamefont {Yu}}]{PhysRevLett.122.080501}%
  \BibitemOpen
  \bibfield  {author} {\bibinfo {author} {\bibfnamefont {T.}~\bibnamefont
  {Yan}}, \bibinfo {author} {\bibfnamefont {B.-J.}\ \bibnamefont {Liu}},
  \bibinfo {author} {\bibfnamefont {K.}~\bibnamefont {Xu}}, \bibinfo {author}
  {\bibfnamefont {C.}~\bibnamefont {Song}}, \bibinfo {author} {\bibfnamefont
  {S.}~\bibnamefont {Liu}}, \bibinfo {author} {\bibfnamefont {Z.}~\bibnamefont
  {Zhang}}, \bibinfo {author} {\bibfnamefont {H.}~\bibnamefont {Deng}},
  \bibinfo {author} {\bibfnamefont {Z.}~\bibnamefont {Yan}}, \bibinfo {author}
  {\bibfnamefont {H.}~\bibnamefont {Rong}}, \bibinfo {author} {\bibfnamefont
  {K.}~\bibnamefont {Huang}}, \bibinfo {author} {\bibfnamefont {M.-H.}\
  \bibnamefont {Yung}}, \bibinfo {author} {\bibfnamefont {Y.}~\bibnamefont
  {Chen}},\ and\ \bibinfo {author} {\bibfnamefont {D.}~\bibnamefont {Yu}},\
  }\bibfield  {title} {\bibinfo {title} {Experimental realization of
  nonadiabatic shortcut to non-abelian geometric gates},\ }\href
  {https://doi.org/10.1103/PhysRevLett.122.080501} {\bibfield  {journal}
  {\bibinfo  {journal} {Phys. Rev. Lett.}\ }\textbf {\bibinfo {volume} {122}},\
  \bibinfo {pages} {080501} (\bibinfo {year} {2019})}\BibitemShut {NoStop}%
\bibitem [{\citenamefont {Chu}\ \emph {et~al.}(2020)\citenamefont {Chu},
  \citenamefont {Li}, \citenamefont {Yang}, \citenamefont {Song}, \citenamefont
  {Han}, \citenamefont {Yang}, \citenamefont {Dong}, \citenamefont {Zheng},
  \citenamefont {Wang}, \citenamefont {Yu}, \citenamefont {Lan}, \citenamefont
  {Tan},\ and\ \citenamefont {Yu}}]{PhysRevApplied.13.064012}%
  \BibitemOpen
  \bibfield  {author} {\bibinfo {author} {\bibfnamefont {J.}~\bibnamefont
  {Chu}}, \bibinfo {author} {\bibfnamefont {D.}~\bibnamefont {Li}}, \bibinfo
  {author} {\bibfnamefont {X.}~\bibnamefont {Yang}}, \bibinfo {author}
  {\bibfnamefont {S.}~\bibnamefont {Song}}, \bibinfo {author} {\bibfnamefont
  {Z.}~\bibnamefont {Han}}, \bibinfo {author} {\bibfnamefont {Z.}~\bibnamefont
  {Yang}}, \bibinfo {author} {\bibfnamefont {Y.}~\bibnamefont {Dong}}, \bibinfo
  {author} {\bibfnamefont {W.}~\bibnamefont {Zheng}}, \bibinfo {author}
  {\bibfnamefont {Z.}~\bibnamefont {Wang}}, \bibinfo {author} {\bibfnamefont
  {X.}~\bibnamefont {Yu}}, \bibinfo {author} {\bibfnamefont {D.}~\bibnamefont
  {Lan}}, \bibinfo {author} {\bibfnamefont {X.}~\bibnamefont {Tan}},\ and\
  \bibinfo {author} {\bibfnamefont {Y.}~\bibnamefont {Yu}},\ }\bibfield
  {title} {\bibinfo {title} {Realization of superadiabatic two-qubit gates
  using parametric modulation in superconducting circuits},\ }\href
  {https://doi.org/10.1103/PhysRevApplied.13.064012} {\bibfield  {journal}
  {\bibinfo  {journal} {Phys. Rev. Applied}\ }\textbf {\bibinfo {volume}
  {13}},\ \bibinfo {pages} {064012} (\bibinfo {year} {2020})}\BibitemShut
  {NoStop}%
\bibitem [{\citenamefont {Qiu}\ \emph {et~al.}(2021)\citenamefont {Qiu},
  \citenamefont {Li}, \citenamefont {Han}, \citenamefont {Zheng}, \citenamefont
  {Yang}, \citenamefont {Dong}, \citenamefont {Song}, \citenamefont {Lan},
  \citenamefont {Tan},\ and\ \citenamefont {Yu}}]{doi:10.1063/5.0049967}%
  \BibitemOpen
  \bibfield  {author} {\bibinfo {author} {\bibfnamefont {L.}~\bibnamefont
  {Qiu}}, \bibinfo {author} {\bibfnamefont {H.}~\bibnamefont {Li}}, \bibinfo
  {author} {\bibfnamefont {Z.}~\bibnamefont {Han}}, \bibinfo {author}
  {\bibfnamefont {W.}~\bibnamefont {Zheng}}, \bibinfo {author} {\bibfnamefont
  {X.}~\bibnamefont {Yang}}, \bibinfo {author} {\bibfnamefont {Y.}~\bibnamefont
  {Dong}}, \bibinfo {author} {\bibfnamefont {S.}~\bibnamefont {Song}}, \bibinfo
  {author} {\bibfnamefont {D.}~\bibnamefont {Lan}}, \bibinfo {author}
  {\bibfnamefont {X.}~\bibnamefont {Tan}},\ and\ \bibinfo {author}
  {\bibfnamefont {Y.}~\bibnamefont {Yu}},\ }\bibfield  {title} {\bibinfo
  {title} {Experimental realization of noncyclic geometric gates with shortcut
  to adiabaticity in a superconducting circuit},\ }\href
  {https://doi.org/10.1063/5.0049967} {\bibfield  {journal} {\bibinfo
  {journal} {Applied Physics Letters}\ }\textbf {\bibinfo {volume} {118}},\
  \bibinfo {pages} {254002} (\bibinfo {year} {2021})}\BibitemShut {NoStop}%
\bibitem [{\citenamefont {Abdumalikov~Jr}\ \emph {et~al.}(2013)\citenamefont
  {Abdumalikov~Jr}, \citenamefont {Fink}, \citenamefont {Juliusson},
  \citenamefont {Pechal}, \citenamefont {Berger}, \citenamefont {Wallraff},\
  and\ \citenamefont {Filipp}}]{abdumalikov2013experimental}%
  \BibitemOpen
  \bibfield  {author} {\bibinfo {author} {\bibfnamefont {A.~A.}\ \bibnamefont
  {Abdumalikov~Jr}}, \bibinfo {author} {\bibfnamefont {J.~M.}\ \bibnamefont
  {Fink}}, \bibinfo {author} {\bibfnamefont {K.}~\bibnamefont {Juliusson}},
  \bibinfo {author} {\bibfnamefont {M.}~\bibnamefont {Pechal}}, \bibinfo
  {author} {\bibfnamefont {S.}~\bibnamefont {Berger}}, \bibinfo {author}
  {\bibfnamefont {A.}~\bibnamefont {Wallraff}},\ and\ \bibinfo {author}
  {\bibfnamefont {S.}~\bibnamefont {Filipp}},\ }\bibfield  {title} {\bibinfo
  {title} {Experimental realization of non-abelian non-adiabatic geometric
  gates},\ }\href {https://doi.org/10.1038/nature12010} {\bibfield  {journal}
  {\bibinfo  {journal} {Nature}\ }\textbf {\bibinfo {volume} {496}},\ \bibinfo
  {pages} {482} (\bibinfo {year} {2013})}\BibitemShut {NoStop}%
\bibitem [{\citenamefont {Xu}\ \emph {et~al.}(2018)\citenamefont {Xu},
  \citenamefont {Cai}, \citenamefont {Ma}, \citenamefont {Mu}, \citenamefont
  {Hu}, \citenamefont {Chen}, \citenamefont {Wang}, \citenamefont {Song},
  \citenamefont {Xue}, \citenamefont {Yin},\ and\ \citenamefont
  {Sun}}]{PhysRevLett.121.110501}%
  \BibitemOpen
  \bibfield  {author} {\bibinfo {author} {\bibfnamefont {Y.}~\bibnamefont
  {Xu}}, \bibinfo {author} {\bibfnamefont {W.}~\bibnamefont {Cai}}, \bibinfo
  {author} {\bibfnamefont {Y.}~\bibnamefont {Ma}}, \bibinfo {author}
  {\bibfnamefont {X.}~\bibnamefont {Mu}}, \bibinfo {author} {\bibfnamefont
  {L.}~\bibnamefont {Hu}}, \bibinfo {author} {\bibfnamefont {T.}~\bibnamefont
  {Chen}}, \bibinfo {author} {\bibfnamefont {H.}~\bibnamefont {Wang}}, \bibinfo
  {author} {\bibfnamefont {Y.~P.}\ \bibnamefont {Song}}, \bibinfo {author}
  {\bibfnamefont {Z.-Y.}\ \bibnamefont {Xue}}, \bibinfo {author} {\bibfnamefont
  {Z.-q.}\ \bibnamefont {Yin}},\ and\ \bibinfo {author} {\bibfnamefont
  {L.}~\bibnamefont {Sun}},\ }\bibfield  {title} {\bibinfo {title} {Single-loop
  realization of arbitrary nonadiabatic holonomic single-qubit quantum gates in
  a superconducting circuit},\ }\href
  {https://doi.org/10.1103/PhysRevLett.121.110501} {\bibfield  {journal}
  {\bibinfo  {journal} {Phys. Rev. Lett.}\ }\textbf {\bibinfo {volume} {121}},\
  \bibinfo {pages} {110501} (\bibinfo {year} {2018})}\BibitemShut {NoStop}%
\bibitem [{\citenamefont {Zhang}\ \emph {et~al.}(2019)\citenamefont {Zhang},
  \citenamefont {Zhao}, \citenamefont {Wang}, \citenamefont {Xiang},
  \citenamefont {Jia}, \citenamefont {Duan}, \citenamefont {Tong},
  \citenamefont {Yin},\ and\ \citenamefont {Guo}}]{Zhang_2019}%
  \BibitemOpen
  \bibfield  {author} {\bibinfo {author} {\bibfnamefont {Z.}~\bibnamefont
  {Zhang}}, \bibinfo {author} {\bibfnamefont {P.~Z.}\ \bibnamefont {Zhao}},
  \bibinfo {author} {\bibfnamefont {T.}~\bibnamefont {Wang}}, \bibinfo {author}
  {\bibfnamefont {L.}~\bibnamefont {Xiang}}, \bibinfo {author} {\bibfnamefont
  {Z.}~\bibnamefont {Jia}}, \bibinfo {author} {\bibfnamefont {P.}~\bibnamefont
  {Duan}}, \bibinfo {author} {\bibfnamefont {D.~M.}\ \bibnamefont {Tong}},
  \bibinfo {author} {\bibfnamefont {Y.}~\bibnamefont {Yin}},\ and\ \bibinfo
  {author} {\bibfnamefont {G.}~\bibnamefont {Guo}},\ }\bibfield  {title}
  {\bibinfo {title} {Single-shot realization of nonadiabatic holonomic gates
  with a superconducting xmon qutrit},\ }\href
  {https://doi.org/10.1088/1367-2630/ab2e26} {\bibfield  {journal} {\bibinfo
  {journal} {New Journal of Physics}\ }\textbf {\bibinfo {volume} {21}},\
  \bibinfo {pages} {073024} (\bibinfo {year} {2019})}\BibitemShut {NoStop}%
\bibitem [{\citenamefont {Li}\ \emph {et~al.}(2021{\natexlab{a}})\citenamefont
  {Li}, \citenamefont {Liu}, \citenamefont {Ni}, \citenamefont {Zhang},
  \citenamefont {Xue}, \citenamefont {Li}, \citenamefont {Yan}, \citenamefont
  {Chen}, \citenamefont {Liu}, \citenamefont {Yung}, \citenamefont {Xu},\ and\
  \citenamefont {Yu}}]{PhysRevApplied.16.064003}%
  \BibitemOpen
  \bibfield  {author} {\bibinfo {author} {\bibfnamefont {S.}~\bibnamefont
  {Li}}, \bibinfo {author} {\bibfnamefont {B.-J.}\ \bibnamefont {Liu}},
  \bibinfo {author} {\bibfnamefont {Z.}~\bibnamefont {Ni}}, \bibinfo {author}
  {\bibfnamefont {L.}~\bibnamefont {Zhang}}, \bibinfo {author} {\bibfnamefont
  {Z.-Y.}\ \bibnamefont {Xue}}, \bibinfo {author} {\bibfnamefont
  {J.}~\bibnamefont {Li}}, \bibinfo {author} {\bibfnamefont {F.}~\bibnamefont
  {Yan}}, \bibinfo {author} {\bibfnamefont {Y.}~\bibnamefont {Chen}}, \bibinfo
  {author} {\bibfnamefont {S.}~\bibnamefont {Liu}}, \bibinfo {author}
  {\bibfnamefont {M.-H.}\ \bibnamefont {Yung}}, \bibinfo {author}
  {\bibfnamefont {Y.}~\bibnamefont {Xu}},\ and\ \bibinfo {author}
  {\bibfnamefont {D.}~\bibnamefont {Yu}},\ }\bibfield  {title} {\bibinfo
  {title} {Superrobust geometric control of a superconducting circuit},\ }\href
  {https://doi.org/10.1103/PhysRevApplied.16.064003} {\bibfield  {journal}
  {\bibinfo  {journal} {Phys. Rev. Applied}\ }\textbf {\bibinfo {volume}
  {16}},\ \bibinfo {pages} {064003} (\bibinfo {year}
  {2021}{\natexlab{a}})}\BibitemShut {NoStop}%
\bibitem [{\citenamefont {Feng}\ \emph {et~al.}(2013)\citenamefont {Feng},
  \citenamefont {Xu},\ and\ \citenamefont {Long}}]{PhysRevLett.110.190501}%
  \BibitemOpen
  \bibfield  {author} {\bibinfo {author} {\bibfnamefont {G.}~\bibnamefont
  {Feng}}, \bibinfo {author} {\bibfnamefont {G.}~\bibnamefont {Xu}},\ and\
  \bibinfo {author} {\bibfnamefont {G.}~\bibnamefont {Long}},\ }\bibfield
  {title} {\bibinfo {title} {Experimental realization of nonadiabatic holonomic
  quantum computation},\ }\href
  {https://doi.org/10.1103/PhysRevLett.110.190501} {\bibfield  {journal}
  {\bibinfo  {journal} {Phys. Rev. Lett.}\ }\textbf {\bibinfo {volume} {110}},\
  \bibinfo {pages} {190501} (\bibinfo {year} {2013})}\BibitemShut {NoStop}%
\bibitem [{\citenamefont {Zu}\ \emph {et~al.}(2014)\citenamefont {Zu},
  \citenamefont {Wang}, \citenamefont {He}, \citenamefont {Zhang},
  \citenamefont {Dai}, \citenamefont {Wang},\ and\ \citenamefont
  {Duan}}]{zu2014experimental}%
  \BibitemOpen
  \bibfield  {author} {\bibinfo {author} {\bibfnamefont {C.}~\bibnamefont
  {Zu}}, \bibinfo {author} {\bibfnamefont {W.-B.}\ \bibnamefont {Wang}},
  \bibinfo {author} {\bibfnamefont {L.}~\bibnamefont {He}}, \bibinfo {author}
  {\bibfnamefont {W.-G.}\ \bibnamefont {Zhang}}, \bibinfo {author}
  {\bibfnamefont {C.-Y.}\ \bibnamefont {Dai}}, \bibinfo {author} {\bibfnamefont
  {F.}~\bibnamefont {Wang}},\ and\ \bibinfo {author} {\bibfnamefont {L.-M.}\
  \bibnamefont {Duan}},\ }\bibfield  {title} {\bibinfo {title} {Experimental
  realization of universal geometric quantum gates with solid-state spins},\
  }\href {https://doi.org/10.1038/nature13729} {\bibfield  {journal} {\bibinfo
  {journal} {Nature}\ }\textbf {\bibinfo {volume} {514}},\ \bibinfo {pages}
  {72} (\bibinfo {year} {2014})}\BibitemShut {NoStop}%
\bibitem [{\citenamefont {Arroyo-Camejo}\ \emph {et~al.}(2014)\citenamefont
  {Arroyo-Camejo}, \citenamefont {Lazariev}, \citenamefont {Hell},\ and\
  \citenamefont {Balasubramanian}}]{arroyo2014room}%
  \BibitemOpen
  \bibfield  {author} {\bibinfo {author} {\bibfnamefont {S.}~\bibnamefont
  {Arroyo-Camejo}}, \bibinfo {author} {\bibfnamefont {A.}~\bibnamefont
  {Lazariev}}, \bibinfo {author} {\bibfnamefont {S.~W.}\ \bibnamefont {Hell}},\
  and\ \bibinfo {author} {\bibfnamefont {G.}~\bibnamefont {Balasubramanian}},\
  }\bibfield  {title} {\bibinfo {title} {Room temperature high-fidelity
  holonomic single-qubit gate on a solid-state spin},\ }\href
  {https://doi.org/10.1038/ncomms5870} {\bibfield  {journal} {\bibinfo
  {journal} {Nat. Commun.}\ }\textbf {\bibinfo {volume} {5}},\ \bibinfo {pages}
  {4870} (\bibinfo {year} {2014})}\BibitemShut {NoStop}%
\bibitem [{\citenamefont {Song}\ \emph {et~al.}(2017)\citenamefont {Song},
  \citenamefont {Zheng}, \citenamefont {Zhang}, \citenamefont {Xu},
  \citenamefont {Zhang}, \citenamefont {Guo}, \citenamefont {Liu},
  \citenamefont {Xu}, \citenamefont {Deng}, \citenamefont {Huang} \emph
  {et~al.}}]{song2017continuous}%
  \BibitemOpen
  \bibfield  {author} {\bibinfo {author} {\bibfnamefont {C.}~\bibnamefont
  {Song}}, \bibinfo {author} {\bibfnamefont {S.-B.}\ \bibnamefont {Zheng}},
  \bibinfo {author} {\bibfnamefont {P.}~\bibnamefont {Zhang}}, \bibinfo
  {author} {\bibfnamefont {K.}~\bibnamefont {Xu}}, \bibinfo {author}
  {\bibfnamefont {L.}~\bibnamefont {Zhang}}, \bibinfo {author} {\bibfnamefont
  {Q.}~\bibnamefont {Guo}}, \bibinfo {author} {\bibfnamefont {W.}~\bibnamefont
  {Liu}}, \bibinfo {author} {\bibfnamefont {D.}~\bibnamefont {Xu}}, \bibinfo
  {author} {\bibfnamefont {H.}~\bibnamefont {Deng}}, \bibinfo {author}
  {\bibfnamefont {K.}~\bibnamefont {Huang}}, \emph {et~al.},\ }\bibfield
  {title} {\bibinfo {title} {Continuous-variable geometric phase and its
  manipulation for quantum computation in a superconducting circuit},\ }\href
  {https://doi.org/10.1038/s41467-017-01156-5} {\bibfield  {journal} {\bibinfo
  {journal} {Nat. Commun.}\ }\textbf {\bibinfo {volume} {8}},\ \bibinfo {pages}
  {1061} (\bibinfo {year} {2017})}\BibitemShut {NoStop}%
\bibitem [{\citenamefont {Sekiguchi}\ \emph {et~al.}(2017)\citenamefont
  {Sekiguchi}, \citenamefont {Niikura}, \citenamefont {Kuroiwa}, \citenamefont
  {Kano},\ and\ \citenamefont {Kosaka}}]{sekiguchi2017optical}%
  \BibitemOpen
  \bibfield  {author} {\bibinfo {author} {\bibfnamefont {Y.}~\bibnamefont
  {Sekiguchi}}, \bibinfo {author} {\bibfnamefont {N.}~\bibnamefont {Niikura}},
  \bibinfo {author} {\bibfnamefont {R.}~\bibnamefont {Kuroiwa}}, \bibinfo
  {author} {\bibfnamefont {H.}~\bibnamefont {Kano}},\ and\ \bibinfo {author}
  {\bibfnamefont {H.}~\bibnamefont {Kosaka}},\ }\bibfield  {title} {\bibinfo
  {title} {Optical holonomic single quantum gates with a geometric spin under a
  zero field},\ }\href {https://doi.org/10.1038/nphoton.2017.40} {\bibfield
  {journal} {\bibinfo  {journal} {Nat. photonics}\ }\textbf {\bibinfo {volume}
  {11}},\ \bibinfo {pages} {309} (\bibinfo {year} {2017})}\BibitemShut
  {NoStop}%
\bibitem [{\citenamefont {Zhou}\ \emph
  {et~al.}(2017{\natexlab{b}})\citenamefont {Zhou}, \citenamefont {Jerger},
  \citenamefont {Shkolnikov}, \citenamefont {Heremans}, \citenamefont
  {Burkard},\ and\ \citenamefont {Awschalom}}]{PhysRevLett.119.140503}%
  \BibitemOpen
  \bibfield  {author} {\bibinfo {author} {\bibfnamefont {B.~B.}\ \bibnamefont
  {Zhou}}, \bibinfo {author} {\bibfnamefont {P.~C.}\ \bibnamefont {Jerger}},
  \bibinfo {author} {\bibfnamefont {V.~O.}\ \bibnamefont {Shkolnikov}},
  \bibinfo {author} {\bibfnamefont {F.~J.}\ \bibnamefont {Heremans}}, \bibinfo
  {author} {\bibfnamefont {G.}~\bibnamefont {Burkard}},\ and\ \bibinfo {author}
  {\bibfnamefont {D.~D.}\ \bibnamefont {Awschalom}},\ }\bibfield  {title}
  {\bibinfo {title} {Holonomic quantum control by coherent optical excitation
  in diamond},\ }\href {https://doi.org/10.1103/PhysRevLett.119.140503}
  {\bibfield  {journal} {\bibinfo  {journal} {Phys. Rev. Lett.}\ }\textbf
  {\bibinfo {volume} {119}},\ \bibinfo {pages} {140503} (\bibinfo {year}
  {2017}{\natexlab{b}})}\BibitemShut {NoStop}%
\bibitem [{\citenamefont {Nagata}\ \emph {et~al.}(2018)\citenamefont {Nagata},
  \citenamefont {Kuramitani}, \citenamefont {Sekiguchi},\ and\ \citenamefont
  {Kosaka}}]{nagata2018universal}%
  \BibitemOpen
  \bibfield  {author} {\bibinfo {author} {\bibfnamefont {K.}~\bibnamefont
  {Nagata}}, \bibinfo {author} {\bibfnamefont {K.}~\bibnamefont {Kuramitani}},
  \bibinfo {author} {\bibfnamefont {Y.}~\bibnamefont {Sekiguchi}},\ and\
  \bibinfo {author} {\bibfnamefont {H.}~\bibnamefont {Kosaka}},\ }\bibfield
  {title} {\bibinfo {title} {Universal holonomic quantum gates over geometric
  spin qubits with polarised microwaves},\ }\href
  {https://doi.org/10.1038/s41467-018-05664-w} {\bibfield  {journal} {\bibinfo
  {journal} {Nat. Commun.}\ }\textbf {\bibinfo {volume} {9}},\ \bibinfo {pages}
  {3227} (\bibinfo {year} {2018})}\BibitemShut {NoStop}%
\bibitem [{\citenamefont {Ishida}\ \emph {et~al.}(2018)\citenamefont {Ishida},
  \citenamefont {Nakamura}, \citenamefont {Tanaka}, \citenamefont {Mishima},
  \citenamefont {Kano}, \citenamefont {Kuroiwa}, \citenamefont {Sekiguchi},\
  and\ \citenamefont {Kosaka}}]{Ishida.18}%
  \BibitemOpen
  \bibfield  {author} {\bibinfo {author} {\bibfnamefont {N.}~\bibnamefont
  {Ishida}}, \bibinfo {author} {\bibfnamefont {T.}~\bibnamefont {Nakamura}},
  \bibinfo {author} {\bibfnamefont {T.}~\bibnamefont {Tanaka}}, \bibinfo
  {author} {\bibfnamefont {S.}~\bibnamefont {Mishima}}, \bibinfo {author}
  {\bibfnamefont {H.}~\bibnamefont {Kano}}, \bibinfo {author} {\bibfnamefont
  {R.}~\bibnamefont {Kuroiwa}}, \bibinfo {author} {\bibfnamefont
  {Y.}~\bibnamefont {Sekiguchi}},\ and\ \bibinfo {author} {\bibfnamefont
  {H.}~\bibnamefont {Kosaka}},\ }\bibfield  {title} {\bibinfo {title}
  {Universal holonomic single quantum gates over a geometric spin with
  phase-modulated polarized light},\ }\href
  {https://doi.org/10.1364/OL.43.002380} {\bibfield  {journal} {\bibinfo
  {journal} {Opt. Lett.}\ }\textbf {\bibinfo {volume} {43}},\ \bibinfo {pages}
  {2380} (\bibinfo {year} {2018})}\BibitemShut {NoStop}%
\bibitem [{\citenamefont {Egger}\ \emph {et~al.}(2019)\citenamefont {Egger},
  \citenamefont {Ganzhorn}, \citenamefont {Salis}, \citenamefont {Fuhrer},
  \citenamefont {M\"uller}, \citenamefont {Barkoutsos}, \citenamefont {Moll},
  \citenamefont {Tavernelli},\ and\ \citenamefont
  {Filipp}}]{PhysRevApplied.11.014017}%
  \BibitemOpen
  \bibfield  {author} {\bibinfo {author} {\bibfnamefont {D.}~\bibnamefont
  {Egger}}, \bibinfo {author} {\bibfnamefont {M.}~\bibnamefont {Ganzhorn}},
  \bibinfo {author} {\bibfnamefont {G.}~\bibnamefont {Salis}}, \bibinfo
  {author} {\bibfnamefont {A.}~\bibnamefont {Fuhrer}}, \bibinfo {author}
  {\bibfnamefont {P.}~\bibnamefont {M\"uller}}, \bibinfo {author}
  {\bibfnamefont {P.}~\bibnamefont {Barkoutsos}}, \bibinfo {author}
  {\bibfnamefont {N.}~\bibnamefont {Moll}}, \bibinfo {author} {\bibfnamefont
  {I.}~\bibnamefont {Tavernelli}},\ and\ \bibinfo {author} {\bibfnamefont
  {S.}~\bibnamefont {Filipp}},\ }\bibfield  {title} {\bibinfo {title}
  {Entanglement generation in superconducting qubits using holonomic
  operations},\ }\href {https://doi.org/10.1103/PhysRevApplied.11.014017}
  {\bibfield  {journal} {\bibinfo  {journal} {Phys. Rev. Applied}\ }\textbf
  {\bibinfo {volume} {11}},\ \bibinfo {pages} {014017} (\bibinfo {year}
  {2019})}\BibitemShut {NoStop}%
\bibitem [{\citenamefont {Ai}\ \emph {et~al.}(2020)\citenamefont {Ai},
  \citenamefont {Li}, \citenamefont {Hou}, \citenamefont {He}, \citenamefont
  {Qian}, \citenamefont {Xue}, \citenamefont {Cui}, \citenamefont {Huang},
  \citenamefont {Li},\ and\ \citenamefont {Guo}}]{PhysRevApplied.14.054062}%
  \BibitemOpen
  \bibfield  {author} {\bibinfo {author} {\bibfnamefont {M.-Z.}\ \bibnamefont
  {Ai}}, \bibinfo {author} {\bibfnamefont {S.}~\bibnamefont {Li}}, \bibinfo
  {author} {\bibfnamefont {Z.}~\bibnamefont {Hou}}, \bibinfo {author}
  {\bibfnamefont {R.}~\bibnamefont {He}}, \bibinfo {author} {\bibfnamefont
  {Z.-H.}\ \bibnamefont {Qian}}, \bibinfo {author} {\bibfnamefont {Z.-Y.}\
  \bibnamefont {Xue}}, \bibinfo {author} {\bibfnamefont {J.-M.}\ \bibnamefont
  {Cui}}, \bibinfo {author} {\bibfnamefont {Y.-F.}\ \bibnamefont {Huang}},
  \bibinfo {author} {\bibfnamefont {C.-F.}\ \bibnamefont {Li}},\ and\ \bibinfo
  {author} {\bibfnamefont {G.-C.}\ \bibnamefont {Guo}},\ }\bibfield  {title}
  {\bibinfo {title} {Experimental realization of nonadiabatic holonomic
  single-qubit quantum gates with optimal control in a trapped ion},\ }\href
  {https://doi.org/10.1103/PhysRevApplied.14.054062} {\bibfield  {journal}
  {\bibinfo  {journal} {Phys. Rev. Applied}\ }\textbf {\bibinfo {volume}
  {14}},\ \bibinfo {pages} {054062} (\bibinfo {year} {2020})}\BibitemShut
  {NoStop}%
\bibitem [{\citenamefont {Leibfried}\ \emph {et~al.}(2003)\citenamefont
  {Leibfried}, \citenamefont {DeMarco}, \citenamefont {Meyer}, \citenamefont
  {Lucas}, \citenamefont {Barrett}, \citenamefont {Britton}, \citenamefont
  {Itano}, \citenamefont {Jelenkovi{\'c}}, \citenamefont {Langer},
  \citenamefont {Rosenband} \emph {et~al.}}]{leibfried2003experimental}%
  \BibitemOpen
  \bibfield  {author} {\bibinfo {author} {\bibfnamefont {D.}~\bibnamefont
  {Leibfried}}, \bibinfo {author} {\bibfnamefont {B.}~\bibnamefont {DeMarco}},
  \bibinfo {author} {\bibfnamefont {V.}~\bibnamefont {Meyer}}, \bibinfo
  {author} {\bibfnamefont {D.}~\bibnamefont {Lucas}}, \bibinfo {author}
  {\bibfnamefont {M.}~\bibnamefont {Barrett}}, \bibinfo {author} {\bibfnamefont
  {J.}~\bibnamefont {Britton}}, \bibinfo {author} {\bibfnamefont {W.~M.}\
  \bibnamefont {Itano}}, \bibinfo {author} {\bibfnamefont {B.}~\bibnamefont
  {Jelenkovi{\'c}}}, \bibinfo {author} {\bibfnamefont {C.}~\bibnamefont
  {Langer}}, \bibinfo {author} {\bibfnamefont {T.}~\bibnamefont {Rosenband}},
  \emph {et~al.},\ }\bibfield  {title} {\bibinfo {title} {Experimental
  demonstration of a robust, high-fidelity geometric two ion-qubit phase
  gate},\ }\href {https://doi.org/10.1038/nature01492} {\bibfield  {journal}
  {\bibinfo  {journal} {Nature}\ }\textbf {\bibinfo {volume} {422}},\ \bibinfo
  {pages} {412} (\bibinfo {year} {2003})}\BibitemShut {NoStop}%
\bibitem [{\citenamefont {Xu}\ \emph {et~al.}(2020)\citenamefont {Xu},
  \citenamefont {Hua}, \citenamefont {Chen}, \citenamefont {Pan}, \citenamefont
  {Li}, \citenamefont {Han}, \citenamefont {Cai}, \citenamefont {Ma},
  \citenamefont {Wang}, \citenamefont {Song}, \citenamefont {Xue},\ and\
  \citenamefont {Sun}}]{PhysRevLett.124.230503}%
  \BibitemOpen
  \bibfield  {author} {\bibinfo {author} {\bibfnamefont {Y.}~\bibnamefont
  {Xu}}, \bibinfo {author} {\bibfnamefont {Z.}~\bibnamefont {Hua}}, \bibinfo
  {author} {\bibfnamefont {T.}~\bibnamefont {Chen}}, \bibinfo {author}
  {\bibfnamefont {X.}~\bibnamefont {Pan}}, \bibinfo {author} {\bibfnamefont
  {X.}~\bibnamefont {Li}}, \bibinfo {author} {\bibfnamefont {J.}~\bibnamefont
  {Han}}, \bibinfo {author} {\bibfnamefont {W.}~\bibnamefont {Cai}}, \bibinfo
  {author} {\bibfnamefont {Y.}~\bibnamefont {Ma}}, \bibinfo {author}
  {\bibfnamefont {H.}~\bibnamefont {Wang}}, \bibinfo {author} {\bibfnamefont
  {Y.~P.}\ \bibnamefont {Song}}, \bibinfo {author} {\bibfnamefont {Z.-Y.}\
  \bibnamefont {Xue}},\ and\ \bibinfo {author} {\bibfnamefont {L.}~\bibnamefont
  {Sun}},\ }\bibfield  {title} {\bibinfo {title} {Experimental implementation
  of universal nonadiabatic geometric quantum gates in a superconducting
  circuit},\ }\href {https://doi.org/10.1103/PhysRevLett.124.230503} {\bibfield
   {journal} {\bibinfo  {journal} {Phys. Rev. Lett.}\ }\textbf {\bibinfo
  {volume} {124}},\ \bibinfo {pages} {230503} (\bibinfo {year}
  {2020})}\BibitemShut {NoStop}%
\bibitem [{\citenamefont {Zhao}\ \emph {et~al.}(2021)\citenamefont {Zhao},
  \citenamefont {Dong}, \citenamefont {Zhang}, \citenamefont {Guo},
  \citenamefont {Tong},\ and\ \citenamefont {Yin}}]{zhao2021experimental}%
  \BibitemOpen
  \bibfield  {author} {\bibinfo {author} {\bibfnamefont {P.}~\bibnamefont
  {Zhao}}, \bibinfo {author} {\bibfnamefont {Z.}~\bibnamefont {Dong}}, \bibinfo
  {author} {\bibfnamefont {Z.}~\bibnamefont {Zhang}}, \bibinfo {author}
  {\bibfnamefont {G.}~\bibnamefont {Guo}}, \bibinfo {author} {\bibfnamefont
  {D.}~\bibnamefont {Tong}},\ and\ \bibinfo {author} {\bibfnamefont
  {Y.}~\bibnamefont {Yin}},\ }\bibfield  {title} {\bibinfo {title}
  {Experimental realization of nonadiabatic geometric gates with a
  superconducting xmon qubit},\ }\href
  {https://doi.org/10.1007/s11433-020-1641-1} {\bibfield  {journal} {\bibinfo
  {journal} {SCIENCE CHINA Physics, Mechanics \& Astronomy}\ }\textbf {\bibinfo
  {volume} {64}},\ \bibinfo {pages} {250362} (\bibinfo {year}
  {2021})}\BibitemShut {NoStop}%
\bibitem [{\citenamefont {Koch}\ \emph {et~al.}(2007)\citenamefont {Koch},
  \citenamefont {Yu}, \citenamefont {Gambetta}, \citenamefont {Houck},
  \citenamefont {Schuster}, \citenamefont {Majer}, \citenamefont {Blais},
  \citenamefont {Devoret}, \citenamefont {Girvin},\ and\ \citenamefont
  {Schoelkopf}}]{PhysRevA.76.042319}%
  \BibitemOpen
  \bibfield  {author} {\bibinfo {author} {\bibfnamefont {J.}~\bibnamefont
  {Koch}}, \bibinfo {author} {\bibfnamefont {T.~M.}\ \bibnamefont {Yu}},
  \bibinfo {author} {\bibfnamefont {J.}~\bibnamefont {Gambetta}}, \bibinfo
  {author} {\bibfnamefont {A.~A.}\ \bibnamefont {Houck}}, \bibinfo {author}
  {\bibfnamefont {D.~I.}\ \bibnamefont {Schuster}}, \bibinfo {author}
  {\bibfnamefont {J.}~\bibnamefont {Majer}}, \bibinfo {author} {\bibfnamefont
  {A.}~\bibnamefont {Blais}}, \bibinfo {author} {\bibfnamefont {M.~H.}\
  \bibnamefont {Devoret}}, \bibinfo {author} {\bibfnamefont {S.~M.}\
  \bibnamefont {Girvin}},\ and\ \bibinfo {author} {\bibfnamefont {R.~J.}\
  \bibnamefont {Schoelkopf}},\ }\bibfield  {title} {\bibinfo {title}
  {Charge-insensitive qubit design derived from the cooper pair box},\ }\href
  {https://doi.org/10.1103/PhysRevA.76.042319} {\bibfield  {journal} {\bibinfo
  {journal} {Phys. Rev. A}\ }\textbf {\bibinfo {volume} {76}},\ \bibinfo
  {pages} {042319} (\bibinfo {year} {2007})}\BibitemShut {NoStop}%
\bibitem [{\citenamefont {Barends}\ \emph {et~al.}(2014)\citenamefont
  {Barends}, \citenamefont {Kelly}, \citenamefont {Megrant}, \citenamefont
  {Veitia}, \citenamefont {Sank}, \citenamefont {Jeffrey}, \citenamefont
  {White}, \citenamefont {Mutus}, \citenamefont {Fowler}, \citenamefont
  {Campbell} \emph {et~al.}}]{barends2014superconducting}%
  \BibitemOpen
  \bibfield  {author} {\bibinfo {author} {\bibfnamefont {R.}~\bibnamefont
  {Barends}}, \bibinfo {author} {\bibfnamefont {J.}~\bibnamefont {Kelly}},
  \bibinfo {author} {\bibfnamefont {A.}~\bibnamefont {Megrant}}, \bibinfo
  {author} {\bibfnamefont {A.}~\bibnamefont {Veitia}}, \bibinfo {author}
  {\bibfnamefont {D.}~\bibnamefont {Sank}}, \bibinfo {author} {\bibfnamefont
  {E.}~\bibnamefont {Jeffrey}}, \bibinfo {author} {\bibfnamefont {T.~C.}\
  \bibnamefont {White}}, \bibinfo {author} {\bibfnamefont {J.}~\bibnamefont
  {Mutus}}, \bibinfo {author} {\bibfnamefont {A.~G.}\ \bibnamefont {Fowler}},
  \bibinfo {author} {\bibfnamefont {B.}~\bibnamefont {Campbell}}, \emph
  {et~al.},\ }\bibfield  {title} {\bibinfo {title} {Superconducting quantum
  circuits at the surface code threshold for fault tolerance},\ }\href
  {https://doi.org/10.1038/nature13171} {\bibfield  {journal} {\bibinfo
  {journal} {Nature}\ }\textbf {\bibinfo {volume} {508}},\ \bibinfo {pages}
  {500} (\bibinfo {year} {2014})}\BibitemShut {NoStop}%
\bibitem [{\citenamefont {Liu}\ \emph {et~al.}(2020)\citenamefont {Liu},
  \citenamefont {Su},\ and\ \citenamefont {Yung}}]{PhysRevResearch.2.043130}%
  \BibitemOpen
  \bibfield  {author} {\bibinfo {author} {\bibfnamefont {B.-J.}\ \bibnamefont
  {Liu}}, \bibinfo {author} {\bibfnamefont {S.-L.}\ \bibnamefont {Su}},\ and\
  \bibinfo {author} {\bibfnamefont {M.-H.}\ \bibnamefont {Yung}},\ }\bibfield
  {title} {\bibinfo {title} {Nonadiabatic noncyclic geometric quantum
  computation in rydberg atoms},\ }\href
  {https://doi.org/10.1103/PhysRevResearch.2.043130} {\bibfield  {journal}
  {\bibinfo  {journal} {Phys. Rev. Research}\ }\textbf {\bibinfo {volume}
  {2}},\ \bibinfo {pages} {043130} (\bibinfo {year} {2020})}\BibitemShut
  {NoStop}%
\bibitem [{\citenamefont {Chen}\ and\ \citenamefont
  {Xue}(2020)}]{PhysRevApplied.14.064009}%
  \BibitemOpen
  \bibfield  {author} {\bibinfo {author} {\bibfnamefont {T.}~\bibnamefont
  {Chen}}\ and\ \bibinfo {author} {\bibfnamefont {Z.-Y.}\ \bibnamefont {Xue}},\
  }\bibfield  {title} {\bibinfo {title} {High-fidelity and robust geometric
  quantum gates that outperform dynamical ones},\ }\href
  {https://doi.org/10.1103/PhysRevApplied.14.064009} {\bibfield  {journal}
  {\bibinfo  {journal} {Phys. Rev. Applied}\ }\textbf {\bibinfo {volume}
  {14}},\ \bibinfo {pages} {064009} (\bibinfo {year} {2020})}\BibitemShut
  {NoStop}%
\bibitem [{\citenamefont {Zhang}\ \emph {et~al.}(2021)\citenamefont {Zhang},
  \citenamefont {Yan}, \citenamefont {Li}, \citenamefont {Ding}, \citenamefont
  {Bu}, \citenamefont {Chen}, \citenamefont {Su}, \citenamefont {Zhou},\ and\
  \citenamefont {Feng}}]{PhysRevLett.127.030502}%
  \BibitemOpen
  \bibfield  {author} {\bibinfo {author} {\bibfnamefont {J.~W.}\ \bibnamefont
  {Zhang}}, \bibinfo {author} {\bibfnamefont {L.-L.}\ \bibnamefont {Yan}},
  \bibinfo {author} {\bibfnamefont {J.~C.}\ \bibnamefont {Li}}, \bibinfo
  {author} {\bibfnamefont {G.~Y.}\ \bibnamefont {Ding}}, \bibinfo {author}
  {\bibfnamefont {J.~T.}\ \bibnamefont {Bu}}, \bibinfo {author} {\bibfnamefont
  {L.}~\bibnamefont {Chen}}, \bibinfo {author} {\bibfnamefont {S.-L.}\
  \bibnamefont {Su}}, \bibinfo {author} {\bibfnamefont {F.}~\bibnamefont
  {Zhou}},\ and\ \bibinfo {author} {\bibfnamefont {M.}~\bibnamefont {Feng}},\
  }\bibfield  {title} {\bibinfo {title} {Single-atom verification of the
  noise-resilient and fast characteristics of universal nonadiabatic noncyclic
  geometric quantum gates},\ }\href
  {https://doi.org/10.1103/PhysRevLett.127.030502} {\bibfield  {journal}
  {\bibinfo  {journal} {Phys. Rev. Lett.}\ }\textbf {\bibinfo {volume} {127}},\
  \bibinfo {pages} {030502} (\bibinfo {year} {2021})}\BibitemShut {NoStop}%
\bibitem [{\citenamefont {Ji}\ \emph {et~al.}(2021)\citenamefont {Ji},
  \citenamefont {Ding}, \citenamefont {Chen},\ and\ \citenamefont
  {Xue}}]{https://doi.org/10.1002/qute.202100019}%
  \BibitemOpen
  \bibfield  {author} {\bibinfo {author} {\bibfnamefont {L.-N.}\ \bibnamefont
  {Ji}}, \bibinfo {author} {\bibfnamefont {C.-Y.}\ \bibnamefont {Ding}},
  \bibinfo {author} {\bibfnamefont {T.}~\bibnamefont {Chen}},\ and\ \bibinfo
  {author} {\bibfnamefont {Z.-Y.}\ \bibnamefont {Xue}},\ }\bibfield  {title}
  {\bibinfo {title} {Noncyclic geometric quantum gates with smooth paths via
  invariant-based shortcuts},\ }\href
  {https://doi.org/https://doi.org/10.1002/qute.202100019} {\bibfield
  {journal} {\bibinfo  {journal} {Advanced Quantum Technologies}\ }\textbf
  {\bibinfo {volume} {4}},\ \bibinfo {pages} {2100019} (\bibinfo {year}
  {2021})}\BibitemShut {NoStop}%
\bibitem [{\citenamefont {Li}\ \emph {et~al.}(2020)\citenamefont {Li},
  \citenamefont {Zhao},\ and\ \citenamefont {Tong}}]{PhysRevResearch.2.023295}%
  \BibitemOpen
  \bibfield  {author} {\bibinfo {author} {\bibfnamefont {K.~Z.}\ \bibnamefont
  {Li}}, \bibinfo {author} {\bibfnamefont {P.~Z.}\ \bibnamefont {Zhao}},\ and\
  \bibinfo {author} {\bibfnamefont {D.~M.}\ \bibnamefont {Tong}},\ }\bibfield
  {title} {\bibinfo {title} {Approach to realizing nonadiabatic geometric gates
  with prescribed evolution paths},\ }\href
  {https://doi.org/10.1103/PhysRevResearch.2.023295} {\bibfield  {journal}
  {\bibinfo  {journal} {Phys. Rev. Research}\ }\textbf {\bibinfo {volume}
  {2}},\ \bibinfo {pages} {023295} (\bibinfo {year} {2020})}\BibitemShut
  {NoStop}%
\bibitem [{\citenamefont {Ding}\ \emph
  {et~al.}(2021{\natexlab{a}})\citenamefont {Ding}, \citenamefont {Ji},
  \citenamefont {Chen},\ and\ \citenamefont {Xue}}]{Ding_2021}%
  \BibitemOpen
  \bibfield  {author} {\bibinfo {author} {\bibfnamefont {C.-Y.}\ \bibnamefont
  {Ding}}, \bibinfo {author} {\bibfnamefont {L.-N.}\ \bibnamefont {Ji}},
  \bibinfo {author} {\bibfnamefont {T.}~\bibnamefont {Chen}},\ and\ \bibinfo
  {author} {\bibfnamefont {Z.-Y.}\ \bibnamefont {Xue}},\ }\bibfield  {title}
  {\bibinfo {title} {Path-optimized nonadiabatic geometric quantum computation
  on superconducting qubits},\ }\href
  {https://doi.org/10.1088/2058-9565/ac3621} {\bibfield  {journal} {\bibinfo
  {journal} {Quantum Science and Technology}\ }\textbf {\bibinfo {volume}
  {7}},\ \bibinfo {pages} {015012} (\bibinfo {year}
  {2021}{\natexlab{a}})}\BibitemShut {NoStop}%
\bibitem [{\citenamefont {Ding}\ \emph
  {et~al.}(2021{\natexlab{b}})\citenamefont {Ding}, \citenamefont {Liang},
  \citenamefont {Yu},\ and\ \citenamefont {Xue}}]{doi:10.1063/5.0071569}%
  \BibitemOpen
  \bibfield  {author} {\bibinfo {author} {\bibfnamefont {C.-Y.}\ \bibnamefont
  {Ding}}, \bibinfo {author} {\bibfnamefont {Y.}~\bibnamefont {Liang}},
  \bibinfo {author} {\bibfnamefont {K.-Z.}\ \bibnamefont {Yu}},\ and\ \bibinfo
  {author} {\bibfnamefont {Z.-Y.}\ \bibnamefont {Xue}},\ }\bibfield  {title}
  {\bibinfo {title} {Nonadiabatic geometric quantum computation with shortened
  path on superconducting circuits},\ }\href
  {https://doi.org/10.1063/5.0071569} {\bibfield  {journal} {\bibinfo
  {journal} {Applied Physics Letters}\ }\textbf {\bibinfo {volume} {119}},\
  \bibinfo {pages} {184001} (\bibinfo {year} {2021}{\natexlab{b}})}\BibitemShut
  {NoStop}%
\bibitem [{\citenamefont {Li}\ \emph {et~al.}(2021{\natexlab{b}})\citenamefont
  {Li}, \citenamefont {Xue}, \citenamefont {Chen},\ and\ \citenamefont
  {Xue}}]{li2021high}%
  \BibitemOpen
  \bibfield  {author} {\bibinfo {author} {\bibfnamefont {S.}~\bibnamefont
  {Li}}, \bibinfo {author} {\bibfnamefont {J.}~\bibnamefont {Xue}}, \bibinfo
  {author} {\bibfnamefont {T.}~\bibnamefont {Chen}},\ and\ \bibinfo {author}
  {\bibfnamefont {Z.-Y.}\ \bibnamefont {Xue}},\ }\bibfield  {title} {\bibinfo
  {title} {High-fidelity geometric quantum gates with short paths on
  superconducting circuits},\ }\href
  {https://doi.org/https://doi.org/10.1002/qute.202000140} {\bibfield
  {journal} {\bibinfo  {journal} {Advanced Quantum Technologies}\ }\textbf
  {\bibinfo {volume} {4}},\ \bibinfo {pages} {2000140} (\bibinfo {year}
  {2021}{\natexlab{b}})}\BibitemShut {NoStop}%
\bibitem [{\citenamefont {McKay}\ \emph {et~al.}(2017)\citenamefont {McKay},
  \citenamefont {Wood}, \citenamefont {Sheldon}, \citenamefont {Chow},\ and\
  \citenamefont {Gambetta}}]{PhysRevA.96.022330}%
  \BibitemOpen
  \bibfield  {author} {\bibinfo {author} {\bibfnamefont {D.~C.}\ \bibnamefont
  {McKay}}, \bibinfo {author} {\bibfnamefont {C.~J.}\ \bibnamefont {Wood}},
  \bibinfo {author} {\bibfnamefont {S.}~\bibnamefont {Sheldon}}, \bibinfo
  {author} {\bibfnamefont {J.~M.}\ \bibnamefont {Chow}},\ and\ \bibinfo
  {author} {\bibfnamefont {J.~M.}\ \bibnamefont {Gambetta}},\ }\bibfield
  {title} {\bibinfo {title} {Efficient $z$ gates for quantum computing},\
  }\href {https://doi.org/10.1103/PhysRevA.96.022330} {\bibfield  {journal}
  {\bibinfo  {journal} {Phys. Rev. A}\ }\textbf {\bibinfo {volume} {96}},\
  \bibinfo {pages} {022330} (\bibinfo {year} {2017})}\BibitemShut {NoStop}%
\bibitem [{\citenamefont {Duan}\ \emph
  {et~al.}(2021{\natexlab{a}})\citenamefont {Duan}, \citenamefont {Chen},
  \citenamefont {Zhou}, \citenamefont {Kong}, \citenamefont {Zhang},\ and\
  \citenamefont {Guo}}]{PhysRevApplied.16.024063}%
  \BibitemOpen
  \bibfield  {author} {\bibinfo {author} {\bibfnamefont {P.}~\bibnamefont
  {Duan}}, \bibinfo {author} {\bibfnamefont {Z.-F.}\ \bibnamefont {Chen}},
  \bibinfo {author} {\bibfnamefont {Q.}~\bibnamefont {Zhou}}, \bibinfo {author}
  {\bibfnamefont {W.-C.}\ \bibnamefont {Kong}}, \bibinfo {author}
  {\bibfnamefont {H.-F.}\ \bibnamefont {Zhang}},\ and\ \bibinfo {author}
  {\bibfnamefont {G.-P.}\ \bibnamefont {Guo}},\ }\bibfield  {title} {\bibinfo
  {title} {Mitigating crosstalk-induced qubit readout error with
  shallow-neural-network discrimination},\ }\href
  {https://doi.org/10.1103/PhysRevApplied.16.024063} {\bibfield  {journal}
  {\bibinfo  {journal} {Phys. Rev. Applied}\ }\textbf {\bibinfo {volume}
  {16}},\ \bibinfo {pages} {024063} (\bibinfo {year}
  {2021}{\natexlab{a}})}\BibitemShut {NoStop}%
\bibitem [{\citenamefont {Motzoi}\ \emph {et~al.}(2009)\citenamefont {Motzoi},
  \citenamefont {Gambetta}, \citenamefont {Rebentrost},\ and\ \citenamefont
  {Wilhelm}}]{PhysRevLett.103.110501}%
  \BibitemOpen
  \bibfield  {author} {\bibinfo {author} {\bibfnamefont {F.}~\bibnamefont
  {Motzoi}}, \bibinfo {author} {\bibfnamefont {J.~M.}\ \bibnamefont
  {Gambetta}}, \bibinfo {author} {\bibfnamefont {P.}~\bibnamefont
  {Rebentrost}},\ and\ \bibinfo {author} {\bibfnamefont {F.~K.}\ \bibnamefont
  {Wilhelm}},\ }\bibfield  {title} {\bibinfo {title} {Simple pulses for
  elimination of leakage in weakly nonlinear qubits},\ }\href
  {https://doi.org/10.1103/PhysRevLett.103.110501} {\bibfield  {journal}
  {\bibinfo  {journal} {Phys. Rev. Lett.}\ }\textbf {\bibinfo {volume} {103}},\
  \bibinfo {pages} {110501} (\bibinfo {year} {2009})}\BibitemShut {NoStop}%
\bibitem [{\citenamefont {Chuang}\ and\ \citenamefont
  {Nielsen}(1997)}]{doi:10.1080/09500349708231894}%
  \BibitemOpen
  \bibfield  {author} {\bibinfo {author} {\bibfnamefont {I.~L.}\ \bibnamefont
  {Chuang}}\ and\ \bibinfo {author} {\bibfnamefont {M.~A.}\ \bibnamefont
  {Nielsen}},\ }\bibfield  {title} {\bibinfo {title} {Prescription for
  experimental determination of the dynamics of a quantum black box},\ }\href
  {https://doi.org/10.1080/09500349708231894} {\bibfield  {journal} {\bibinfo
  {journal} {Journal of Modern Optics}\ }\textbf {\bibinfo {volume} {44}},\
  \bibinfo {pages} {2455} (\bibinfo {year} {1997})}\BibitemShut {NoStop}%
\bibitem [{\citenamefont {O'Brien}\ \emph {et~al.}(2004)\citenamefont
  {O'Brien}, \citenamefont {Pryde}, \citenamefont {Gilchrist}, \citenamefont
  {James}, \citenamefont {Langford}, \citenamefont {Ralph},\ and\ \citenamefont
  {White}}]{PhysRevLett.93.080502}%
  \BibitemOpen
  \bibfield  {author} {\bibinfo {author} {\bibfnamefont {J.~L.}\ \bibnamefont
  {O'Brien}}, \bibinfo {author} {\bibfnamefont {G.~J.}\ \bibnamefont {Pryde}},
  \bibinfo {author} {\bibfnamefont {A.}~\bibnamefont {Gilchrist}}, \bibinfo
  {author} {\bibfnamefont {D.~F.~V.}\ \bibnamefont {James}}, \bibinfo {author}
  {\bibfnamefont {N.~K.}\ \bibnamefont {Langford}}, \bibinfo {author}
  {\bibfnamefont {T.~C.}\ \bibnamefont {Ralph}},\ and\ \bibinfo {author}
  {\bibfnamefont {A.~G.}\ \bibnamefont {White}},\ }\bibfield  {title} {\bibinfo
  {title} {Quantum process tomography of a controlled-not gate},\ }\href
  {https://doi.org/10.1103/PhysRevLett.93.080502} {\bibfield  {journal}
  {\bibinfo  {journal} {Phys. Rev. Lett.}\ }\textbf {\bibinfo {volume} {93}},\
  \bibinfo {pages} {080502} (\bibinfo {year} {2004})}\BibitemShut {NoStop}%
\bibitem [{\citenamefont {Chow}\ \emph {et~al.}(2009)\citenamefont {Chow},
  \citenamefont {Gambetta}, \citenamefont {Tornberg}, \citenamefont {Koch},
  \citenamefont {Bishop}, \citenamefont {Houck}, \citenamefont {Johnson},
  \citenamefont {Frunzio}, \citenamefont {Girvin},\ and\ \citenamefont
  {Schoelkopf}}]{PhysRevLett.102.090502}%
  \BibitemOpen
  \bibfield  {author} {\bibinfo {author} {\bibfnamefont {J.~M.}\ \bibnamefont
  {Chow}}, \bibinfo {author} {\bibfnamefont {J.~M.}\ \bibnamefont {Gambetta}},
  \bibinfo {author} {\bibfnamefont {L.}~\bibnamefont {Tornberg}}, \bibinfo
  {author} {\bibfnamefont {J.}~\bibnamefont {Koch}}, \bibinfo {author}
  {\bibfnamefont {L.~S.}\ \bibnamefont {Bishop}}, \bibinfo {author}
  {\bibfnamefont {A.~A.}\ \bibnamefont {Houck}}, \bibinfo {author}
  {\bibfnamefont {B.~R.}\ \bibnamefont {Johnson}}, \bibinfo {author}
  {\bibfnamefont {L.}~\bibnamefont {Frunzio}}, \bibinfo {author} {\bibfnamefont
  {S.~M.}\ \bibnamefont {Girvin}},\ and\ \bibinfo {author} {\bibfnamefont
  {R.~J.}\ \bibnamefont {Schoelkopf}},\ }\bibfield  {title} {\bibinfo {title}
  {Randomized benchmarking and process tomography for gate errors in a
  solid-state qubit},\ }\href {https://doi.org/10.1103/PhysRevLett.102.090502}
  {\bibfield  {journal} {\bibinfo  {journal} {Phys. Rev. Lett.}\ }\textbf
  {\bibinfo {volume} {102}},\ \bibinfo {pages} {090502} (\bibinfo {year}
  {2009})}\BibitemShut {NoStop}%
\bibitem [{\citenamefont {Knill}\ \emph {et~al.}(2008)\citenamefont {Knill},
  \citenamefont {Leibfried}, \citenamefont {Reichle}, \citenamefont {Britton},
  \citenamefont {Blakestad}, \citenamefont {Jost}, \citenamefont {Langer},
  \citenamefont {Ozeri}, \citenamefont {Seidelin},\ and\ \citenamefont
  {Wineland}}]{PhysRevA.77.012307}%
  \BibitemOpen
  \bibfield  {author} {\bibinfo {author} {\bibfnamefont {E.}~\bibnamefont
  {Knill}}, \bibinfo {author} {\bibfnamefont {D.}~\bibnamefont {Leibfried}},
  \bibinfo {author} {\bibfnamefont {R.}~\bibnamefont {Reichle}}, \bibinfo
  {author} {\bibfnamefont {J.}~\bibnamefont {Britton}}, \bibinfo {author}
  {\bibfnamefont {R.~B.}\ \bibnamefont {Blakestad}}, \bibinfo {author}
  {\bibfnamefont {J.~D.}\ \bibnamefont {Jost}}, \bibinfo {author}
  {\bibfnamefont {C.}~\bibnamefont {Langer}}, \bibinfo {author} {\bibfnamefont
  {R.}~\bibnamefont {Ozeri}}, \bibinfo {author} {\bibfnamefont
  {S.}~\bibnamefont {Seidelin}},\ and\ \bibinfo {author} {\bibfnamefont
  {D.~J.}\ \bibnamefont {Wineland}},\ }\bibfield  {title} {\bibinfo {title}
  {Randomized benchmarking of quantum gates},\ }\href
  {https://doi.org/10.1103/PhysRevA.77.012307} {\bibfield  {journal} {\bibinfo
  {journal} {Phys. Rev. A}\ }\textbf {\bibinfo {volume} {77}},\ \bibinfo
  {pages} {012307} (\bibinfo {year} {2008})}\BibitemShut {NoStop}%
\bibitem [{\citenamefont {Magesan}\ \emph {et~al.}(2012)\citenamefont
  {Magesan}, \citenamefont {Gambetta}, \citenamefont {Johnson}, \citenamefont
  {Ryan}, \citenamefont {Chow}, \citenamefont {Merkel}, \citenamefont
  {da~Silva}, \citenamefont {Keefe}, \citenamefont {Rothwell}, \citenamefont
  {Ohki}, \citenamefont {Ketchen},\ and\ \citenamefont
  {Steffen}}]{PhysRevLett.109.080505}%
  \BibitemOpen
  \bibfield  {author} {\bibinfo {author} {\bibfnamefont {E.}~\bibnamefont
  {Magesan}}, \bibinfo {author} {\bibfnamefont {J.~M.}\ \bibnamefont
  {Gambetta}}, \bibinfo {author} {\bibfnamefont {B.~R.}\ \bibnamefont
  {Johnson}}, \bibinfo {author} {\bibfnamefont {C.~A.}\ \bibnamefont {Ryan}},
  \bibinfo {author} {\bibfnamefont {J.~M.}\ \bibnamefont {Chow}}, \bibinfo
  {author} {\bibfnamefont {S.~T.}\ \bibnamefont {Merkel}}, \bibinfo {author}
  {\bibfnamefont {M.~P.}\ \bibnamefont {da~Silva}}, \bibinfo {author}
  {\bibfnamefont {G.~A.}\ \bibnamefont {Keefe}}, \bibinfo {author}
  {\bibfnamefont {M.~B.}\ \bibnamefont {Rothwell}}, \bibinfo {author}
  {\bibfnamefont {T.~A.}\ \bibnamefont {Ohki}}, \bibinfo {author}
  {\bibfnamefont {M.~B.}\ \bibnamefont {Ketchen}},\ and\ \bibinfo {author}
  {\bibfnamefont {M.}~\bibnamefont {Steffen}},\ }\bibfield  {title} {\bibinfo
  {title} {Efficient measurement of quantum gate error by interleaved
  randomized benchmarking},\ }\href
  {https://doi.org/10.1103/PhysRevLett.109.080505} {\bibfield  {journal}
  {\bibinfo  {journal} {Phys. Rev. Lett.}\ }\textbf {\bibinfo {volume} {109}},\
  \bibinfo {pages} {080505} (\bibinfo {year} {2012})}\BibitemShut {NoStop}%
\bibitem [{\citenamefont {Didier}\ \emph {et~al.}(2018)\citenamefont {Didier},
  \citenamefont {Sete}, \citenamefont {da~Silva},\ and\ \citenamefont
  {Rigetti}}]{PhysRevA.97.022330}%
  \BibitemOpen
  \bibfield  {author} {\bibinfo {author} {\bibfnamefont {N.}~\bibnamefont
  {Didier}}, \bibinfo {author} {\bibfnamefont {E.~A.}\ \bibnamefont {Sete}},
  \bibinfo {author} {\bibfnamefont {M.~P.}\ \bibnamefont {da~Silva}},\ and\
  \bibinfo {author} {\bibfnamefont {C.}~\bibnamefont {Rigetti}},\ }\bibfield
  {title} {\bibinfo {title} {Analytical modeling of parametrically modulated
  transmon qubits},\ }\href {https://doi.org/10.1103/PhysRevA.97.022330}
  {\bibfield  {journal} {\bibinfo  {journal} {Phys. Rev. A}\ }\textbf {\bibinfo
  {volume} {97}},\ \bibinfo {pages} {022330} (\bibinfo {year}
  {2018})}\BibitemShut {NoStop}%
\bibitem [{\citenamefont {Caldwell}\ \emph {et~al.}(2018)\citenamefont
  {Caldwell}, \citenamefont {Didier}, \citenamefont {Ryan}, \citenamefont
  {Sete}, \citenamefont {Hudson}, \citenamefont {Karalekas}, \citenamefont
  {Manenti}, \citenamefont {da~Silva}, \citenamefont {Sinclair}, \citenamefont
  {Acala}, \citenamefont {Alidoust}, \citenamefont {Angeles}, \citenamefont
  {Bestwick}, \citenamefont {Block}, \citenamefont {Bloom}, \citenamefont
  {Bradley}, \citenamefont {Bui}, \citenamefont {Capelluto}, \citenamefont
  {Chilcott}, \citenamefont {Cordova}, \citenamefont {Crossman}, \citenamefont
  {Curtis}, \citenamefont {Deshpande}, \citenamefont {Bouayadi}, \citenamefont
  {Girshovich}, \citenamefont {Hong}, \citenamefont {Kuang}, \citenamefont
  {Lenihan}, \citenamefont {Manning}, \citenamefont {Marchenkov}, \citenamefont
  {Marshall}, \citenamefont {Maydra}, \citenamefont {Mohan}, \citenamefont
  {O'Brien}, \citenamefont {Osborn}, \citenamefont {Otterbach}, \citenamefont
  {Papageorge}, \citenamefont {Paquette}, \citenamefont {Pelstring},
  \citenamefont {Polloreno}, \citenamefont {Prawiroatmodjo}, \citenamefont
  {Rawat}, \citenamefont {Reagor}, \citenamefont {Renzas}, \citenamefont
  {Rubin}, \citenamefont {Russell}, \citenamefont {Rust}, \citenamefont
  {Scarabelli}, \citenamefont {Scheer}, \citenamefont {Selvanayagam},
  \citenamefont {Smith}, \citenamefont {Staley}, \citenamefont {Suska},
  \citenamefont {Tezak}, \citenamefont {Thompson}, \citenamefont {To},
  \citenamefont {Vahidpour}, \citenamefont {Vodrahalli}, \citenamefont
  {Whyland}, \citenamefont {Yadav}, \citenamefont {Zeng},\ and\ \citenamefont
  {Rigetti}}]{PhysRevApplied.10.034050}%
  \BibitemOpen
  \bibfield  {author} {\bibinfo {author} {\bibfnamefont {S.~A.}\ \bibnamefont
  {Caldwell}}, \bibinfo {author} {\bibfnamefont {N.}~\bibnamefont {Didier}},
  \bibinfo {author} {\bibfnamefont {C.~A.}\ \bibnamefont {Ryan}}, \bibinfo
  {author} {\bibfnamefont {E.~A.}\ \bibnamefont {Sete}}, \bibinfo {author}
  {\bibfnamefont {A.}~\bibnamefont {Hudson}}, \bibinfo {author} {\bibfnamefont
  {P.}~\bibnamefont {Karalekas}}, \bibinfo {author} {\bibfnamefont
  {R.}~\bibnamefont {Manenti}}, \bibinfo {author} {\bibfnamefont {M.~P.}\
  \bibnamefont {da~Silva}}, \bibinfo {author} {\bibfnamefont {R.}~\bibnamefont
  {Sinclair}}, \bibinfo {author} {\bibfnamefont {E.}~\bibnamefont {Acala}},
  \bibinfo {author} {\bibfnamefont {N.}~\bibnamefont {Alidoust}}, \bibinfo
  {author} {\bibfnamefont {J.}~\bibnamefont {Angeles}}, \bibinfo {author}
  {\bibfnamefont {A.}~\bibnamefont {Bestwick}}, \bibinfo {author}
  {\bibfnamefont {M.}~\bibnamefont {Block}}, \bibinfo {author} {\bibfnamefont
  {B.}~\bibnamefont {Bloom}}, \bibinfo {author} {\bibfnamefont
  {A.}~\bibnamefont {Bradley}}, \bibinfo {author} {\bibfnamefont
  {C.}~\bibnamefont {Bui}}, \bibinfo {author} {\bibfnamefont {L.}~\bibnamefont
  {Capelluto}}, \bibinfo {author} {\bibfnamefont {R.}~\bibnamefont {Chilcott}},
  \bibinfo {author} {\bibfnamefont {J.}~\bibnamefont {Cordova}}, \bibinfo
  {author} {\bibfnamefont {G.}~\bibnamefont {Crossman}}, \bibinfo {author}
  {\bibfnamefont {M.}~\bibnamefont {Curtis}}, \bibinfo {author} {\bibfnamefont
  {S.}~\bibnamefont {Deshpande}}, \bibinfo {author} {\bibfnamefont {T.~E.}\
  \bibnamefont {Bouayadi}}, \bibinfo {author} {\bibfnamefont {D.}~\bibnamefont
  {Girshovich}}, \bibinfo {author} {\bibfnamefont {S.}~\bibnamefont {Hong}},
  \bibinfo {author} {\bibfnamefont {K.}~\bibnamefont {Kuang}}, \bibinfo
  {author} {\bibfnamefont {M.}~\bibnamefont {Lenihan}}, \bibinfo {author}
  {\bibfnamefont {T.}~\bibnamefont {Manning}}, \bibinfo {author} {\bibfnamefont
  {A.}~\bibnamefont {Marchenkov}}, \bibinfo {author} {\bibfnamefont
  {J.}~\bibnamefont {Marshall}}, \bibinfo {author} {\bibfnamefont
  {R.}~\bibnamefont {Maydra}}, \bibinfo {author} {\bibfnamefont
  {Y.}~\bibnamefont {Mohan}}, \bibinfo {author} {\bibfnamefont
  {W.}~\bibnamefont {O'Brien}}, \bibinfo {author} {\bibfnamefont
  {C.}~\bibnamefont {Osborn}}, \bibinfo {author} {\bibfnamefont
  {J.}~\bibnamefont {Otterbach}}, \bibinfo {author} {\bibfnamefont
  {A.}~\bibnamefont {Papageorge}}, \bibinfo {author} {\bibfnamefont {J.-P.}\
  \bibnamefont {Paquette}}, \bibinfo {author} {\bibfnamefont {M.}~\bibnamefont
  {Pelstring}}, \bibinfo {author} {\bibfnamefont {A.}~\bibnamefont
  {Polloreno}}, \bibinfo {author} {\bibfnamefont {G.}~\bibnamefont
  {Prawiroatmodjo}}, \bibinfo {author} {\bibfnamefont {V.}~\bibnamefont
  {Rawat}}, \bibinfo {author} {\bibfnamefont {M.}~\bibnamefont {Reagor}},
  \bibinfo {author} {\bibfnamefont {R.}~\bibnamefont {Renzas}}, \bibinfo
  {author} {\bibfnamefont {N.}~\bibnamefont {Rubin}}, \bibinfo {author}
  {\bibfnamefont {D.}~\bibnamefont {Russell}}, \bibinfo {author} {\bibfnamefont
  {M.}~\bibnamefont {Rust}}, \bibinfo {author} {\bibfnamefont {D.}~\bibnamefont
  {Scarabelli}}, \bibinfo {author} {\bibfnamefont {M.}~\bibnamefont {Scheer}},
  \bibinfo {author} {\bibfnamefont {M.}~\bibnamefont {Selvanayagam}}, \bibinfo
  {author} {\bibfnamefont {R.}~\bibnamefont {Smith}}, \bibinfo {author}
  {\bibfnamefont {A.}~\bibnamefont {Staley}}, \bibinfo {author} {\bibfnamefont
  {M.}~\bibnamefont {Suska}}, \bibinfo {author} {\bibfnamefont
  {N.}~\bibnamefont {Tezak}}, \bibinfo {author} {\bibfnamefont {D.~C.}\
  \bibnamefont {Thompson}}, \bibinfo {author} {\bibfnamefont {T.-W.}\
  \bibnamefont {To}}, \bibinfo {author} {\bibfnamefont {M.}~\bibnamefont
  {Vahidpour}}, \bibinfo {author} {\bibfnamefont {N.}~\bibnamefont
  {Vodrahalli}}, \bibinfo {author} {\bibfnamefont {T.}~\bibnamefont {Whyland}},
  \bibinfo {author} {\bibfnamefont {K.}~\bibnamefont {Yadav}}, \bibinfo
  {author} {\bibfnamefont {W.}~\bibnamefont {Zeng}},\ and\ \bibinfo {author}
  {\bibfnamefont {C.}~\bibnamefont {Rigetti}},\ }\bibfield  {title} {\bibinfo
  {title} {Parametrically activated entangling gates using transmon qubits},\
  }\href {https://doi.org/10.1103/PhysRevApplied.10.034050} {\bibfield
  {journal} {\bibinfo  {journal} {Phys. Rev. Applied}\ }\textbf {\bibinfo
  {volume} {10}},\ \bibinfo {pages} {034050} (\bibinfo {year}
  {2018})}\BibitemShut {NoStop}%
\bibitem [{\citenamefont {Samuel}\ and\ \citenamefont
  {Bhandari}(1988)}]{PhysRevLett.60.2339}%
  \BibitemOpen
  \bibfield  {author} {\bibinfo {author} {\bibfnamefont {J.}~\bibnamefont
  {Samuel}}\ and\ \bibinfo {author} {\bibfnamefont {R.}~\bibnamefont
  {Bhandari}},\ }\bibfield  {title} {\bibinfo {title} {General setting for
  berry's phase},\ }\href {https://doi.org/10.1103/PhysRevLett.60.2339}
  {\bibfield  {journal} {\bibinfo  {journal} {Phys. Rev. Lett.}\ }\textbf
  {\bibinfo {volume} {60}},\ \bibinfo {pages} {2339} (\bibinfo {year}
  {1988})}\BibitemShut {NoStop}%
\bibitem [{\citenamefont {Korotkov}(2013)}]{korotkov2013error}%
  \BibitemOpen
  \bibfield  {author} {\bibinfo {author} {\bibfnamefont {A.~N.}\ \bibnamefont
  {Korotkov}},\ }\bibfield  {title} {\bibinfo {title} {Error matrices in
  quantum process tomography},\ }\bibfield  {journal} {\bibinfo  {journal}
  {arXiv preprint arXiv:1309.6405}\ }\href
  {https://doi.org/10.48550/ARXIV.1309.6405} {10.48550/ARXIV.1309.6405}
  (\bibinfo {year} {2013})\BibitemShut {NoStop}%
\bibitem [{\citenamefont {Wood}\ and\ \citenamefont
  {Gambetta}(2018)}]{PhysRevA.97.032306}%
  \BibitemOpen
  \bibfield  {author} {\bibinfo {author} {\bibfnamefont {C.~J.}\ \bibnamefont
  {Wood}}\ and\ \bibinfo {author} {\bibfnamefont {J.~M.}\ \bibnamefont
  {Gambetta}},\ }\bibfield  {title} {\bibinfo {title} {Quantification and
  characterization of leakage errors},\ }\href
  {https://doi.org/10.1103/PhysRevA.97.032306} {\bibfield  {journal} {\bibinfo
  {journal} {Phys. Rev. A}\ }\textbf {\bibinfo {volume} {97}},\ \bibinfo
  {pages} {032306} (\bibinfo {year} {2018})}\BibitemShut {NoStop}%
\bibitem [{\citenamefont {Sung}\ \emph {et~al.}(2021)\citenamefont {Sung},
  \citenamefont {Ding}, \citenamefont {Braum\"uller}, \citenamefont
  {Veps\"al\"ainen}, \citenamefont {Kannan}, \citenamefont {Kjaergaard},
  \citenamefont {Greene}, \citenamefont {Samach}, \citenamefont {McNally},
  \citenamefont {Kim}, \citenamefont {Melville}, \citenamefont {Niedzielski},
  \citenamefont {Schwartz}, \citenamefont {Yoder}, \citenamefont {Orlando},
  \citenamefont {Gustavsson},\ and\ \citenamefont
  {Oliver}}]{PhysRevX.11.021058}%
  \BibitemOpen
  \bibfield  {author} {\bibinfo {author} {\bibfnamefont {Y.}~\bibnamefont
  {Sung}}, \bibinfo {author} {\bibfnamefont {L.}~\bibnamefont {Ding}}, \bibinfo
  {author} {\bibfnamefont {J.}~\bibnamefont {Braum\"uller}}, \bibinfo {author}
  {\bibfnamefont {A.}~\bibnamefont {Veps\"al\"ainen}}, \bibinfo {author}
  {\bibfnamefont {B.}~\bibnamefont {Kannan}}, \bibinfo {author} {\bibfnamefont
  {M.}~\bibnamefont {Kjaergaard}}, \bibinfo {author} {\bibfnamefont
  {A.}~\bibnamefont {Greene}}, \bibinfo {author} {\bibfnamefont {G.~O.}\
  \bibnamefont {Samach}}, \bibinfo {author} {\bibfnamefont {C.}~\bibnamefont
  {McNally}}, \bibinfo {author} {\bibfnamefont {D.}~\bibnamefont {Kim}},
  \bibinfo {author} {\bibfnamefont {A.}~\bibnamefont {Melville}}, \bibinfo
  {author} {\bibfnamefont {B.~M.}\ \bibnamefont {Niedzielski}}, \bibinfo
  {author} {\bibfnamefont {M.~E.}\ \bibnamefont {Schwartz}}, \bibinfo {author}
  {\bibfnamefont {J.~L.}\ \bibnamefont {Yoder}}, \bibinfo {author}
  {\bibfnamefont {T.~P.}\ \bibnamefont {Orlando}}, \bibinfo {author}
  {\bibfnamefont {S.}~\bibnamefont {Gustavsson}},\ and\ \bibinfo {author}
  {\bibfnamefont {W.~D.}\ \bibnamefont {Oliver}},\ }\bibfield  {title}
  {\bibinfo {title} {Realization of high-fidelity cz and $zz$-free iswap gates
  with a tunable coupler},\ }\href {https://doi.org/10.1103/PhysRevX.11.021058}
  {\bibfield  {journal} {\bibinfo  {journal} {Phys. Rev. X}\ }\textbf {\bibinfo
  {volume} {11}},\ \bibinfo {pages} {021058} (\bibinfo {year}
  {2021})}\BibitemShut {NoStop}%
\bibitem [{\citenamefont {Wallman}\ \emph {et~al.}(2015)\citenamefont
  {Wallman}, \citenamefont {Granade}, \citenamefont {Harper},\ and\
  \citenamefont {Flammia}}]{Wallman_2015}%
  \BibitemOpen
  \bibfield  {author} {\bibinfo {author} {\bibfnamefont {J.}~\bibnamefont
  {Wallman}}, \bibinfo {author} {\bibfnamefont {C.}~\bibnamefont {Granade}},
  \bibinfo {author} {\bibfnamefont {R.}~\bibnamefont {Harper}},\ and\ \bibinfo
  {author} {\bibfnamefont {S.~T.}\ \bibnamefont {Flammia}},\ }\bibfield
  {title} {\bibinfo {title} {Estimating the coherence of noise},\ }\href
  {https://doi.org/10.1088/1367-2630/17/11/113020} {\bibfield  {journal}
  {\bibinfo  {journal} {New Journal of Physics}\ }\textbf {\bibinfo {volume}
  {17}},\ \bibinfo {pages} {113020} (\bibinfo {year} {2015})}\BibitemShut
  {NoStop}%
\bibitem [{\citenamefont {Yan}\ \emph {et~al.}(2018)\citenamefont {Yan},
  \citenamefont {Krantz}, \citenamefont {Sung}, \citenamefont {Kjaergaard},
  \citenamefont {Campbell}, \citenamefont {Orlando}, \citenamefont
  {Gustavsson},\ and\ \citenamefont {Oliver}}]{PhysRevApplied.10.054062}%
  \BibitemOpen
  \bibfield  {author} {\bibinfo {author} {\bibfnamefont {F.}~\bibnamefont
  {Yan}}, \bibinfo {author} {\bibfnamefont {P.}~\bibnamefont {Krantz}},
  \bibinfo {author} {\bibfnamefont {Y.}~\bibnamefont {Sung}}, \bibinfo {author}
  {\bibfnamefont {M.}~\bibnamefont {Kjaergaard}}, \bibinfo {author}
  {\bibfnamefont {D.~L.}\ \bibnamefont {Campbell}}, \bibinfo {author}
  {\bibfnamefont {T.~P.}\ \bibnamefont {Orlando}}, \bibinfo {author}
  {\bibfnamefont {S.}~\bibnamefont {Gustavsson}},\ and\ \bibinfo {author}
  {\bibfnamefont {W.~D.}\ \bibnamefont {Oliver}},\ }\bibfield  {title}
  {\bibinfo {title} {Tunable coupling scheme for implementing high-fidelity
  two-qubit gates},\ }\href {https://doi.org/10.1103/PhysRevApplied.10.054062}
  {\bibfield  {journal} {\bibinfo  {journal} {Phys. Rev. Appl.}\ }\textbf
  {\bibinfo {volume} {10}},\ \bibinfo {pages} {054062} (\bibinfo {year}
  {2018})}\BibitemShut {NoStop}%
\bibitem [{\citenamefont {Bruzewicz}\ \emph {et~al.}(2019)\citenamefont
  {Bruzewicz}, \citenamefont {Chiaverini}, \citenamefont {McConnell},\ and\
  \citenamefont {Sage}}]{doi:10.1063/1.5088164}%
  \BibitemOpen
  \bibfield  {author} {\bibinfo {author} {\bibfnamefont {C.~D.}\ \bibnamefont
  {Bruzewicz}}, \bibinfo {author} {\bibfnamefont {J.}~\bibnamefont
  {Chiaverini}}, \bibinfo {author} {\bibfnamefont {R.}~\bibnamefont
  {McConnell}},\ and\ \bibinfo {author} {\bibfnamefont {J.~M.}\ \bibnamefont
  {Sage}},\ }\bibfield  {title} {\bibinfo {title} {Trapped-ion quantum
  computing: Progress and challenges},\ }\href
  {https://doi.org/10.1063/1.5088164} {\bibfield  {journal} {\bibinfo
  {journal} {Applied Physics Reviews}\ }\textbf {\bibinfo {volume} {6}},\
  \bibinfo {pages} {021314} (\bibinfo {year} {2019})}\BibitemShut {NoStop}%
\bibitem [{\citenamefont {Zhang}\ \emph {et~al.}(2018)\citenamefont {Zhang},
  \citenamefont {Li}, \citenamefont {Cao}, \citenamefont {Xiao}, \citenamefont
  {Guo},\ and\ \citenamefont {Guo}}]{10.1093/nsr/nwy153}%
  \BibitemOpen
  \bibfield  {author} {\bibinfo {author} {\bibfnamefont {X.}~\bibnamefont
  {Zhang}}, \bibinfo {author} {\bibfnamefont {H.-O.}\ \bibnamefont {Li}},
  \bibinfo {author} {\bibfnamefont {G.}~\bibnamefont {Cao}}, \bibinfo {author}
  {\bibfnamefont {M.}~\bibnamefont {Xiao}}, \bibinfo {author} {\bibfnamefont
  {G.-C.}\ \bibnamefont {Guo}},\ and\ \bibinfo {author} {\bibfnamefont {G.-P.}\
  \bibnamefont {Guo}},\ }\bibfield  {title} {\bibinfo {title} {{Semiconductor
  quantum computation}},\ }\href {https://doi.org/10.1093/nsr/nwy153}
  {\bibfield  {journal} {\bibinfo  {journal} {National Science Review}\
  }\textbf {\bibinfo {volume} {6}},\ \bibinfo {pages} {32} (\bibinfo {year}
  {2018})}\BibitemShut {NoStop}%
\bibitem [{Ori()}]{OriginAIO}%
  \BibitemOpen
  \href@noop {} {\bibinfo {title} {Origin quantum inc., quantum computer
  control system}},\ \bibinfo {howpublished}
  {\url{https://qcloud.originqc.com.cn/en/product/chipEquipment/15}}\BibitemShut
  {NoStop}%
\bibitem [{\citenamefont {Duan}\ \emph
  {et~al.}(2021{\natexlab{b}})\citenamefont {Duan}, \citenamefont {Jia},
  \citenamefont {Zhang}, \citenamefont {Du}, \citenamefont {Tao}, \citenamefont
  {Yang}, \citenamefont {Guo}, \citenamefont {Chen}, \citenamefont {Zhang},
  \citenamefont {Peng}, \citenamefont {Kong}, \citenamefont {Li}, \citenamefont
  {Cao},\ and\ \citenamefont {Guo}}]{Duan_2021}%
  \BibitemOpen
  \bibfield  {author} {\bibinfo {author} {\bibfnamefont {P.}~\bibnamefont
  {Duan}}, \bibinfo {author} {\bibfnamefont {Z.}~\bibnamefont {Jia}}, \bibinfo
  {author} {\bibfnamefont {C.}~\bibnamefont {Zhang}}, \bibinfo {author}
  {\bibfnamefont {L.}~\bibnamefont {Du}}, \bibinfo {author} {\bibfnamefont
  {H.}~\bibnamefont {Tao}}, \bibinfo {author} {\bibfnamefont {X.}~\bibnamefont
  {Yang}}, \bibinfo {author} {\bibfnamefont {L.}~\bibnamefont {Guo}}, \bibinfo
  {author} {\bibfnamefont {Y.}~\bibnamefont {Chen}}, \bibinfo {author}
  {\bibfnamefont {H.}~\bibnamefont {Zhang}}, \bibinfo {author} {\bibfnamefont
  {Z.}~\bibnamefont {Peng}}, \bibinfo {author} {\bibfnamefont {W.}~\bibnamefont
  {Kong}}, \bibinfo {author} {\bibfnamefont {H.-O.}\ \bibnamefont {Li}},
  \bibinfo {author} {\bibfnamefont {G.}~\bibnamefont {Cao}},\ and\ \bibinfo
  {author} {\bibfnamefont {G.-P.}\ \bibnamefont {Guo}},\ }\bibfield  {title}
  {\bibinfo {title} {Broadband flux-pumped josephson parametric amplifier with
  an on-chip coplanar waveguide impedance transformer},\ }\href
  {https://doi.org/10.35848/1882-0786/abf029} {\bibfield  {journal} {\bibinfo
  {journal} {Applied Physics Express}\ }\textbf {\bibinfo {volume} {14}},\
  \bibinfo {pages} {042011} (\bibinfo {year} {2021}{\natexlab{b}})}\BibitemShut
  {NoStop}%
\bibitem [{\citenamefont {Johansson}\ \emph
  {et~al.}(2012{\natexlab{b}})\citenamefont {Johansson}, \citenamefont
  {Nation},\ and\ \citenamefont {Nori}}]{JOHANSSON20121760}%
  \BibitemOpen
  \bibfield  {author} {\bibinfo {author} {\bibfnamefont {J.}~\bibnamefont
  {Johansson}}, \bibinfo {author} {\bibfnamefont {P.}~\bibnamefont {Nation}},\
  and\ \bibinfo {author} {\bibfnamefont {F.}~\bibnamefont {Nori}},\ }\bibfield
  {title} {\bibinfo {title} {Qutip: An open-source python framework for the
  dynamics of open quantum systems},\ }\href
  {https://doi.org/https://doi.org/10.1016/j.cpc.2012.02.021} {\bibfield
  {journal} {\bibinfo  {journal} {Computer Physics Communications}\ }\textbf
  {\bibinfo {volume} {183}},\ \bibinfo {pages} {1760} (\bibinfo {year}
  {2012}{\natexlab{b}})}\BibitemShut {NoStop}%
\bibitem [{\citenamefont {Johansson}\ \emph {et~al.}(2013)\citenamefont
  {Johansson}, \citenamefont {Nation},\ and\ \citenamefont
  {Nori}}]{JOHANSSON20131234}%
  \BibitemOpen
  \bibfield  {author} {\bibinfo {author} {\bibfnamefont {J.}~\bibnamefont
  {Johansson}}, \bibinfo {author} {\bibfnamefont {P.}~\bibnamefont {Nation}},\
  and\ \bibinfo {author} {\bibfnamefont {F.}~\bibnamefont {Nori}},\ }\bibfield
  {title} {\bibinfo {title} {Qutip 2: A python framework for the dynamics of
  open quantum systems},\ }\href
  {https://doi.org/https://doi.org/10.1016/j.cpc.2012.11.019} {\bibfield
  {journal} {\bibinfo  {journal} {Computer Physics Communications}\ }\textbf
  {\bibinfo {volume} {184}},\ \bibinfo {pages} {1234} (\bibinfo {year}
  {2013})}\BibitemShut {NoStop}%
\bibitem [{\citenamefont {McKay}\ \emph {et~al.}(2016)\citenamefont {McKay},
  \citenamefont {Filipp}, \citenamefont {Mezzacapo}, \citenamefont {Magesan},
  \citenamefont {Chow},\ and\ \citenamefont
  {Gambetta}}]{PhysRevApplied.6.064007}%
  \BibitemOpen
  \bibfield  {author} {\bibinfo {author} {\bibfnamefont {D.~C.}\ \bibnamefont
  {McKay}}, \bibinfo {author} {\bibfnamefont {S.}~\bibnamefont {Filipp}},
  \bibinfo {author} {\bibfnamefont {A.}~\bibnamefont {Mezzacapo}}, \bibinfo
  {author} {\bibfnamefont {E.}~\bibnamefont {Magesan}}, \bibinfo {author}
  {\bibfnamefont {J.~M.}\ \bibnamefont {Chow}},\ and\ \bibinfo {author}
  {\bibfnamefont {J.~M.}\ \bibnamefont {Gambetta}},\ }\bibfield  {title}
  {\bibinfo {title} {Universal gate for fixed-frequency qubits via a tunable
  bus},\ }\href {https://doi.org/10.1103/PhysRevApplied.6.064007} {\bibfield
  {journal} {\bibinfo  {journal} {Phys. Rev. Applied}\ }\textbf {\bibinfo
  {volume} {6}},\ \bibinfo {pages} {064007} (\bibinfo {year}
  {2016})}\BibitemShut {NoStop}%
\bibitem [{\citenamefont {Zheng}\ \emph {et~al.}(2016)\citenamefont {Zheng},
  \citenamefont {Yang},\ and\ \citenamefont {Nori}}]{PhysRevA.93.032313}%
  \BibitemOpen
  \bibfield  {author} {\bibinfo {author} {\bibfnamefont {S.-B.}\ \bibnamefont
  {Zheng}}, \bibinfo {author} {\bibfnamefont {C.-P.}\ \bibnamefont {Yang}},\
  and\ \bibinfo {author} {\bibfnamefont {F.}~\bibnamefont {Nori}},\ }\bibfield
  {title} {\bibinfo {title} {Comparison of the sensitivity to systematic errors
  between nonadiabatic non-abelian geometric gates and their dynamical
  counterparts},\ }\href {https://doi.org/10.1103/PhysRevA.93.032313}
  {\bibfield  {journal} {\bibinfo  {journal} {Phys. Rev. A}\ }\textbf {\bibinfo
  {volume} {93}},\ \bibinfo {pages} {032313} (\bibinfo {year}
  {2016})}\BibitemShut {NoStop}%
\bibitem [{\citenamefont {Jing}\ \emph {et~al.}(2017)\citenamefont {Jing},
  \citenamefont {Lam},\ and\ \citenamefont {Wu}}]{PhysRevA.95.012334}%
  \BibitemOpen
  \bibfield  {author} {\bibinfo {author} {\bibfnamefont {J.}~\bibnamefont
  {Jing}}, \bibinfo {author} {\bibfnamefont {C.-H.}\ \bibnamefont {Lam}},\ and\
  \bibinfo {author} {\bibfnamefont {L.-A.}\ \bibnamefont {Wu}},\ }\bibfield
  {title} {\bibinfo {title} {Non-abelian holonomic transformation in the
  presence of classical noise},\ }\href
  {https://doi.org/10.1103/PhysRevA.95.012334} {\bibfield  {journal} {\bibinfo
  {journal} {Phys. Rev. A}\ }\textbf {\bibinfo {volume} {95}},\ \bibinfo
  {pages} {012334} (\bibinfo {year} {2017})}\BibitemShut {NoStop}%
\bibitem [{\citenamefont {Liu}\ \emph {et~al.}(2019)\citenamefont {Liu},
  \citenamefont {Song}, \citenamefont {Xue}, \citenamefont {Wang},\ and\
  \citenamefont {Yung}}]{PhysRevLett.123.100501}%
  \BibitemOpen
  \bibfield  {author} {\bibinfo {author} {\bibfnamefont {B.-J.}\ \bibnamefont
  {Liu}}, \bibinfo {author} {\bibfnamefont {X.-K.}\ \bibnamefont {Song}},
  \bibinfo {author} {\bibfnamefont {Z.-Y.}\ \bibnamefont {Xue}}, \bibinfo
  {author} {\bibfnamefont {X.}~\bibnamefont {Wang}},\ and\ \bibinfo {author}
  {\bibfnamefont {M.-H.}\ \bibnamefont {Yung}},\ }\bibfield  {title} {\bibinfo
  {title} {Plug-and-play approach to nonadiabatic geometric quantum gates},\
  }\href {https://doi.org/10.1103/PhysRevLett.123.100501} {\bibfield  {journal}
  {\bibinfo  {journal} {Phys. Rev. Lett.}\ }\textbf {\bibinfo {volume} {123}},\
  \bibinfo {pages} {100501} (\bibinfo {year} {2019})}\BibitemShut {NoStop}%
\bibitem [{\citenamefont {Li}\ \emph {et~al.}(2022)\citenamefont {Li},
  \citenamefont {Dong}, \citenamefont {Zheng}, \citenamefont {Zhang},
  \citenamefont {Ma}, \citenamefont {Liu}, \citenamefont {Wang}, \citenamefont
  {Li}, \citenamefont {Liu}, \citenamefont {Zhao}, \citenamefont {Lan},
  \citenamefont {Li}, \citenamefont {Tan},\ and\ \citenamefont
  {Yu}}]{https://doi.org/10.1002/pssb.202200040}%
  \BibitemOpen
  \bibfield  {author} {\bibinfo {author} {\bibfnamefont {Y.}~\bibnamefont
  {Li}}, \bibinfo {author} {\bibfnamefont {Y.}~\bibnamefont {Dong}}, \bibinfo
  {author} {\bibfnamefont {W.}~\bibnamefont {Zheng}}, \bibinfo {author}
  {\bibfnamefont {Y.}~\bibnamefont {Zhang}}, \bibinfo {author} {\bibfnamefont
  {Z.}~\bibnamefont {Ma}}, \bibinfo {author} {\bibfnamefont {Q.}~\bibnamefont
  {Liu}}, \bibinfo {author} {\bibfnamefont {J.}~\bibnamefont {Wang}}, \bibinfo
  {author} {\bibfnamefont {Y.}~\bibnamefont {Li}}, \bibinfo {author}
  {\bibfnamefont {Y.}~\bibnamefont {Liu}}, \bibinfo {author} {\bibfnamefont
  {J.}~\bibnamefont {Zhao}}, \bibinfo {author} {\bibfnamefont {D.}~\bibnamefont
  {Lan}}, \bibinfo {author} {\bibfnamefont {S.}~\bibnamefont {Li}}, \bibinfo
  {author} {\bibfnamefont {X.}~\bibnamefont {Tan}},\ and\ \bibinfo {author}
  {\bibfnamefont {Y.}~\bibnamefont {Yu}},\ }\bibfield  {title} {\bibinfo
  {title} {Nonadiabatic geometric gates with a shortened loop in a
  superconducting circuit},\ }\href
  {https://doi.org/https://doi.org/10.1002/pssb.202200040} {\bibfield
  {journal} {\bibinfo  {journal} {physica status solidi (b)}\ }\textbf
  {\bibinfo {volume} {259}},\ \bibinfo {pages} {2200040} (\bibinfo {year}
  {2022})}\BibitemShut {NoStop}%
\bibitem [{\citenamefont {Chen}\ \emph {et~al.}(2020)\citenamefont {Chen},
  \citenamefont {Shen},\ and\ \citenamefont {Xue}}]{PhysRevApplied.14.034038}%
  \BibitemOpen
  \bibfield  {author} {\bibinfo {author} {\bibfnamefont {T.}~\bibnamefont
  {Chen}}, \bibinfo {author} {\bibfnamefont {P.}~\bibnamefont {Shen}},\ and\
  \bibinfo {author} {\bibfnamefont {Z.-Y.}\ \bibnamefont {Xue}},\ }\bibfield
  {title} {\bibinfo {title} {Robust and fast holonomic quantum gates with
  encoding on superconducting circuits},\ }\href
  {https://doi.org/10.1103/PhysRevApplied.14.034038} {\bibfield  {journal}
  {\bibinfo  {journal} {Phys. Rev. Applied}\ }\textbf {\bibinfo {volume}
  {14}},\ \bibinfo {pages} {034038} (\bibinfo {year} {2020})}\BibitemShut
  {NoStop}%
\end{thebibliography}%

\end{document}